\documentclass[11pt,a4paper]{article}
\pdfoutput=1
\usepackage[top=25mm,bottom=30mm,left=29mm,right=29mm]{geometry}
\usepackage{tocloft}

\usepackage[nottoc,notlot,notlof]{tocbibind}

\usepackage[pdfencoding=auto]{hyperref}
\hypersetup{
    colorlinks,
    linkcolor={red!50!black},
    citecolor={blue!50!black},
    urlcolor={blue!80!black},
    breaklinks=true 
}

\usepackage[utf8]{inputenc}

\usepackage{amssymb,amsmath,amsthm}
\usepackage{enumerate}

\usepackage[export]{adjustbox}

\usepackage{subcaption}
\usepackage{graphicx}
\usepackage{cite}

\usepackage{mathtools}
\usepackage{xcolor}
\usepackage{tikz}
\numberwithin{equation}{section}

\def\ket#1{|{#1}\rangle}

\newcommand{\bndstate}[1]{|\hspace{-1pt}\ket{#1}\!\rangle}

\def\Fsym#1#2#3#4#5#6{F_{#1#2}\!\begin{bmatrix}#3 & #4 \\ #5 & #6 \end{bmatrix}}

\newcommand{\one}{\mathbf{1}} 

\begin{document} 
\thispagestyle{empty}
\phantom{.}
\vspace{.7cm}
\begin{center}{\Large \textbf{
Lattice models from CFT on surfaces with holes II:\\[.5em]
Cloaking boundary conditions and loop models}}
\end{center}

\begin{center}
Enrico M. Brehm\textsuperscript{$\beta$},
Ingo Runkel\textsuperscript{$\mu$}
\end{center}

\begin{center}
	{ ${}^{\beta}$} CQTA, Deutsches Elektronen-Synchrotron DESY\\
    Platanenallee 6\\
    15738 Zeuthen, Germany\\[.5em]
{\small \sf enrico.brehm@desy.de}
\\[1em]
	{${}^{\mu}$} 
Fachbereich Mathematik,\\
Universität Hamburg,\\
Bundesstraße 55,\\
20146 Hamburg, Germany\\[.5em]
{\small \sf Ingo.Runkel@uni-hamburg.de}
\end{center}


\section*{Abstract}
In this paper we continue to investigate the lattice models obtained from 2d\,CFTs via the construction introduced in \cite{Brehm:2021wev}. 
On the side of the 2d\,CFT we consider the cloaking boundary condition relative to a fixed
fusion category $\mathcal{F}$ of topological line defects. The resulting lattice model realises the topological symmetry $\mathcal{F}$ exactly. We compute the state spaces and Boltzmann weights of these lattice model in the example of unitary Virasoro minimal models. We work directly with amplitudes, rather than with normalised correlators, and we provide a careful treatment of the Weyl anomaly factor in terms of the Liouville action. 

We numerically evaluate the Ising CFT on the torus with one hole and cloaking boundary condition in two channels, and illustrate in this example that the anomaly factors are essential to obtain matching results for the amplitudes. We show that lattice models obtained from Virasoro minimal models at lowest non-trivial cutoff can be exactly mapped to loop models. This provides a first non-trivial check that our lattice models can contain the 2d\,CFT they were constructed from in their phase diagram, and we propose a condition on the cloaking boundary condition for $\mathcal{F}$ under which we expect this to happen in general.

\newpage
\setcounter{tocdepth}{2}
\tableofcontents

\newpage

\section{Introduction and overview}

When studying lattice models from the point of view of statistical mechanics, one usually starts from a given microscopic description of the local degrees of freedom and their interaction. From these one tries to deduce the behaviour of the model for large sample sizes, such as phase transitions, critical exponents, and correlation functions. Continuum field theory, and in particular conformal field theory, arises as an effective description of the long-range behaviour of lattice correlators. At a critical point, the field theory is by construction scale invariant, and often conformally invariant. Conversely, a given conformal field theory can arise at critical points of many different lattice models, and in this sense describes a \emph{universality class of critical behaviour}.

One important property of a universality class given by a continuum field theory is its \emph{generalised symmetry}, by which we mean its collection of topological defects (see e.g.\ \cite{Carqueville:2023jhb} for an overview and references). In this series of papers we mostly work with two-dimensional euclidean conformal field theories $C$. The topological defects form a pivotal monoidal category $\mathcal{E}(C)$ (``endomorphisms of $C$''), see \cite{Fuchs:2002cm,Frohlich:2006ch,Davydov:2011kb,Carqueville:2012dk,Bhardwaj:2017xup,Chang:2018iay,Thorngren:2021yso}. This generalises a group symmetry in two ways: firstly, the fusion of topological defects -- described by the tensor product of $\mathcal{E}(C)$ -- need not by invertible, and secondly, $\mathcal{E}(C)$ also encodes the behaviour of topological field insertions on line defects and at defect junctions.

The symmetry $\mathcal{E}(C)$ will typically be different from that of the lattice model itself, and in general will be larger. From the point of view of universality classes, one may intuitively think of $\mathcal{E}(C)$ as containing ``all symmetries of all lattice models which realise this universality class.''
In fact, knowing $\mathcal{E}(C)$ fully is currently out of reach except for in special cases (these are Virasoro-minimal models \cite{Petkova:2000ip,Fuchs:2002cm} and unitary $c=1$ theories with discrete spectrum \cite{Fuchs:2007tx,Thorngren:2021yso}). 

A much more tractable question is whether a given monoidal category $\mathcal{M}$ is contained in $\mathcal{E}(C)$, i.e.\ if the CFT $C$ realises the generalised symmetry $\mathcal{M}$. 
The simplest form of generalised symmetry is described by so-called pivotal fusion categories $\mathcal{F}$. These are semisimple with only a finite number of simple objects, and can be thought of as a generalisation of a finite group symmetry. To find and study universality classes with a given symmetry $\mathcal{F}$, one approach is to use the data contained in $\mathcal{F}$ directly to build lattice systems which have topological line defects that realises $\mathcal{F}$ exactly \cite{PhysRevLett.98.160409,Kitaev:2011dxc,Buican:2017rxc,Vanhove:2018wlb,Aasen:2020jwb,Lootens:2020mso,Huang:2021nvb,Vanhove:2021zop}. However, this (potentially) constructs \emph{one specific} universality class realising $\mathcal{F}$ out of an infinity of possible choices.

\medskip

In this series of papers, we study the opposite approach:
\begin{center}
\fbox{~\begin{minipage}{0.96\textwidth}
\phantom{.}\\[-.2em]
From a universality class described by a 2d\,CFT $C$ and from a fusion category $\mathcal{F}$ realised by topological defects of $C$, i.e.\ $\mathcal{F} \subset \mathcal{E}(C)$, build a lattice model which realises $\mathcal{F}$ and which contains $C$ in its phase diagram.
\\[-.6em]
\end{minipage}~}
\end{center}

\begin{figure}
    \centering
    \includegraphics[width=\textwidth]{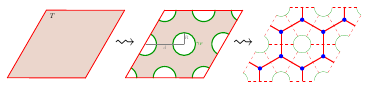}
    \caption{Constructing the lattice model from a CFT: cut a regular lattice of holes into the torus $T$ (a triangular lattice in this example), with distance $d$ and radius $R$; perform a sum over intermediate boundary states at the dashed lines; treat a (truncated) basis of boundary states as states of the lattice model assigned to the edges of the dual lattice (a hexagonal lattice in this example), and the disc amplitudes arising after cutting along the dashed lines as Boltzmann weights assigned to the vertices of the dual lattice.}
    \label{fig:intro-lattice-construction}
\end{figure}

The idea of the construction is very simple (Figure~\ref{fig:intro-lattice-construction}). Fix a CFT $C$ which has discrete spectrum and which for the sake of the introduction we assume to be unitary. Starting from a two-dimensional torus $T$, one cuts a lattice of circular holes into $T$ -- this introduces two parameters, the distance $d$ of the hole centres, and the radius $R$ of each hole. 
Then on fixes a particular conformal boundary condition $\gamma_\mathcal{F}(\delta)$ for each of these new circular boundaries which we call an $\mathcal{F}$-cloaking boundary condition. Next one performs a sum over intermediate boundary states at the intervals indicated by the dashed lines.
Finally, one chooses a cutoff $h_\text{max}$ for the boundary spectrum.
The resulting lattice model has states associated to its edges given by a basis of the boundary state space on $\gamma_\mathcal{F}(\delta)$ of conformal weight $\le h_\text{max}$. The Boltzmann weights are associated to the vertices. Altogether, we obtain the lattice model partition function
\begin{equation}
  Z^{C,\mathcal{F}}_{d,R,h_{\text{max}}}(M,N) \ , 
\end{equation}
where $M,N$ denote the dimensions of the lattice. This family of lattice models has two important properties:
\begin{enumerate}
    \item Thanks to the choice of boundary condition $\gamma_\mathcal{F}(\delta)$, the lattice model realises the topological symmetry $\mathcal{F}$ exactly for all choices of $d,R,h_{\text{max}}$.

    \item For fixed $M,N$, taking first the limit $h_{\text{max}} \to \infty$ and then the limit $R \to 0$, one recovers the partition function of the CFT $C$ on $T$ exactly.
\end{enumerate}
We will explain how these properties arise when we review the construction in more detail and in greater generality in Section~\ref{sec:lattice-construction}. In particular, we will allow for more general quantum field theories and for more general geometries than a triangular lattice on the torus. Related constructions have recently been considered in \cite{Sopenko:2023utk,Cheng:2023kxh,Hung:2024gma}.

\smallskip
The limit in property 2 is different from the continuum limit of the lattice model, where $h_\text{max}, R, d$ are fixed but $M,N \to \infty$. We propose a condition for when our lattice model reproduces the original CFT $C$ for a fixed but large enough $h_\text{max}$, and for some finite range of $R$ near $R=0$. The condition is that opening a lattice of small holes amounts to an irrelevant perturbation of the CFT, i.e.\ that the universality class is stable under adding this lattice of holes. In terms of boundary states, this can be phrased as:
\begin{center}
\fbox{~\begin{minipage}{0.96\textwidth}
\phantom{.}\\[-.2em]
Proposed stability condition:
\\
If the boundary state $\gamma_\mathcal{F}(\delta)$ of the cloaking boundary condition for $\mathcal{F}$ is a sum of irrelevant bulk fields of $C$, then there is a fixed $h_\text{max} \gg 0$ and some finite range $R \in (0,\varepsilon)$ such that the universality class of the corresponding lattice model is $C$.
\\[-.6em]
\end{minipage}~}
\end{center}
Note that this does not mean that the lattice model is critical for all hole radii $R \in (0,\epsilon)$, but that its long-distance behaviour is given by $C$ for these values, i.e.\ that the infrared fixed point of the renormalisation group flow is $C$.
We discuss the stability condition in more detail in Section~\ref{sec:CFT-hexagonal} and look at examples in Section~\ref{sec:cloaking-bcs}. Roughly speaking, making $\mathcal{F}$ larger reduces the number of primary bulk fields contributing to the boundary state $\gamma_\mathcal{F}(\delta)$. Thus for a given $C$ one has to choose $\mathcal{F}$ ``large enough'' to satisfy the stability condition.

\medskip

In \cite{Brehm:2021wev}, the first paper of this series, we introduced the discretisation idea outlined above, and worked out the necessary uniformisation maps to express the Boltzmann weights in terms of boundary three-point correlators on the unit disc. 
Building on these results, in this second paper we make five main new contributions:

\smallskip

\noindent
\textbf{Section~\ref{sec:anomaly-factor}:}
We extend the formalism to allow for a direct comparison of partition functions, i.e.\ of unnormalised amplitudes, rather than just ratios. This is achieved by a detailed study of the Weyl anomaly of amplitudes with and without boundaries via the Liouville action. While this transformation behaviour is standard, we have not found a treatment as explicit and detailed as we provide here. In Appendices~\ref{app:AnomalyAction} and \ref{app:state-sum-examples} we work through several examples to illustrate the formalism.

\smallskip

\noindent
\textbf{Section~\ref{sec:cloak-minimal}:} We give explicit expressions for the Boltzmann weights of the lattice model in the case of A-series Virasoro minimal models, allowing for different choices of fusion category symmetry $\mathcal{F}$. The answer is expressed in terms of chiral data of the minimal model, specifically in terms of the modular $S$-matrix and the fusion matrices $F$, see \eqref{eq:Nijkabc-expression} and \eqref{eq:triangle-amplitude-cloaking}. We also describe how to correctly normalise the ON-basis of boundary (changing) fields on the $\mathcal{F}$-cloaking boundary condition in the sum over intermediate states.

\smallskip

\noindent
\textbf{Section~\ref{sec:one-hole-comparison}:}
We carry out the numerical computation of the torus partition function with one hole for the cloaking boundary condition (rather than elementary boundary conditions as in \cite{Brehm:2021wev}) and we compare the partition functions in the numerical approximation of the two channels directly (rather than just comparing ratios as in \cite{Brehm:2021wev}). 
The Weyl factors depend on $d$ and $R$ and need to be evaluated numerically, too. It is remarkable how different the two channels are from each other without including the anomaly factor, and how well they agree across the whole parameter range with the anomaly factor, even though all ingredients are evaluated numerically (Figure~\ref{fig:compare1}).

\smallskip

\noindent
\textbf{Section~\ref{sec:Ising-lattice}:} We compute the lattice model for the Ising CFT with fusion category symmetry $\mathcal{F}$ given by the $\mathbb{Z}_2$-symmetry of the Ising CFT. We use the smallest non-trivial cutoff $h_\text{max}$ and recover the Ising lattice model on a triangular lattice. The cloaking boundary state does not satisfy the stability condition, and indeed in this example there is no region near $R=0$ where the universality class of the lattice model is the Ising CFT. Instead, the Ising CFT is obtained at an unstable critical value $R_C>0$.

\smallskip

\noindent
\textbf{Section~\ref{sec:loop-models}:} We investigate the lattice model for the unitary A-type Virasoro minimal model $M(p,p+1)$ with fusion category symmetry $\mathcal{F}$ given by the first row $(1,s)$ of the Kac-table. In this case, the cloaking boundary state satisfies the stability condition.
For $h_\text{max}$ we again take the lowest non-trivial cutoff, and we show that the resulting lattice model can be mapped to a loop model on the same hexagonal lattice. The known (but largely still conjectural) phase diagram of the loop model gives a critical ratio $(R/d)_\text{critical}$ of radius and hole-centre distance, such that in the region $R/d < (R/d)_\text{critical}$ the continuum theory has the same central charge as the minimal model $M(p)$ we started from. 

\subsubsection*{Acknowledgements}

We would like to thank Andr\'e Henriques, Jesper Lykke Jacobsen, Sven Möller, and Tobias Osborne for helpful discussions.
IR is supported in part by the Deutsche Forschungsgemeinschaft (DFG, German Research Foundation) under Germany`s Excellence Strategy - EXC 2121 ``Quantum Universe" - 390833306, and via the Collaborative Research Centre CRC 1624 ``Higher structures, moduli spaces and integrability'' - 506632645.

\newpage

\section{Lattice discretisations with topological symmetry}\label{sec:lattice-construction}

In this section we present the construction introduced in \cite{Brehm:2021wev} in a more general setting. Starting from a 2d\,QFT $Q$ and a pivotal fusion category $\mathcal{F}$ of topological defect lines for $Q$, we construct a family of lattice models which realise  $\mathcal{F}$ exactly via topological defects on the lattice. We then specialise to rational conformal field theories, which are also used in the explicit examples in Sections~\ref{sec:one-hole-comparison}--\ref{sec:loop-models} later on. 

Our approach is opposite in spirit to constructions which start from $\mathcal{F}$ alone and build a lattice model realising $\mathcal{F}$ with the aim of finding interesting continuum theories, see e.g.\ \cite{PhysRevLett.98.160409,Kitaev:2011dxc,Vanhove:2018wlb,Lootens:2020mso,Aasen:2020jwb,Huang:2021nvb,Vanhove:2021zop}. A given fusion category $\mathcal{F}$ can be realised in infinitely many different QFTs, and one way to look at our construction is as an attempt to identify the additional information on top of $\mathcal{F}$ needed to recover a specific $Q$. 

However, whether or not the lattice model recovers the field theory one started from will depend on the choice of $\mathcal{F}$.
Concretely, in the case of CFTs we propose a stability condition on the symmetry $\mathcal{F}$ under which we expect the universality class of the lattice model to be the CFT it was constructed from for some finite region of hole radii near $R=0$.

\subsection{Lattice models from 2d QFTs}\label{sec:lattice-from-QFT}

Consider a euclidean 2d\,QFT $Q$ which is defined at least on flat geometries, possibly with boundaries. More assumptions on $Q$ and the types of boundary conditions we consider will be added as the construction progresses.

Topological line defects in $Q$ form a pivotal monoidal category \cite[Sec.\,2.4]{Davydov:2011kb}, see also \cite{Fuchs:2002cm,Frohlich:2006ch,Carqueville:2012dk,Bhardwaj:2017xup,Chang:2018iay,Thorngren:2021yso,Carqueville:2023jhb}.
Fix a pivotal fusion category $\mathcal{F}$. We suppose that $\mathcal{F}$ is a (not necessarily full) pivotal subcategory in the category of topological line defects of $Q$. In more physical terms, this means that the objects of $\mathcal{F}$ correspond to topological line defects for $Q$, and the morphisms of $\mathcal{F}$ to topological defect fields. However, there may be more topological defects and more topological defect fields than those indexed by $\mathcal{F}$.

\medskip

The \textbf{first ingredient} in the construction of the lattice model is the \textit{$\mathcal{F}$-cloaking defect} $\gamma_{\mathcal{F}}(\delta)$ associated to $\mathcal{F}$:
\begin{equation}\label{eq:cloaking-defect-def}
\hbox{\includegraphics[scale=0.9,valign=c]{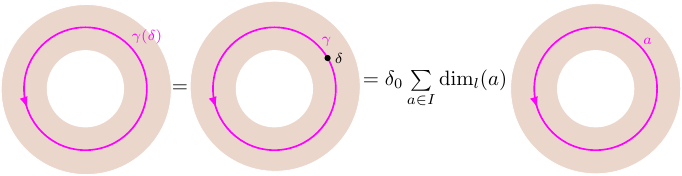}} \quad .
\end{equation}
Here, the area surrounded by the defect line can include any collection of fields, defects, or non-trivial topology of the surface. 
The equalities are understood to hold in all amplitudes of the QFT $Q$, where field and defect insertions are kept fixed outside the surface patch shown in the picture.
Thus, the notation $\gamma_{\mathcal{F}}(\delta)$ represents the superposition $\bigoplus_{a \in \mathrm{Irr}(\mathcal{F})} a$ of the topological defects corresponding to the simple objects $\mathrm{Irr}(\mathcal{F})$ of $\mathcal{F}$, and $\delta$ is the topological defect field given by $\delta_0 \sum_{a \in \mathrm{Irr}(\mathcal{F})} \dim_r(q) 1_a$. The field $1_a$ denotes the identity on the summand $a$ and $\dim_r(a)$ is the right quantum dimension of $a$. 
It does not matter where on the defect loop the field $\delta$ is inserted.
In pictures, the left and right quantum dimensions are given by
\begin{equation}
\includegraphics[scale=0.9,valign=c]{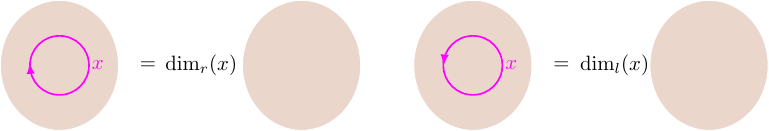} 
~~.
\end{equation}
In many examples, we have $\dim_l(a)=\dim_r(a)$ for all $a \in \mathrm{Irr}(\mathcal{F})$ -- in this case $\mathcal{F}$ is called \textit{spherical}.
The \textit{global dimension} of $\mathcal{F}$ is given by
\begin{equation}
    \mathrm{Dim}(\mathcal{F}) = 
    \sum_{a \in \mathrm{Irr}(\mathcal{F})} 
    \dim_l(a)\dim_r(a)~,
\end{equation}
which is a real number $\ge 1$ \cite[Thm.\,2.3]{Etingof:2002}.
The overall constant $\delta_0$ in \eqref{eq:cloaking-defect-def} can be chosen for convenience later in the application at hand. For example,
if we take $\delta_0 = 1/\mathrm{Dim}(\mathcal{F})$, then a contractible loop of cloaking defect with no other defects or field insertions inside it evaluates to $1$,
\begin{equation}\label{eq:cloaking-defect-normalisation}
\includegraphics[scale=1.0,valign=c]{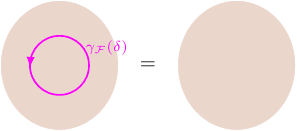}
~~.
\end{equation}
The key property of $\gamma_{\mathcal{F}}(\delta)$ is that surrounding anything in the QFT $Q$ (fields, defects, non-trivial geometry) with the cloaking defect renders it invisible to all topological defects in $\mathcal{F}$. I.e., for all $x \in \mathcal{F}$ we have
\begin{equation}\label{eq:cloaking-defect-main}
\includegraphics[scale=0.9,valign=c]{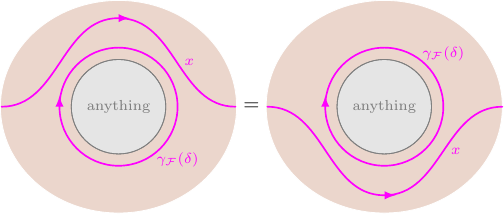}
~~.
\end{equation}
The definition of $\gamma_\mathcal{F}$ and the proof of the above equality is the same as in \cite[Sec.\,2.1]{Brehm:2021wev}, except that there all (chiral symmetry preserving) topological defects of a rational diagonal CFT were used, while here we restrict to $\mathcal{F}$ and we distinguish left and right dimension. To see that the dimension factors work out correctly also in the pivotal (but possibly non-spherical) case, compare to the corresponding computation in \cite[Eqn.\,(4.18)]{Runkel:2019vze} in the context of string-net models. 

Combining \eqref{eq:cloaking-defect-normalisation} and \eqref{eq:cloaking-defect-main} one finds that, up to an overall constant, a $\gamma_\mathcal{F}$ defect loop acts as an idempotent. Namely, fusing two parallel $\gamma_\mathcal{F}$ defect loops gives
\begin{equation}\label{eq:gammaF-idempot}
    \gamma_\mathcal{F}(\delta)\,\gamma_\mathcal{F}(\delta)
    = \delta_0 \, \mathrm{Dim}(\mathcal{F}) \, \gamma_\mathcal{F}(\delta) \ .    
\end{equation}
Accordingly, when acting on bulk fields, $\gamma_\mathcal{F}$ will project to some subspace of the space of bulk fields.

\medskip

The \textbf{second ingredient} is a boundary condition $b$ for $Q$. We only consider circular boundaries of some radius $R$, and we assume that for the given boundary condition $b$ there is a function $f(R)$ which compensates potential singularities as $R$ goes to zero, and that in this limit the leading contribution is the identity field, 
\begin{equation}\label{eq:zero-radius-limit-is-id}
\includegraphics[scale=0.9,valign=c]{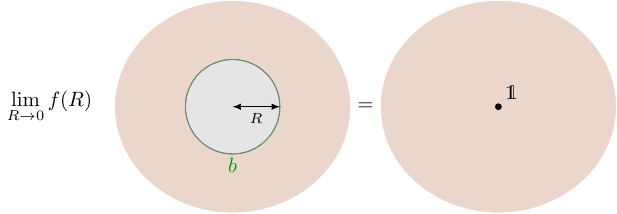}
~~.
\end{equation}
In practice, this will mean that we work with a unitary QFT with a discrete energy spectrum, so that the vacuum is the lowest energy state and will give the leading contribution, with all other contributions parametrically suppressed due to the finite energy gap over the vacuum. 

\begin{figure}[tb]
    \centering
    \begin{subfigure}{.45\textwidth}
    (a)
        \includegraphics[scale=0.9,valign=t]{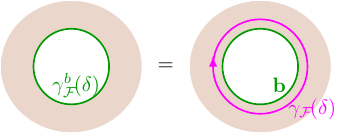}
    \end{subfigure}
\hspace{1em}
    \begin{subfigure}{.45\textwidth}
        (b)
        \includegraphics[scale=0.9,valign=t]{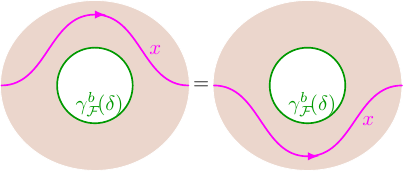}       
    \end{subfigure}
    \caption{a) Definition of the $\mathcal{F}$-cloaking boundary condition $\gamma_{\mathcal{F}}^b(\delta)$ in terms of the cloaking defect. b) The cloaking boundary condition is transparent to topological defects $x \in \mathcal{F}$.}
    \label{fig:cloaking-boundary-condition}
\end{figure}

One can now wrap the cloaking defect around a hole in the surface with boundary condition $b$ and fuse it with the boundary to obtain the \textit{cloaking boundary condition} $\gamma_{\mathcal{F}}^b(\delta)$ for $\mathcal{F}$, where $\delta$ now refers to a topological boundary field, see Figure~\ref{fig:cloaking-boundary-condition}\,a). Part b) of that figure illustrates that due to \eqref{eq:cloaking-defect-main}, holes labelled by $\gamma_{\mathcal{F}}^b(\delta)$ are transparent to all topological defects $x \in \mathcal{F}$ (but not in general to all topological defects of $Q$). If we choose $\delta_0 = 1/\mathrm{Dim}(\mathcal{F})$, then using \eqref{eq:cloaking-defect-normalisation} we see that \eqref{eq:zero-radius-limit-is-id} still holds with $b$ replaced by $\gamma_{\mathcal{F}}^b(\delta)$.

\medskip

\begin{figure}[tb]
    \centering
    \begin{subfigure}{0.45\textwidth}
    (a)
        \includegraphics[scale=0.9,valign=t]{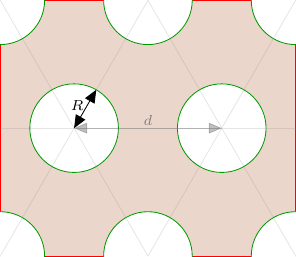}
    \end{subfigure}%
    \hspace*{\fill}   
    \begin{subfigure}{0.45\textwidth}
    (b)
        \includegraphics[scale=0.9,valign=t]{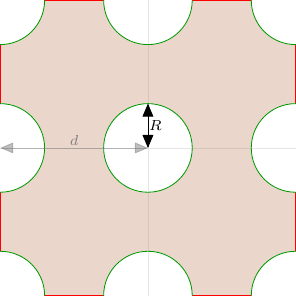}
    \end{subfigure}%
    \hspace*{\fill} 
    \\[1em]
    \hspace*{\fill}   
    \begin{subfigure}{0.45\textwidth}
    (c)
        \includegraphics[scale=1.3,valign=t]{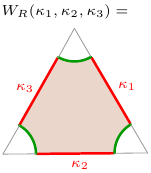}
    \end{subfigure}
    \hspace*{\fill}
    \begin{subfigure}{0.45\textwidth}
    (d)
        \includegraphics[scale=1.3,valign=t]{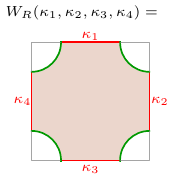}
    \end{subfigure}
    \hspace*{\fill}
    \caption{Example of regular lattices on the torus with (a) triangular cells and (b) square cells, together with holes of radius $R$ at each vertex, equipped with the cloaking boundary condition $\gamma_\mathcal{F}^b(\delta)$. Figures (c) and (d) show the corresponding elementary cells, whose amplitudes $W_{d,R}(\kappa_1,\dots)$ define the vertex weights of the lattice model.}
    \label{fig:lattice-model-from-punching-holes}
\end{figure}

For the \textbf{third ingredient}, we consider $Q$ on the torus, together with a regular lattice, for example with equilateral triangles or squares as cells as shown in Figure~\ref{fig:lattice-model-from-punching-holes}\,a),b). Let $d$ be the distance of any two vertices in these lattices. Around each vertex we punch a hole of radius $R \in (0,d/2)$ into the surface and place the cloaking boundary condition $\gamma_{\mathcal{F}}^b(\delta)$ on each boundary circle. 

Denote by $\mathcal{H}_{\gamma\gamma}(d,R)$ the space of states on an interval connecting two boundary circles along one of the edges of the original lattice (the interval thus has length $d-2R$). 
We assume that $b$ has been chosen such that $\mathcal{H}_{\gamma\gamma}(d,R)$ has discrete (and finitely degenerate) energy spectrum which is bounded from below.
By inserting sums over a complete set of intermediate states, we see that the full partition function can be recovered if one knows the amplitudes for the elementary cells, see Figure~\ref{fig:lattice-model-from-punching-holes}\,c),\,d). 
Denote the amplitude obtained by evaluating $Q$ on an elementary cell with states $\kappa_1,\dots,\kappa_n \in \mathcal{H}_{\gamma\gamma}(d,R)$ by
\begin{equation}\label{eq:vertex-amplitudes}
    W_{d,R}(\kappa_1,\dots,\kappa_n)  ~.
\end{equation}
Here $n$ is the valency of a vertex of the dual lattice. In the examples in Figure~\ref{fig:lattice-model-from-punching-holes}\,c),d) we have $n=3$ and $n=4$, respectively.
The amplitudes \eqref{eq:vertex-amplitudes} are the third ingredient.

\medskip

So far the lattice model has an infinite number of degrees of freedom, namely a basis of $\mathcal{H}_{\gamma\gamma}(d,R)$. The \textbf{fourth ingredient} is a choice of energy cutoff $E_\mathrm{max}$.
By our assumptions on the energy spectrum, the truncated state spaces $\mathcal{H}_{\gamma\gamma}^{\le E_{\mathrm{max}}}(d,R)$ are finite dimensional.

\medskip

We have now collected all the ingredients to define the lattice model for the QFT $Q$. The parameters were:
\begin{itemize}
    \item $\mathcal{F}$ -- the pivotal fusion category giving the collection of topological defects in $Q$ that are to be exactly realised in the lattice model,
    \item $b$ -- the boundary condition of $Q$ from which the cloaking boundary condition for $\mathcal{F}$ is constructed,
    \item $d,R$ -- the spacing of the centres of the holes cut into the surface, and the radius of these holes,
    \item $E_\mathrm{max}$ -- the energy cutoff to be applied to each of the interval state spaces $\mathcal{H}_{\gamma\gamma}(d,R)$ .
\end{itemize}
As shown in Figure~\ref{fig:intro-lattice-construction},
the lattice model defined by this data lives on the dual lattice $\Gamma$ relative to the one shown in Figure~\ref{fig:lattice-model-from-punching-holes}. Denote the set of vertices of the dual lattice $\Gamma$ by $V$ and the set of edges by $E$. Let $\{\kappa_1,\dots,\kappa_N\}$ be an ON-basis of $\mathcal{H}_{\gamma\gamma}^{\le E_{\mathrm{max}}}(d,R)$. 
The degrees of freedom live on the edges, and a spin configuration on the lattice consists of a function $\alpha : E \to \{1,\dots,N\}$. Denote by $v_1,\dots,v_n \in E$ the edges around a vertex $v$ in counter-clockwise order. The partition function reads
\begin{equation}
    Z(d,R,E_\mathrm{max}) = \sum_\alpha \prod_{v \in V} W_{d,R}(\kappa_{\alpha(v_1)}, \dots , \kappa_{\alpha(v_n)}) ~,
\end{equation}
where the sum runs over all configurations $\alpha : E \to \{1,\dots,N\}$. This lattice model has the following appealing properties:
\begin{enumerate}
    \item The lattice model realises the topological fusion category symmetry $\mathcal{F}$ exactly. This is true by construction before the cutoff as shown in Figure~\ref{fig:cloaking-boundary-condition}\,b), but it remains true even with the cutoff in place by the same argument as given in \cite[Sec.\,2.3]{Brehm:2021wev} in the case of a 2d\,CFT. 
    Namely, by integrating multiple insertions of the stress tensor parallel to the interval where the surface is cut into elementary cells, one can insert a polynomial of the Hamiltonian that approximates a step function in the energy. The stress tensor insertions are transparent to the topological defects, and so the latter can still move freely on the surface.

    \item The continuum torus partition function $Z_Q$ of the original QFT $Q$ can be recovered as a limit from the lattice model, independent of the size of lattice:
    \begin{equation}
        Z_Q = \lim_{R \to 0} \Big( \lim_{E_{\mathrm{max}}\to\infty} f(R)^{\#(\mathrm{faces})} \, Z(d,R,E_\mathrm{max}) \Big)~,
    \end{equation}
    where $\#(\mathrm{faces})$ denotes the number of faces of the lattice $\Gamma$. The inner limit recovers the QFT partition amplitude on the torus with holes, and the outer limit removes the holes due to \eqref{eq:zero-radius-limit-is-id} (with $\gamma_\mathcal{F}^b(\delta)$ in place of $b$).

    \item 
In the $R \to \frac d2$ limit, the lattice model becomes a 2d topological field theory in state sum formulation. The reason is that in this limit only the vacuum can propagate through the intervals connecting the different elementary cells, and for all-vacuum insertions, the amplitudes of elementary cells define an associative algebra, 
which is automatically semisimple (and hence Frobenius) by unitarity.
    See \cite[Sec.\,3]{Brehm:2021wev} for more details in the context of 2d\,CFTs.
\end{enumerate}

If we work with closed compact surfaces, regular lattices as in Figure~\ref{fig:lattice-model-from-punching-holes} only exist if the underlying surface is a torus. 
Surfaces of arbitrary genus, rather than just genus one, can be included by appealing to a beautiful piece of combinatorial geometry called \textit{circle packings},\footnote{We would like to thank Andr\'e Henriques and Tobias Osborne for pointing out the relevance of the theory of circle packings in this context.}
see e.g.\ \cite{Stephenson:2003} for a non-technical introduction. One has the following surprising result (the \textit{Discrete Uniformisation Theorem}):

\begin{figure}[tb]
    \centering
    \begin{subfigure}{0.45\textwidth}
    (a)
    \includegraphics[scale=0.6,valign=t]{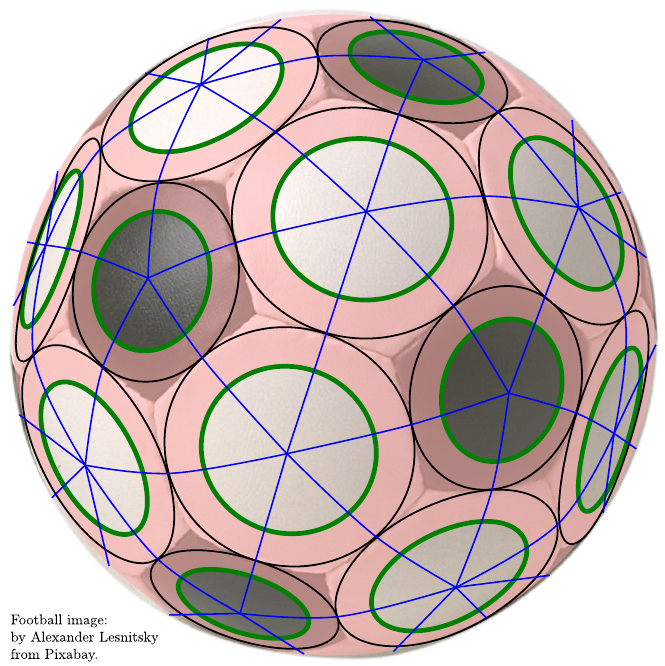}  
    \end{subfigure}
    \hspace{\fill}
    \begin{subfigure}{0.45\textwidth}
    (b)
    \includegraphics[scale=1.1,valign=t]{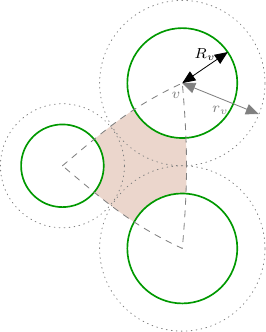}  
    \end{subfigure}
    \caption{a) Example of a circle packing on the sphere determined by a triangulation, together with some of the holes of radii $R_v = \rho r_v$ cut into the sphere. b) Example of an elementary cell given by a geodesic triangle with clipped edges as dictated by the radii $r_v$ of the circles in the circle packing and the ratio $\rho = R_v/r_v$.}
    \label{fig:lattice-on-other-surfaces}
\end{figure}

Pick an oriented compact surface $\Sigma$ of arbitrary genus $g$. Pick a triangulation of $\Sigma$. At this point there is no metric on $\Sigma$ and the triangulation is just a combinatorial object. Then there exist 
\begin{itemize}
\item
a constant curvature metric on $\Sigma$ (with curvature $>0$ for genus zero, $=0$ for genus one, and $<0$ for genus two and more),
\item 
a radius $r_v$ for each vertex $v$ of the triangulation and
a way to draw the triangulation on the surface such that the edges are geodesic and the circles of radius $r_v$ around $v$ for the three vertices of a triangle touch tangentially.
\end{itemize}
See Figure~\ref{fig:lattice-on-other-surfaces}\,a) for an illustration.
Moreover, the complex structure induced by the metric is \textit{uniquely determined by the triangulation}, and the centres of the circles and their radii are \textit{unique up to conformal mappings}. This is remarkable, as a purely combinatorial choice (the abstract triangulation) completely determines the geometry. 

Returning to the construction of lattice models, we see that we can include arbitrary surfaces by allowing for elementary cells which are geodesic triangles specified by three touching circles. One can control the hole radius by specifying a ratio $\rho \in (0,1)$ between the radius of the cut-out hole and the radius of the touching circles, see Figure~\ref{fig:lattice-on-other-surfaces}\,b). 
The fact that we start from a circle packing guarantees that in the $\rho \to 0$ limit the lattice model reduces to a state sum 2d\,TFT, since all intervals joining different elementary cells approach length zero simultaneously.

Will we not investigate lattice models for these more general geometries in the present paper and hope to return to this generalisation in the future. But we already see from the quick excursion into circle packings that our approach to construct lattice models from QFTs seems to be particularly well suited to conformal field theories, to which we turn next.

\subsection{Specialisation to 2d CFTs and the stability condition}
\label{sec:CFT-hexagonal}

In the reminder of the paper we focus on the case that the QFT $Q$ is a unitary rational conformal field theory, and that the original lattice is triangular and hence that dual lattice is hexagonal as in Figure~\ref{fig:lattice-model-from-punching-holes}\,a) (other lattices have been considered in \cite{Cheng:2023kxh}).
This results in a number of simplifications which make the weights \eqref{eq:vertex-amplitudes} of the lattice model explicitly computable:
\begin{itemize}
    \item The state space $\mathcal{H}_{\gamma\gamma}(R)$ is independent of $R$ in the following sense: The $L_0$-spectrum does not change with $R$, and the energy spectrum is related to the $L_0$-spectrum by a simple $R$-dependent shift and rescaling. For CFTs we will directly work with a cutoff $h_\mathrm{max}$ in $L_0$-weight, rather than an energy cutoff.

    \item The function $f(R)$ in \eqref{eq:zero-radius-limit-is-id} is simply given by $R^{\frac{c}{6}}$, where $c$ is the central charge, see \cite[Sec.\,5.1]{Brehm:2021wev}
    and Section~\ref{sec:one-hole-comparison} below.

    \item For the equilateral triangle with edges clipped off by a circle of radius $R$ (and three half-discs glued to the three intervals), one can explicitly write down the uniformising map to the unit disc in terms of hypergeometric functions. 
    The conformal transformation affects the amplitude by acting on the states labelling the three intervals, and by an exponentiated anomaly factor proportional to the central charge. Both can be computed explicitly, see Section~\ref{sec:anomaly-factor}.

    \item The resulting three-point function of boundary fields on the disc is fixed on primary fields by the OPE-coefficients and on descendant fields by a recursion relation. It is therefore explicitly computable to arbitrary order, and we explain the procedure in Section~\ref{sec:cloak-minimal}.

\end{itemize}

Let us fix a unitary rational CFT $C$, a conformal boundary condition $b$, and a pivotal fusion category $\mathcal{F}$ of topological line defects. 
We make the assumption that the topological defects in $\mathcal{F}$ are transparent to the rational chiral symmetry of $C$ (this is automatic only for Virasoro minimal models).

In this setting, we can also describe in more detail the stability condition proposed in the introduction. Namely, to $b$ there belongs a boundary state $\bndstate{b}$ in (a completion of) the space of bulk fields which describes a circular hole of radius one in the plane around the origin with boundary condition $b$,
\begin{equation}
    \bndstate{b} = \sum_{i \in \mathcal{I}} \sum_\alpha \varphi_{i,\alpha}(0) \ .
\end{equation}
Here the $\varphi_{i,\alpha}$ are (suitably normalised) bulk fields with $i\in \mathcal{I}$ denoting the primary family (possibly with multiplicity) and $\alpha$ denoting a descendant in that family. All $\varphi_{i,\alpha}$ are spinless, $h_{i,\alpha} = \bar h_{i,\alpha}$, as the boundary circle is rotation invariant.
A loop of the cloaking defect for $\mathcal{F}$ will in general project out some of the $\varphi_{i,\alpha}$, cf.\ \eqref{eq:gammaF-idempot}. There is thus some subset $\widetilde{\mathcal{I}}_{\mathcal{F}} \subset \mathcal{I}$ of all spin-less primary bulk fields $\varphi_i$ which describes the image of the projection. In the normalisation $\delta_0 = 1/\mathrm{Dim}(\mathcal{F})$ we get
\begin{equation}
    \gamma_\mathcal{F}(\delta)(\varphi_{i,\alpha}(0)) ~=~ \delta_{i \in \widetilde{\mathcal{I}}_{\mathcal{F}}} \, \varphi_{i,\alpha}(0) \ .
\end{equation}
Here we used the assumption that defects in $\mathcal{F}$ commute with modes of the chiral algebra and so act in the same way on all descendants $\varphi_{i,\alpha}$ of $\varphi_i$.

Write $\varphi_{1,\alpha}$ for the identity field and its descendants. Note that $i=1$ is necessarily contained in $\widetilde{\mathcal{I}}_{\mathcal{F}}$. Since we are working with a unitary theory, only the vacuum representation can contain descendants which are not irrelevant, and these then need to have weights $h_{0,\alpha} = \bar h_{0,\alpha} = 1$. The stability condition from the introduction can now be rephrased as:
\begin{center}
\fbox{~\begin{minipage}{0.96\textwidth}
\phantom{.}\\[-.2em]
Proposed stability condition:\\
Suppose the bulk identity field has no descendants of weight $h=\bar h = 1$, and that for all $i \in \widetilde{\mathcal{I}}_{\mathcal{F}}$ the primary field $\varphi_i$ has weights $h_i = \bar h_i > 1$. 
Then there is a fixed cutoff $h_\text{max}$ and radius $R_C>0$ such that for all $R \in (0,R_C)$ the universality class of the corresponding lattice model is given by $C$.
\\[-.6em]
\end{minipage}~}
\end{center}
In particular, for models with current-algebra symmetry such as WZW models one has to pass to a coset description in order to have a large enough $\mathcal{F}$ at one's disposal.
\footnote{
    For rational CFTs this reduction of chiral symmetry in order to have a large enough $\mathcal{F}$ is always possible: 
Namely, every rational VOA $V$ (technically: strongly rational and of CFT type) has a rational sub-VOA of the form $W \otimes_{\mathbb{C}} L$ (with the same stress tensor), where $L$ is a lattice VOA and $W$ has no weight-one fields, see \cite[Prop.\,4.3]{Hohn:2023auw}. 
In the factor $L$ there are still two sources of weight-one fields: lattice vectors of length squared 2 in the even lattice $\Lambda$ underlying $L$, and the weight-one space of the Heisenberg sub-VOA of $L$. 
To remove the first source, one can simply pass to twice the underlying lattice, i.e.\ one can consider the lattice VOA for $2\Lambda$. We assume this has been done. Then the weight-one space of $L$ is that of the Heisenberg sub-VOA.
One can now pass to the fixed-point VOA $L'$ under the involution of $L$ which acts as $-1$ on the weight-one space. The VOA $L'$ has no weight-one fields and is again rational \cite{Miyamoto:2013,Carnahan:2016}. Altogether, $U := W \otimes_{\mathbb{C}} L' \subset V$ is a rational sub-VOA without weight-one fields. We thank Sven Möller for providing this argument.
If one works relative to the rational VOA $U$, the resulting CFT has enough topological defects transparent to $U$ so that the stability condition can be satisfied.
}
A related stability mechanism in terms of compatibility with topological defects has been observed in the context of anyon chains and their continuum limits \cite{PhysRevLett.98.160409,Buican:2017rxc}.

The idea behind this stability condition was already explained in \cite[Sec.\,1.2]{Brehm:2021wev}, and we recall it here in the present notation. The boundary state for a hole of radius $R$ is given by
(see also Appendix~\ref{eq:radius-bnd-state})
\begin{equation}
    \bndstate{b,R} 
    \,=\,
    R^{L_0+\bar L_0- \frac{c}{6}} \bndstate{b} 
    \,=\,
    R^{-c/6} \sum_{i \in \mathcal{I}} \sum_\alpha R^{2 h_{i,\alpha}} \varphi_{i,\alpha}(0) \ .
\end{equation}
A regular lattice of holes in the torus as in the middle of Figure~\ref{fig:intro-lattice-construction} corresponds to inserting $\bndstate{b,R}$ at each vertex of the triangular lattice. 
Consider the resulting expression order by order in $R$. The leading order comes from $\varphi_1$, and the lowest subleading order from the field $\varphi_{i,\alpha}$, $i \in \widetilde{\mathcal{I}}_{\mathcal{F}}$ which has lowest weight above the vacuum. As shown in \cite[Sec.\,1.2]{Brehm:2021wev}, again to first order this amounts to perturbing the CFT $C$ by
\begin{equation}
    \text{(const)}\, \tfrac{R^2}{d^2} \, R^{2 h_{i,\alpha}-2} \, \varphi_{i,\alpha} \ .
\end{equation}
But by construction, $\varphi_{i,\alpha}$ is already irrelevant. In this sense, creating a lattice of small holes can be interpreted as an irrelevant perturbation of $C$.
As the hole radius $R$ increases, states of high energy will be suppressed exponentially to a stronger degree (cf.\ Section~\ref{sec:clipped-amplitude}), so that one may think of the cutoff $h_\text{max}$ as effectively working at finite radius $R$.

\smallskip
If the assumptions in the stability condition are satisfied, we would expect the following phase diagram for the lattice model (with fixed cutoff $h_\text{max}$):
\begin{equation}\label{eq:general-phase-diagram}
  \includegraphics[width=.88\linewidth,valign=c]{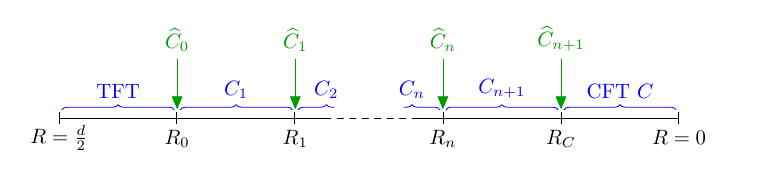}
\end{equation}
Namely, for $R<R_C$ the lattice model flows to the initial CFT $C$, and as the values of $R$ increase from $R_C$ to $d/2$, the lattice model may cross other universality classes, until near $d/2$ it is topological. At a point of phase transition $R=R_i$ between two universality classes $C_i$ and $C_{i+1}$, we expect an unstable critical point $\widehat C_i$ of a higher central charge.

The CFT $C$ at $R=0$ can be assumed to have a unique vacuum (if this is not the case, by unitarity $C$ can be decomposed into a direct sum of CFTs with unique vacuum). 
However, depending on the choice of initial boundary condition $b$ in the $\mathcal{F}$-cloaking boundary condition (Figure~\ref{fig:cloaking-boundary-condition}\,a), the TFT reached at $R \to d/2$ may have many degenerate vacua, up to the number of simple objects in $\mathcal{F}$ (see \cite[Sec.\,3]{Brehm:2021wev}).

\smallskip

To support the proposed phase diagram \eqref{eq:general-phase-diagram}, 
in Sections~\ref{sec:Ising-lattice} and~\ref{sec:loop-models} we study two examples, one where the assumptions in the proposal do not hold, and one where they hold. 

Namely, in Sections~\ref{sec:Ising-lattice} we take $\mathcal{F}$ to be the $\mathbb{Z}_2$-symmetry of the Ising CFT, in which case the boundary state still contains a relevant field. Working at lowest non-trivial cutoff $h_\mathrm{max}$, we recover the usual Ising lattice model on a triangular lattice. 
The resulting phase diagram is thus different from \eqref{eq:general-phase-diagram} and consists of a topological (i.e.\ massive) theory for all $0<R<\frac d2$ except at one critical value $R_0$, where it produces the critical Ising lattice model as an unstable fixed point.

In Section~\ref{sec:loop-models} we investigate unitary A-type Virasoro minimal models $C = M(p,p+1)$ with $\mathcal{F}$ being spanned by all Virasoro representations with Kac-labels of the form $(1,s)$. We find that already for the lowest non-trivial cutoff, the phase diagram is of the form \eqref{eq:general-phase-diagram}. Namely, there is a value $R_C$ such that for all $R<R_C$ the continuum model has the central charge of $C$. We conjecture that the corresponding CFT is indeed $C$, but we do not investigate correlation functions or the spectrum of the transfer matrix.
In these examples it turns out that $R_0=R_C$, so that the transition is directly between the 2d\,CFT $C$ and a 2d\,TFT. The corresponding CFT $\widehat C_0$ at the unstable fixed point separating $C$ and the TFT has the central charge of $M(p+1,p+2)$, and we conjecture that $\widehat C_0 = M(p+1,p+2)$.

\smallskip
By construction, the symmetry $\mathcal{F}$ persists for all values of $R$ in \eqref{eq:general-phase-diagram}. On the other hand, the surrounding symmetry as given by the full pivotal monoidal category of topological line defects may vary as one passes through the range of $R$'s. 
An example is provided by the models just discussed (assuming the fixed point CFTs are indeed as conjectured): For $R<R_C$ the full topological symmetry is that of $C$, namely the fusion category $\mathcal{M}_p$ spanned by all Kac-labels $(r,s)$, $1\le r\le p-1$, $1 \le s \le p$ for some $p \ge 3$, and with $\mathcal{F} \hookrightarrow \mathcal{M}_{p}$ the subcategory spanned by $(1,s)$. At $R=R_C$, the full symmetry is that of $\widehat C$, which is $\mathcal{M}_{p+1}$ where the Kac-labels $(r',s')$ now run over $1\le r\le p$, $1 \le s \le p+1$. The symmetry $\mathcal{F} \hookrightarrow \mathcal{M}_{p+1}$ is realised (fully faithfully) via the Kac-labels $(r',1)$.\footnote{
    The transition of defect labels from $(a,1)$ in $\mathcal{M}_{p+1}$ to $(1,a)$ in $\mathcal{M}_p$ also occurs in the RG-flow induced by the $\phi_{(1,3)}$ bulk perturbation of unitary minimal models \cite[Sec.\,4.1]{Fredenhagen:2009tn}, and as a property of the corresponding conformal RG-interface \cite{Gaiotto:2012np}. The change of the full topological symmetry under such RG-flows was investigated in \cite{Kikuchi:2021qxz}. 
}
For $R>R_C$ one obtains the TFT of \cite[Sec.\,3]{Brehm:2021wev}, which has $p$ degenerate vacua. Its pivotal fusion category of topological defects is $\mathsf{Vect}^{\oplus p}$, i.e.\ $p$ copies of the category of vector spaces, with $\mathcal{F} \hookrightarrow \mathsf{Vect}^{\oplus p}$ still faithful but no longer full -- see e.g.\ \cite[Sec.\,3]{Carqueville:2023jhb} for a discussion of fullness and faithfulness in topological symmetries given by pivotal fusion categories. 

We note that there are no monoidal functors $\mathcal{M}_p \to \mathcal{M}_{p+1}$ or $\mathcal{M}_p \to \mathsf{Vect}^{\oplus p}$ at all (for $p \ge 4$ -- the Ising case $p=3$ is an exception as there $\mathcal{F} = \mathcal{M}_3$). So in this sense we cannot say that the total symmetry of $\mathcal{M}_p$ is enlarged, but instead the symmetry $\mathcal{F}$ is realised in different ways.

\section{Amplitudes and the anomaly factor}\label{sec:anomaly-factor}

When considering ratios of amplitudes, as we did in \cite{Brehm:2021wev}, it is not necessary to know the explicit form of the anomaly factor that appears in the conformal transformations of the triangle amplitude. 
Accordingly, in \cite{Brehm:2021wev} we computed the triangle amplitude, and hence the Boltzmann weights, of the lattice model only up to an overall function of $R$ and $d$.
In this paper we will improve on this by comparing amplitudes for the cloaking boundary condition directly, rather than just comparing ratios, see Section~\ref{sec:one-hole-comparison}. To prepare the ground, we will first recall the Liouville action and our notion of surfaces with boundaries and field insertions. Then we consider general amplitudes and four rules that relate them to each other -- one of which involves the anomaly factor. This section can also be regarded as a general prescription on how to explicitly compute arbitrary amplitudes in 2d CFT.  

\subsection{Preliminaries}

\subsubsection{Liouville action} 

Let $\Sigma_{g}$ be a compact Riemannian surface with piecewise smooth boundary $\partial\Sigma$ and  metric $g$. For a smooth, real function $\Omega$ on $\Sigma_g$ the Liouville action is given by 
(see e.g.~\cite{Fateev:2000ik,Teschner:2000md,Bautista:2021ogd,Zamolodchikov:2007})
\begin{equation}
\begin{split}
    L_{\Sigma_g}(\Omega) \,= &~~~\,\frac{1}{4\pi} \int_\Sigma \sqrt{g} \left( q \,R\, \Omega + g^{\mu\nu}\partial_\mu\Omega \partial_\nu \Omega  - 2\pi\lambda\, e^{2\beta \,\Omega}\right)\,d^2x\\
    &+ \frac{1}{2\pi} \int_{\partial\Sigma} 
\Big(
    q \,k_g(l)\, \Omega\!\left(x^\mu(l)\right) - \lambda_B \,  e^{\beta\Omega} g^{\frac{1}{4}}
\Big)    
    \, dl \ ,
\end{split}
\end{equation}
where $R$ is the Ricci scalar of the metric $g$, $k_g$ is the geodesic curvature of the boundary, $q$ is the so-called background charge, $\lambda$ is the cosmological constant, $\lambda_B$ is the boundary cosmological constant, and $\beta$ is the Liouville coupling constant. 
The integral $\int_{\partial\Sigma}$ is taken over the smooth pieces of the boundary, i.e.\ over $\partial\Sigma \setminus \{ \text{corners} \}$.

For our purposes we set $q=2$, $\lambda = 0$, and $\lambda_B = 0$, resulting in 
\begin{equation}\label{eq:anomaly-action}
\begin{split}
    L_{\Sigma_g}(\Omega) \,= &~~~\,\frac{1}{2\pi} \int_\Sigma \sqrt{g} \left( R\, \Omega + \frac12 \, g^{\mu\nu}\partial_\mu\Omega \partial_\nu \Omega  \right)\,d^2x\\
    &+ \frac{1}{\pi} \int_{\partial\Sigma} \,k_g(l)\, \Omega\!\left(x^\mu(l)\right)\, dl \ ,
\end{split}
\end{equation}
which we also call the \emph{anomaly action}. If $\Sigma$ is contained in the complex plane with its standard metric, in complex coordinates the anomaly action simplifies further to 
%
%
%
%
%
%
%
%
\begin{equation}
    L_\Sigma(\Omega) \,=\, \frac{1}{2\pi} \int_\Sigma  \partial_z\Omega \partial_{\bar{z}} \Omega \,dzd\bar{z}+ \frac{1}{\pi} \int_{\partial\Sigma} k(l)\, \Omega\!\left(z(l)\right) dl \,.
\end{equation}
Some useful properties of the anomaly action are listed in Appendix~\ref{app:LProperties}.

Some care has to be taken with the treatment of corners on the physical boundary. Following \cite{Cardy:1988tk}, we propose that such corners produces an extra contribution to the anomaly action which we discuss in Appendix~\ref{app:nonSmoothBdy}. However, in this paper all physical boundaries will be smooth so that this additional corner term will not be needed.

\smallskip

For the sake of computing $L_{\Sigma_g}(\Omega)$, we will allow to decompose $\Sigma_g$ into a union of polygons $\Sigma_g = P_1 \cup \cdots \cup P_n$, where the $P_i$ have piecewise smooth boundary and pairwise disjoint interior. 
The anomaly action \eqref{eq:anomaly-action} has the property 
\begin{equation}\label{eq:anomaly-sum-patches}
    L_{\Sigma_g}(\Omega) = L_{P_1}(\Omega|_{P_1}) + \cdots + L_{P_n}(\Omega|_{P_n})  \ .
\end{equation}
This follows as the contributions of internal boundaries cancel between neighbouring patches.

We stress that even though the $P_i$ have corners, the identity \eqref{eq:anomaly-sum-patches} holds without adding an extra corner term to the action \eqref{eq:anomaly-action} (and would in fact not hold if one were to include the corner term proposed in Appendix~\ref{app:nonSmoothBdy}).
On the other hand, in our setting it will not make sense to evaluate a conformal field theory amplitude on the individual patches $P_i$, but only on the total surface $\Sigma_g$ which itself has a smooth physical boundary.

\subsubsection{Field insertions and local coordinates}

Let $\Sigma$ be a compact surface with boundary $\partial\Sigma$ (possibly empty), equipped with a metric $g$ compatible with the complex structure. 
For each $p \in \Sigma$ there is a local complex coordinate chart on some open neighbourhood in $\mathbb{C}$ or $\mathbb{H}$, where $\mathbb{H} := \{ z \in \mathbb{C}\,|\, \mathrm{im}(z) \ge 0 \}$ denotes the closure of the upper half-plane. In this chart, the metric is of the form $\rho(z) dzd\bar{z}$ with some positive smooth function $\rho$. By the \textit{bulk} of $\Sigma$ we will refer to the interior $\Sigma/\partial\Sigma$ of the surface.

\smallskip
The space $\mathcal{H}$ of bulk states of a conformal field theory carries an action of two commuting copies the Virasoro algebra, with modes $L_m$ and $\overline{L}_m$ and central charges $c, \bar c \in \mathbb{C}$. Below we will exclusively work with field insertions, but we will appeal to the state-field correspondence and use the terms ``state'' and ``field'' interchangeably, e.g.\ ``space of states'' instead of ``space of fields'' and ``sum over intermediate states''.

We place the following restrictions on the conformal field theories we consider: Firstly, the holomorphic and anti-holomorphic central charges agree, $c=\bar c$. Secondly, $\mathcal{H}$ decomposes into a direct sum of finite-dimensional common $L_0$ and $\overline{L}_0$ eigenspaces (or possibly generalised eigenspaces) of real eigenvalues. Thirdly, the eigenvalues are bounded from below.

\smallskip
A \textit{bulk field insertion} in $\Sigma$ is a triple 
\begin{equation}\label{eq:field-insertion}
    \psi(p;\varphi)\,,
\end{equation}
where $\psi\in \mathcal{H}$, $p$ is a point in the bulk and $\varphi: U_0\to\Sigma$ is a holomorphic map from a neighbourhood $U_0\subset\mathbb{C}$ of $0$ to $\Sigma$ with $\varphi(0) = p$ 
and $\partial\varphi(0) \neq 0$ 
(and so $\varphi$ is injective in some neighbourhood of $0$). 
We refer to $\varphi$ as the \textit{local coordinate} of $\psi$. 

\smallskip
Next we turn to the boundary and boundary field insertions. By a \textit{boundary segment} we mean a connected component of the boundary minus the field insertion points.  
At the boundary we introduce specific boundary conditions to the fields of the theory. Without going into the details, we assume that we are dealing with conformal boundary condition and that we can label different boundary segments by different boundary conditions. For the labeling we choose small Latin letters, and we call a segment of the boundary with some boundary condition a \textit{physical boundary} $b$. To every physical boundary $b$ we associate a space of boundary states $\mathcal{H}_b$, and to every two boundary conditions $a$ and $b$ we associate a space of boundary changing states $\mathcal{H}_{ab}$. Then $\mathcal{H}_b = \mathcal{H}_{bb}$, and below we will use ``boundary fields'' (or states) to refer to both, boundary changing fields and fields that preserve the boundary condition.

We assume that each $\mathcal{H}_{ab}$ carries an action of the Virasoro algebra with the same central charge $c$ as for $\mathcal{H}$, and that $\mathcal{H}_{ab}$ decomposes into (generalised) $L_0$-eigenspaces with real eigenvalues which are bounded from below.

\smallskip
A \textit{boundary field insertion} at the junction point of physical boundaries $a,b$
is a triple as in \eqref{eq:field-insertion}, where now
$\psi\in\mathcal{H}_{ab}$, $p \in \partial\Sigma$, and $\varphi: U_0\to\Sigma$ is a holomorphic map from a neighbourhood $U_0\subset\mathbb{H}$ of $0$ to $\Sigma$ with $\varphi(0) = p$ and $\partial\varphi(0) \neq 0$. 

\subsection{Transformation rules for amplitudes} \label{sec:amplitute}

Let
\begin{equation}
    \mathcal{A}\!\left(\Sigma_{g};b_k;\psi_i(p_i;\varphi_i)\right)\in \mathbb{C}
\end{equation}
be the (non-normalised) CFT amplitude on a Riemann surface $\Sigma$ with metric $g$, physical boundaries $b_k$, and (bulk and boundary) field insertions $\psi_i(p_i;\varphi_i)$\,. An amplitude without any field insertion is also called a \textit{partition function}. We stress that the local coordinates $\varphi_k$ and the metric $g$ are chosen independently, except that both are required to be compatible with the complex structure on $\Sigma$.

\smallskip
Amplitudes can be related to each other by applying the following four rules. 
\begin{enumerate}[(i)]
    \item \textbf{Invariance under isometries}: 
    Let $\Sigma_g$ and $\Sigma'_{g'}$ be surfaces and let $F:\Sigma\to\Sigma'$ be an orientation preserving isometric bijection: $g = F^* g'$. For any CFT amplitude we have
    \begin{equation}
        \mathcal{A}\!\left(\Sigma_{g};b_k;\psi_i(p_i;\varphi_i)\right) 
        \,=\, 
        \mathcal{A}\!\left(\Sigma'_{g'};b_k';\psi_i(p'_i;\varphi'_i)\right) \ ,
    \end{equation}
    where $\varphi'_i = F\circ \varphi_i$, $p'_i= F(p_i)$, and if $F$ maps a boundary segment $\beta \subset \partial\Sigma$ with boundary condition $b$ to a boundary segment $\beta'\subset\partial\Sigma'$ with boundary condition $b'$, then $b=b'$.
    
    \item \textbf{Field transformation under change of local coordinates}: 
    Consider a state $\psi$, a point $p \in \Sigma$ and a local coordinate $\varphi$. 
    Let $G : \mathbb{C} \to \mathbb{C}$ 
    be a holomorphic function such that $G(0) = 0$ and $\partial G(0) \neq 0$ (and such that $G$ preserves $\mathbb{H}$ for $p \in \partial\Sigma$, in particular $G'(0) \in \mathbb{R}_{> 0}$). 
    Then inside amplitudes we can trade a change of local coordinate at $p$ for a change of the state inserted at $p$ via
    \begin{equation}\label{eq:rule-ii-local-coord}
        \psi(p;\tilde\varphi) \,=\, \tilde{\psi}(p;\varphi) \ ,
    \end{equation}
    with 
    $\tilde\varphi = \varphi \circ G$ and
    $\tilde{\psi}$=$\Gamma_G\psi$. Here, $\Gamma_G$ is the action of the local coordinate change on the state space. Up to level 2 descendants $\Gamma_G$ is given by, for $G(z)=a_1 z+ a_2 z^2 + \cdots$, 
\begin{equation}\label{eq:Gamma_G-2nd-order}
    \Gamma_G = \Bigg(1 + \frac{a_2}{a_1^{\,2}} L_1
    + \frac12 \frac{a_2^{\,2}}{a_1^{\,4}} L_1 L_1
    + \Big(\frac{a_3}{a_1^{\,3}} - \frac{a_2^{\,2}}{a_1^{\,4}} \Big)L_2
    + \cdots \Bigg) a_1^{L_0} \cdot \big(\text{conj.}\big) \ ,
\end{equation}    
where in $(\text{conj.})$ all $L_m$ are replaced by $\bar L_m$ and all $a_j$ are replaced by their complex conjugates $a_j^*$.
The recursive procedure to compute $\Gamma_G$ is reviewed in \cite[Sec.\,6.2]{Brehm:2021wev}. 

    \item \textbf{Anomaly factor under Weyl transformation}: 
    The anomaly action \eqref{eq:anomaly-action} quantifies the response of an amplitude to a change of the metric within a given conformal class: if two metrics $g$ and $\hat{g}$ on $\Sigma$ are related by
    a Weyl transformation
    \begin{equation}
        g = e^\Omega \hat{g}
    \end{equation}
    for a smooth real function $\Omega$ on $\Sigma$,
    then we get the relation
    \begin{equation}\label{eq:AmplLiou} 
        \mathcal{A}(\Sigma_{g},\dots) \,=\, \exp\!\big( \tfrac{c}{24}L_{\Sigma_{\hat{g}}}(\Omega)\big) \,\mathcal{A}(\Sigma_{\hat g},\dots)\,.
    \end{equation}
    
    \item \textbf{Cutting the surface}: Loosely speaking, one can cut the surface $\Sigma$ along an embedded circle, or along an embedded arc with endpoints on the physical boundary. Gluing (half-)discs to the cuts results in a new surface $\Sigma'$. The amplitude on $\Sigma$ can be recovered by an appropriate ``sum over intermediate states'' from amplitudes on $\Sigma'$. The details are given in the next section.
\end{enumerate}
 
For more details on rule (ii) we refer to \cite{Gaberdiel:1994fs} and \cite[Sec.\,5]{Frenkel:2004} or also \cite[Sec.\,2.2]{Brehm:2020zri}.
For rule (iii) see for example \cite{Friedan:1986ua,Gawedzki:1989,Zamolodchikov:2007} -- we review the motivation from the path integral point of view in Appendix~\ref{app:LiouvilleMotivation}.

\smallskip
We can use the four rules above to successively relate amplitudes on any surface to amplitudes on simple surfaces with a standard choice of conformal structure and metric: a sphere with three bulk field insertions, a disc with one bulk and one boundary insertion, and a disc with three boundary insertions.

The behaviour under conformal transformations follows from rules (i)--(iii). For example, if $E : \Sigma'_{h} \to \Sigma_g$ is an orientation preserving conformal bijection, then by definition $E^* g = e^{\Omega} h$ for some smooth function $\Omega : \Sigma' \to \mathbb{R}$, and we get
\begin{align}
        \mathcal{A}\!\left(\Sigma_{g};b_k;\psi_i(p_i;\varphi_i)\right) 
        &\,\overset{\text{(i)}}=\, 
        \mathcal{A}\!\left(\Sigma'_{E^*g};b_k';\psi_i(p'_i;\varphi'_i)\right)
\nonumber\\
        &\overset{\text{(iii)}}= 
        e^{\frac{c}{24}L_{\Sigma_{h}}(\Omega)}
        \mathcal{A}\!\left(\Sigma'_{h};b_k';\psi_i(p'_i;\varphi'_i)\right) \ .
\label{eq:example-amplitude-conformal-transformation}
\end{align}
Here, $b_k'$, $p_i'$ and $\varphi_i'$ are related to $b_k$, $p_i$ and $\varphi_i$ as in rule (i) with $F = E^{-1}$.

As a further illustration, in Appendix~\ref{app:anomalyExamples} and Appendix~\ref{app:state-sum-examples} we give some examples for applying rules (i)--(iv) in simple cases, including computations involving anomaly factors.

\subsection{Cutting properties of amplitudes}\label{sec:cutting-amplitude}

Here we review the procedure of cutting along an embedded arc. Cutting along an embedded circle is only needed for the examples in Appendix~\ref{app:state-sum-examples} and is discussed there.
We start by defining orthonormal bases in the spaces of boundary fields, then set up the geometry of the cutting procedure, and finally give the formula for the sum over intermediate states. Except for the details on the anomaly factors, this is already described in \cite[Sec.\,7.1]{Brehm:2021wev}, so we will be brief.

\subsubsection{Orthonormal boundary fields}

\begin{figure}[tb]
    \centering
    \includegraphics[scale=1.4]{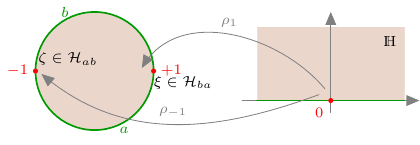}
    \caption{The unit disc $D$ with field insertion points and local coordinates. The corresponding amplitude defines the pairing $(\zeta,\xi)_{ab}$.}
    \label{fig:disc-defining-pairing}
\end{figure}

Consider the unit disc $D \subset \mathbb{C}$ with standard metric $g_{ij} = \delta_{ij}$, with boundary insertions at $1$ and $-1$, and 
with boundary condition $a$ in the lower half plane, and $b$ in the upper half plane. In our conventions, fields inserted at $1$ are taken from $\mathcal{H}_{ba}$ and fields inserted at $-1$ are from $\mathcal{H}_{ab}$. The local coordinates at $1$ and $-1$ are given by the Möbius transformations (cf.\ \cite[Eqn.\,7.4]{Brehm:2021wev})
\begin{equation}\label{eq:rho+-1}
\rho_{\pm 1} : \mathbb{H} \to D \setminus \{ \mp 1 \}
\quad , \quad
\rho_{\pm 1}(z) = \pm \frac{i-z}{i+z} \ ,
\end{equation}
which satisfy $\rho_{\pm 1}(0) = \pm 1$. 
The CFT amplitude defines a pairing (see Figure~\ref{fig:disc-defining-pairing}),
\begin{equation}\label{eq:disc-pairing}
(-,-)_{ab} : \mathcal{H}_{ab} \times \mathcal{H}_{ba} \to \mathbb{C}
\quad,\quad
(\zeta,\xi)_{ab} = \mathcal{A}\big(D;a,b;\zeta(-1;\rho_{-1}),\xi(1;\rho_{1})\big)
\ ,
\end{equation}
which we assume to be non-degenerate in the CFTs we consider (otherwise not all fields can be distinguished in correlators). This pairing satisfies
\begin{equation}\label{eq:disc-pairing-properties}
(\zeta,\xi)_{ab} = (\xi,\zeta)_{ba}
\quad , \quad
(L_m \zeta,\xi)_{ab} = (-1)^m (\xi,L_{-m} \zeta)_{ba}
~,~~ m \in \mathbb{Z}
\ .
\end{equation}
The first property follows from transformation rule (i), applied to a $180^\circ$  rotation of the disc, and the second property is a standard contour integral computation.

In each space $\mathcal{H}_{ab}$ we fix a basis $(\kappa^{(ab)}_i)_{i \in \mathbb{N}}$ of $L_0$-eigenvectors (or possibly generalised $L_0$-eigenvectors), subject to the orthonormality condition
\begin{equation}\label{eq:kappa-ON-condition}
(\kappa^{(ab)}_i,\kappa^{(ba)}_j)_{ab} = \delta_{ij} \ .
\end{equation}
This is consistently possible due to the above symmetry property of $(-,-)_{ab}$.

Below we will need to consider discs $D(\Omega)$ with the more general metric $g = e^{\Omega} \delta_{ij}$. By the transformation rule (iii) we have
\begin{equation}
\mathcal{A}\big(D(\Omega);a,b;\zeta(-1;\rho_{-1}),\xi(1;\rho_{1})\big)
~=~ e^{\frac{c}{24}L_{D}(\Omega)} \, (\zeta,\xi)_{ab} 
\ .
\end{equation}
It follows that the basis
\begin{equation}\label{eq:kappa-ON-condition-Omega}
    \kappa(\Omega)^{(ab)}_i \,:=\, e^{-\frac{c}{48}L_{D}(\Omega)}  \, \kappa^{(ab)}_i 
\end{equation}
of $\mathcal{H}_{ab}$ is orthonormal when paired with $\kappa(\Omega)^{(ba)}_j$ via the amplitude for the disc $D(\Omega)$. Note that even though $D(\Omega)$ may itself not be invariant under rotations by $180^\circ$, the resulting pairing is still symmetric as the Liouville factors for $\Omega$ and for the rotated version of $\Omega$ are the same.  

\subsubsection{Cutting of surfaces and sum over ON-bases}\label{sec:open-sum-states}

Let $I \subset D$ be the interval from $-i$ to $i$ on the imaginary axis and let $U \subset D$ be an open neighbourhood of $I$ in $D$.
Let $\Sigma_g$ be a surface with metric $g$ and let $\gamma : U \to \Sigma$ be a boundary and orientation preserving conformal map, which is a diffeomorphism onto its image.
Then $\gamma(I)$ is an embedded interval in $\Sigma$ whose endpoints lie on the boundary of $\Sigma$.

\begin{figure}[tb]
    \centering
    \includegraphics[scale=1.0]{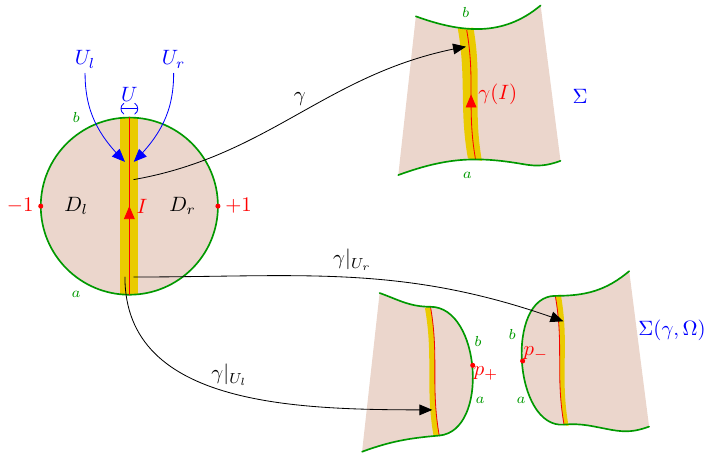}
    \caption{Cutting a surface $\Sigma$ along the interval $\gamma(I)$ and gluing half-discs to the resulting cuts produces the surface $\Sigma(\gamma,\Omega)$.
Here, $D_l = \{z \in D \,|\, \mathrm{Re}(z)\le 0 \}$ and $D_r = \{z \in D \,|\, \mathrm{Re}(z)\ge 0 \}$ are the closed left and right unit half discs; $U$ is an open neighbourhood of $I$ in $D$, $U_l = U \setminus D_r$ and $U_r = U \setminus D_l$ are open sets such that $U \setminus I = U_l \sqcup U_r$. The restrictions $\gamma|_{U_l}$ and $\gamma|_{U_r}$ are used to to glue $D_r \cup U$ and $D_l \cup U$ to $\Sigma \setminus \gamma(I)$.    
The points $p_\pm \in \Sigma(\gamma,\Omega)$ denote $1 \in D_r$, $-1 \in D_l$, seen as part of the new surface $\Sigma(\gamma,\Omega)$, and $\sigma_\pm$ denote the corresponding local coordinates at $p_\pm$ (not shown in the figure).
    }
    \label{fig:cut-surface-def}
\end{figure}

We will now cut $\Sigma$ along $\gamma(I)$ and turn the result into a new surface $\Sigma(\gamma,\Omega)$ with smooth boundaries by gluing half-discs to the cuts as explained in Figure~\ref{fig:cut-surface-def}. The metric on $\Sigma(\gamma,\Omega)$ involves a choice, namely we need to fix a smooth function $\Omega : D \to \mathbb{R}$ such that the pullback metric of $g$ on $U$ is given by $\gamma^*g = e^{\Omega} \delta_{ij}$. 
In other words, $e^{\Omega} \delta_{ij}$ is a smooth continuation of the pullback metric from $U$ to $D$.

With these ingredients, we can finally state the behaviour of amplitudes under cutting of surfaces:
\begin{align}
     \mathcal{A}(\Sigma;\dots) 
     &=  \sum_{i=1}^\infty \mathcal{A}\big(\Sigma(\gamma,\Omega); \kappa(\Omega)_i^{(ab)}(p_-,\sigma_-), \kappa(\Omega)_i^{(ba)}(p_+,\sigma_+) , \dots \big) 
     \nonumber\\
     &=
     e^{-\frac{c}{24}L_{D}(\Omega)} 
     \sum_{i=1}^\infty \mathcal{A}\big(\Sigma(\gamma,\Omega); \kappa^{(ab)}_i(p_-,\sigma_-), \kappa^{(ba)}_i(p_+,\sigma_+) , \dots \big) \ .
    \label{eq:cut-amplitude-via-sum}
\end{align}
The second line makes explicit that the right hand side does not depend on the choice of extension of $\Omega$ from $U$ to $D$.

\smallskip
In Appendix~\ref{app:state-sum-examples} we work through several examples in detail, namely the torus partition function, and the two ways to compute the cylinder partition function. These examples illustrate how the well-known $c$-dependence of these partition functions arises in terms of the anomaly action.

 \subsection{Application to the torus with holes}

As an illustration of how the transformation rules (i)--(iv) can be used to reduce amplitudes on complicated Riemann surfaces to sums over simple ones, we express the torus with holes arranged in a triangular lattice (Figure~\ref{fig:lattice-model-from-punching-holes}\,(a)) as a sum over disc amplitudes with three boundary field insertions. To this end, we cut the torus into clipped triangles as in Figure~\ref{fig:lattice-model-from-punching-holes}\,(c), insert a sum over intermediate states as described above, and use the conformal mapping given in \cite[Sec.\,6]{Brehm:2021wev} to map the result to a disc. We summarise the setup and supplement it by a careful treatment of the anomaly factor.

\subsubsection{Uniformisation map for the clipped triangle}

\begin{figure}[tb]
    \centering
    \includegraphics[scale=1.0]{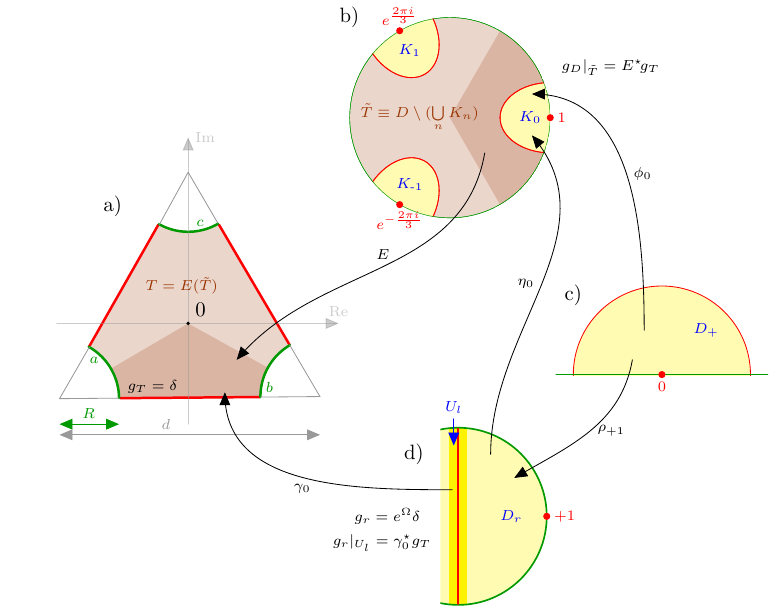}
    \caption{The maps involved in the uniformisation and the computation of the anomaly factor for the clipped triangle.}
    \label{fig:clipped-triangle-and-uniform}
\end{figure}

Consider a clipped triangle with edge length of the corresponding unclipped triangle given by $d$ (this describes the distance between the centres of the holes), and hole radius $R$. Let $L = d/\sqrt{3}$ be the distance between the centre of the unclipped triangle and its corners. Place the triangle in the complex plane so that its centre is at zero and its bottom edge is parallel to the real axis (Figure~\ref{fig:clipped-triangle-and-uniform}\,(a)). 

After gluing half-discs to the cutting boundaries, the overall geometry is that of a disc, and we now describe its uniformisation. 
The uniformisation map depends on a parameter $t \in \mathbb{R}_{>0}$, which determines the ratio $R/d$ via the auxiliary functions
\begin{align}
    R(t) &= 
    \frac{\sqrt{3\pi}}{\cosh^2(\pi t)-\frac{3}{4}} \cdot
    \frac{\Gamma\!\left(\frac{7}{6}\right)}{\Gamma\!\left(\frac{5}{6}-i t\right)\Gamma\!\left(\frac{5}{6}+i t\right)}\,,
\nonumber
    \\
    d(t) &= 2\cosh(\pi t) \, R(t)\,. 
\label{eq:d(t)R(t)}
\end{align}
Thus $R(t)/d(t) = (2 \cosh(\pi t))^{-1}$, which varies from $\frac12$ (for $t\to 0$, the touching hole limit) to $0$ (for $t \to \infty$, the vanishing hole limit).

Let $D$ be a unit disc with field insertion points at $e^{2 \pi i n/3}$ for $n \in \{0,\pm1\}$. 
Let us start with the coordinate neighbourhoods $\phi_n$. These are defined on the closed unit half disc $D_+ = D \cap \mathbb{H}$, $\phi_n : D_+ \to D$ (and in fact extend to an open neighbourhood of $D_+$, but we will not need that here) and satisfy $\phi_n(0) = e^{2 \pi i n/3}$, $n \in \{0,\pm1\}$ (Figure~\ref{fig:clipped-triangle-and-uniform}\,(b,c)). The inverse of $\phi_0$ can be expressed explicitly via hypergeometric functions as \cite[Eqn.\,(6.15)]{Brehm:2021wev}
\begin{equation}\label{eq:phi-inv}
    \phi_0^{-1}(u) =  X\, \frac{s}{2} \,\exp\!\left(\frac{1}{2 i t} \log\frac{_2F_1\!\left(\frac{5}{12}+\frac{it}2,\frac{1}{12}+\frac{i t}2,1+i t; s^2\right)}{_2F_1\!\left(\frac{5}{12}-\frac{i t}2,\frac{1}{12}-\frac{i t}2,1-i t; s^2\right)}\right) \ ,
\end{equation}
where 
\begin{equation}
    X(t) = \exp\!\left(\frac{1}{2i t} \log\!\left(\frac{\Gamma\!\left(\frac{1}{6}+i t\right)
    \Gamma\!\left(-it\right)}{\Gamma\!\left(\frac{1}{6}-i t\right)
    \Gamma\!\left(it\right)}\right)\right)
    ~,~~
    s = \frac{1}{2i} \left( u^{\frac32} - u^{-\frac32} \right)
    \ .
\end{equation}
For $\phi_0(z)$ itself one can compute the power series expansion order by order, and the first few terms are  \cite[Eqn.\,(6.18)]{Brehm:2021wev}
\begin{equation}\label{eq:phi0-expansion}
    \phi_0(z) = 1 + \frac{4}{3} \frac{i z}{X(t)} + \frac{8}{9}\left(\frac{i z}{X(t)}\right)^{\!2} + \frac{68 t^2+53}{81(t^2+1)} \left(\frac{i z}{X(t)}\right)^{\!3} + \mathcal{O}\!\left(z^4\right)\,.
\end{equation}
The maps $\phi_{\pm 1}(z)$ are obtained from $\phi_0$ by rotation: $\phi_n(z) = e^{2 \pi i n/3} \phi_0(z)$. Write $K_n = \phi_n(D_+) \subset D$ for the image of $D_+$. We do not have an explicit expression for the shape of $K_n$, and for the computation in the Ising CFT in Section~\ref{sec:one-hole-comparison} it was determined numerically.

The uniformising map $E$ is a bijective holomorphic map from the red shaded area in Figure~\ref{fig:clipped-triangle-and-uniform}\,(b) to the clipped triangle. In the wedge $D_\triangleleft = \{ z \in D \,| \, - \tfrac{\pi}3 \le \arg z \le  \tfrac{\pi}3 \}$ it is given by, for $z \in D_\triangleleft \setminus K_0$,
\begin{equation}\label{eq:uniform}   
    E(z) ~=~ \frac{d}{d(t)} \, \frac{1}{i}\left(\frac{z^{-\frac32}-z^{\frac{3}{2}}}{2}\right)^{\!\!-\frac{2}{3}}
    ~
    \frac{_2F_1\!\left(\frac{5}{12}+\frac{it}{2},\frac{5}{12}-\frac{it}{2},\frac{4}{3};
    -4\,\big(z^{-\frac32}-z^{\frac{3}{2}}\big)^{-2}
    \right)}{_2F_1\!\left(\frac{1}{12}+\frac{it}{2},\frac{1}{12}-\frac{it}{2},\frac{2}{3};
    -4\,\big(z^{-\frac32}-z^{\frac{3}{2}}\big)^{-2}
    \right)} ~.
\end{equation}
This is the map $\tilde F$ in \cite[Eqn.\,(6.10)\,\&\,Fig.\,15]{Brehm:2021wev} up to an overall rescaling of the edge length to $d$, i.e.\ $E(z) = d/d(t) \, \tilde F(z)$.
The map extends to the other wedges on $D$ by conjugation with the rotations $e^{2 \pi i /3}$: if $z \in e^{2 \pi i n/3} ( D_\triangleleft \setminus K_0 )$, then $E(z) = e^{2 \pi i n/3} E(e^{-2 \pi i n/3} z)$. 

\subsubsection{Anomaly factor}

The starting point is the clipped triangle $T$ in Figure~\ref{fig:clipped-triangle-and-uniform}\,(a). We have applied the cutting procedure (transformation rule (iv)) three times and glued three half discs $D_r$ (Figure~\ref{fig:clipped-triangle-and-uniform}\,(d)) to the cuts. Denote the resulting surface by $T_{\text{glue}}$. Due to the rotational symmetry, the anomaly factor will be three times the factor obtained for the wedge-shaped region highlighted in Figure~\ref{fig:clipped-triangle-and-uniform}\,(a). The gluing map for the cutting boundary in that region is denoted by $\gamma_0 : U_l \to T$ in  Figure~\ref{fig:clipped-triangle-and-uniform}.

The metric $g_T$ on $T$ is simply the flat metric induced by the embedding into $\mathbb{C}$, $(g_T)_{ij} = \delta_{ij}$. 
Write $g_r = e^{\Omega} \delta$ for the metric on $D_r \cup U$ given by a smooth function $\Omega : (D_r \cup U) \to \mathbb{R}$. We require that the pullback metric on $U_l$ is $(\gamma_0^* g_T)_{ij} = (g_r)_{ij}|_{U_l} = e^{\Omega} \delta_{ij}$, or, in other words, $\Omega(z) = \log |\partial\gamma_0(z)|^2$ for $z \in U_l$. 
The triangle amplitude is given by
\begin{align}
    &T_{d,R}^{abc}(\kappa^{(ba)}_i,\kappa^{(cb)}_j,\kappa^{(ac)}_k)
    \nonumber\\
    &:=
    \mathcal{A}(T_{\text{glue}};\kappa(\Omega)_i^{(ba)}(p_0,\sigma_0),\kappa(\Omega)_j^{(cb)}(p_1,\sigma_1),\kappa(\Omega)_k^{(ac)}(p_{-1},\sigma_{-1})) \ .
\end{align}
Here, $p_0, p_1, p_{-1}$ are the different copies of the point $1 \in D_r$ obtained from gluing the three half discs $D_r$, and $\sigma_{0,\pm 1}$ are the different copies of the local coordinate $\rho_{+1}$. On the l.h.s.\ we use the ON-basis $\kappa^{(ba)}_i$ introduced in \eqref{eq:kappa-ON-condition} which does not depend on the choice of extension of the metric to the half-discs. On the r.h.s.\ we use  the ON-basis $\kappa(\Omega)_i^{ba}$ defined with respect to the disc with extension of the pullback metric as in \eqref{eq:kappa-ON-condition-Omega}. As already noted below \eqref{eq:cut-amplitude-via-sum}, the dependence on the choice of extension cancels, and we will see this explicitly below.

\medskip

To compute $T_{d,R}^{abc}$, we uniformise the surface $T_{\text{glue}}$ by a holomorphic bijection with the unit disc (Figure~\ref{fig:clipped-triangle-and-uniform}\,(b)). The bijection is described in four patches. On $\widetilde T := \overline{D \setminus (\bigcup_n K_n)}$ it is given by $E$, and on each $K_n$ it is given by $\eta_n^{-1}$, where $\eta_n$ is defined via $\eta_n \circ \rho_{+1} = \phi_n$. In the figure we only show $\eta_0$. 

We denote the pullback metric on $D$ by $g_D$. On $\widetilde T$ it is given by $g_D = E^* g_T = |\partial E|^2 \delta$, and on $K_n$ by $(\eta_n^{-1})^* g_r = |\partial\eta_n|^{-2} e^{\Omega} \delta$. 
By construction, the metric $g_D$ is of the form $g_D = e^{\tilde\Omega} \delta$, with $\tilde \Omega$ smooth on all of $D$. By transformation rules (i) and (iii) we have (see also \eqref{eq:example-amplitude-conformal-transformation}),
\begin{align}
    &T_{d,R}^{abc}(\kappa^{(ba)}_i,\kappa^{(cb)}_j,\kappa^{(ac)}_k)
\nonumber \\
    &\overset{\text{(i)}}=
    \mathcal{A}(D_{g_D};\kappa(\Omega)_i^{(ba)}(1,\phi_0),\kappa(\Omega)_j^{(cb)}(e^{\frac{2 \pi i}{3}},\phi_1),\kappa(\Omega)_k^{(ac)}(e^{-\frac{2 \pi i}{3}},\phi_{-1}))  
\nonumber \\    
    &\overset{\text{(iii)}}=
    e^{\frac{c}{24}( L_D(\tilde\Omega) - 3 L_{D_r}(\Omega) ) }
    \mathcal{A}(D;\kappa^{(ba)}_i(1,\phi_0),\kappa^{(cb)}_j(e^{\frac{2 \pi i}{3}},\phi_1),\kappa^{(ac)}_k(e^{-\frac{2 \pi i}{3}},\phi_{-1}))  \ .
    \label{eq:triangle-aux1}
\end{align}
In the last line we also pulled out the normalisation factors \eqref{eq:kappa-ON-condition-Omega} included in $\kappa(\Omega)_i^{(ba)}$, etc.\ (note that $D_r$ is only the half-disc, leading to an additional factor of $2$ relative to \eqref{eq:kappa-ON-condition-Omega}). 

It remains to compute the anomaly factor $L_D(\tilde\Omega) - 3 L_{D_r}(\Omega)$. For $L_D(\tilde\Omega)$, we decompose $D$ into six patches, namely $K_0$, $\widetilde T_\triangleleft := \overline{D_\triangleleft \setminus K_0}$, and the corresponding rotated versions obtained by multiplying with $e^{ \pm 2 \pi i /3}$. By \eqref{eq:anomaly-sum-patches}, we can compute the contribution on each patch separately. By rotational symmetry, we have
\begin{equation}
    L_D(\tilde\Omega) - 3 L_{D_r}(\Omega)
    =
    3 \big(  L_{\widetilde T_\triangleleft}(\tilde\Omega) + L_{K_0}(\tilde\Omega) - L_{D_r}(\Omega) \big) \ .
\end{equation}
From the explicit form of $g_D$ on $\widetilde T$ we get
$L_{\widetilde T_\triangleleft}(\tilde\Omega) = L_{\widetilde T_\triangleleft}(\log |\partial E|^2)$. 
Since $e^\Omega \delta = \eta_0^* e^{\tilde\Omega} \delta = |\partial \eta_0|^2 e^{\tilde\Omega} \delta$ we get, for $u \in D_r$,
\begin{equation}
    \Omega(u)
    =
    \tilde\Omega(\eta_0(u)) 
    + 
    \log |\partial \eta_0(u)|^2 
    \ .
\end{equation}
In this case we can apply the transformation property \eqref{eq:Lid3} to $D_r$ to obtain
\begin{equation}
    L_{D_r}(\Omega) \,=\,
    L_{K_0}(\tilde\Omega) + L_{D_r}(\log |\partial \eta_0|^2) \ .
\end{equation}
Furthermore, by \eqref{eq:Lid1} we have
\begin{equation}
        L_{D_+}(\log |\partial \phi_0|^2) = L_{D_r}(\log |\partial \eta_0|^2) + L_{D_+}(\log |\partial \rho_{+1}|^2)\ .
\end{equation}
In Appendix~\ref{sec:MoebiusUDLD} we check that $L_{D_+}(\log |\partial \rho_{+1}|^2)=0$. 
Altogether we arrive at
\begin{equation}\label{eq:triangle-anomaly-factor}
    A \,:=\, \frac{c}{24}\Big( L_D(\tilde\Omega) - 3 L_{D_r}(\Omega) \Big)
    \,=\,
    \frac{c}{8}\Big(
    \, L_{\widetilde T_\triangleleft}(\log |\partial E|^2) - L_{D_+}(\log |\partial \phi_0|^2) \, \Big) \ .
\end{equation}
This expression is now indeed independent of the choice of $\Omega$ on $D_r$. In both terms on the r.h.s., the anomaly action needs to be computed with all boundary contributions, in particular including the boundary terms arising from the boundary between $\widetilde T_\triangleleft$ and $K_0$. 

\smallskip
The triangle amplitude 
in \eqref{eq:triangle-aux1}
has now been rewritten as
\begin{align}\label{eq:triangle-amplitude-with-anomaly}
    &T_{d,R}^{abc}(\kappa^{(ba)}_i,\kappa^{(cb)}_j,\kappa^{(ac)}_k)
    \nonumber\\
    & \qquad
    = e^A \, \mathcal{A}(D;\kappa^{(ba)}_i(1,\phi_0),\kappa^{(cb)}_j(e^{\frac{2 \pi i}{3}},\phi_1),\kappa^{(ac)}_k(e^{-\frac{2 \pi i}{3}},\phi_{-1}))  \ ,
\end{align}
where $A$ was defined in \eqref{eq:triangle-anomaly-factor}.
The amplitude $\mathcal{A}(D;\cdots)$ on the right hand side only involves the unit disc $D$ with standard metric $\delta$ (but with non-standard local coordinates). A recursive procedure how to reduce the descendant contributions to those of primary fields is given in \cite[Sec.\,6]{Brehm:2021wev} and is recalled in Section~\ref{sec:clipped-amplitude} below.

\section{Cloaking boundary condition in minimal models}\label{sec:cloak-minimal}

In this section we present our conventions for structure constant in minimal models and provide an explicit expression for the triangle amplitude $T_{d,R}^{abc}$ evaluated on primary boundary fields. We combine these ingredients to obtain the triangle amplitude for a cloaking boundary condition relative to a  choice of fusion subcategory $\mathcal{F}$, and we check the stability condition for the cloaking boundary condition for some choices of $\mathcal{F}$.

\subsection{Virasoro minimal models}\label{sec:vir-min-mod}

We will focus on the A-series Virasoro minimal models, which we denote by $M(p,q)$ with $p,q \in \mathbb{Z}_{\ge 3}$ coprime. 
While in Sections~\ref{sec:one-hole-comparison} and~\ref{sec:Ising-lattice} we specialise to the Ising model and in Section~\ref{sec:loop-models} to unitary models, here we do not restrict $p,q$ further. 

\smallskip
The bulk primary fields $\phi_i$ of $M(p,q)$ are all spinless, i.e. their left and right conformal weight are equal, $h_i = \bar{h}_i$. They are indexed by Kac-labels $(r,s)$, $1\le r \le p-1$ and $1\le s \le q-1$, 
subject to the identification $(r,s) \sim (p-r,q-s)$.
The central charge and the conformal weights are given by
\begin{align}\label{eq:MM-c-h}
    c_{p,q} = 1 - 6 \frac{(1-t)^2}{t} ~,~~
    h_{(r,s)} &= \frac{(r-st)^2 -(1-t)^2}{4t}~,~~ \text{with}\quad t=p/q\,.
\end{align}
Let us denote the full set of Kac-labels modulo the above identification by $\mathcal{I}$. Equivalently, $\mathcal{I}$ indexes the irreducible modules $\mathfrak{R}_i$, $i \in \mathcal{I}$, of the simple quotient of the Virasoro VOA at that central charge. We write $1 := (1,1) \in \mathcal{I}$ for the index of vacuum representation.

\begin{figure}[tb]
    \centering
    \includegraphics[scale=1.5]{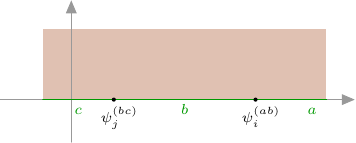}
    \caption{The arrangement of boundary fields and boundary conditions on the upper half plane $\mathbb{H}$ used in the definition of the boundary structure constants $C^{(abc)k}_{ij}$.}
    \label{fig:boundary-structure-constants}
\end{figure}

\smallskip
The elementary conformal boundary conditions of $M(p,q)$ are equally indexed by $a \in \mathcal{I}$ \cite{Cardy:1989ir}, and for $a,b \in \mathcal{I}$, the space of boundary states $\mathcal{H}_{ab}$ decomposes as
\begin{equation}\label{eq:min-mod-bnd-fields}
    \mathcal{H}_{ab} = \bigoplus_{i \in \mathcal{I}} N_{ab}^{~i} 
    \, \mathfrak{R}_i \ , 
\end{equation}
where $N_{xy}^{~z}$ denotes the minimal model fusion rules. Let
$\psi_i^{(ab)} \in \mathcal{H}_{ab}$
denote the primary boundary field between physical boundaries $a,b$ in the irreducible representation $\mathfrak{R}_i$. Their OPE on the upper half plane is determined by structure constants $C^{(abc)k}_{ij}$ defined via (see Figure~\ref{fig:boundary-structure-constants})
\begin{equation} \label{eq:primary-boundary-OPE-minmod}
    \psi_i^{(ab)}(x) \, \psi_j^{(bc)}(y) \,=\, \sum_k C^{(abc)k}_{ij} \, \psi_k^{(ac)}(y) \, (x-y)^{h_k-h_i-h_j} + \cdots \quad; \, x>y\,, 
\end{equation}
where $\dots$ contains the descendant contributions. The structure constants are given by F-symbols \cite{Runkel:1998he},
\begin{equation}\label{eq:stConst}
    C^{(abc)k}_{ij} = F_{bk}\!\begin{bmatrix}a & c \\i & j\end{bmatrix}\,.
\end{equation}
The simple relation between the structure constant and the F-symbol is only possible for a specific normalisation of the conformal four-point blocks. Namely, for a basis with insertion points $\infty$, $1$, $x$, $0$, in which the asymptotic behaviour for $x \to 0$ has to be $1 \cdot x^r$, where $r$ is computed from the conformal weights, see \cite{Runkel:1998he} for details.

We note that typically $C^{(aba)1}_{ii} \neq 1$, so that the boundary fields are not canonically normalised (and, as we will see in a moment, they cannot all be canonically normalised at the same time).

\smallskip
The amplitude of the empty unit disc $D^a$ with boundary condition $a \in \mathcal{I}$ is
\begin{equation}\label{eq:discA}
    \mathcal{A}(D^a) \,=\, \frac{S_{a1}}{\sqrt{S_{11}}} \ .
\end{equation}
Here, $S_{ij}$ is the minimal model $S$-matrix (see e.g.\ \cite[\S\,10.6]{DiFrancesco:1997nk}),
\begin{equation}\label{eq:min-mod-Smatrix}
  S_{(r,s),(x,y)} = \sqrt{8/(pq)} \,\, (-1)^{1+ry+sx} \,
  \sin( \pi r x /t) \, \sin( \pi s y t) 
  \quad , \quad t = p/q \ .
\end{equation}
The $S$-matrix also determines the quantum dimensions via, for $a \in \mathcal{I}$,
\begin{equation}
    \dim(a) = \frac{S_{a1}}{S_{11}} \ .
\end{equation}

\subsection{Pairing boundary fields}

For $\vartheta \in \mathbb{R}$ consider the local coordinate $\rho_{e^{i\vartheta}}(z) = e^{i \vartheta} \rho_1(z)$ with $\rho_1$ as in \eqref{eq:rho+-1}. 
Consider the disc $D^{ab}$ with boundary insertions at $e^{i\vartheta}$ and $1$, and boundary condition $a$ just below the insertion $1$ and boundary condition $b$ just above (cf.\ Figure~\ref{fig:disc-defining-pairing}). 
The disc two-point amplitude is
\begin{align}\label{eq:disc-2pt-amplitude}
    &\mathcal{A}\big(D^{ab};\psi_i^{(ab)}(e^{i\vartheta},\rho_{e^{i\vartheta}})\psi_i^{(ba)}(1,\rho_1)\big)
\nonumber\\
    &= \,
    \mathcal{A}(D^a)\,
    C^{(aba)1}_{ii} \big( \sin \tfrac{\vartheta}2 \big)^{-2 h_i} 
    \,=\,
    \frac{S_{a1}}{\sqrt{S_{11}}}\,
    F_{b1}\!\begin{bmatrix}a & a \\i & i\end{bmatrix}
    \big( \sin \tfrac{\vartheta}2 \big)^{-2 h_i} \ .
\end{align}
Here $\vartheta \in (0,2\pi)$, and 
for $\vartheta = \pi$ this reproduces the situation in Figure~\ref{fig:disc-defining-pairing}. Applying transformation rule (i) for a rotation by $e^{-i\vartheta}$ yields the identity
\begin{align}
    &\mathcal{A}\big(D^{ab};\psi_i^{(ab)}(e^{i\vartheta},\rho_{e^{i\vartheta}})\psi_i^{(ba)}(1,\rho_1)\big)
    \nonumber \\
    &=\,
    \mathcal{A}\big(D^{ba};\psi_i^{(ba)}(e^{i(2\pi -\vartheta)},\rho_{e^{i(2\pi -\vartheta)}})\psi_i^{(ab)}(1,\rho_1)\big) \ .
\end{align}
That this is compatible with the explicit expression in \eqref{eq:disc-2pt-amplitude} follows from the identity
\begin{equation}\label{eq:F-matrix-S-identity}
     F_{b1}\!\begin{bmatrix}a & a \\i & i\end{bmatrix} S_{a1} \,=\,  F_{a1}\!\begin{bmatrix}b & b \\i & i\end{bmatrix} S_{b1}
\end{equation}
satisfied by the minimal model F-symbols (see e.g.\ the appendix of \cite{Runkel:1998he} for this and more F-matrix identities). This also implies the following relation, which illustrates that one cannot consistently set $C^{(aba)1}_{ii} = 1$ for all $a,b,i$:
\begin{equation}\label{eq:2-point-sc-ratio}
    \frac{C^{(aba)1}_{ii}}{C^{(bab)1}_{ii}} 
    = \frac{S_{b1}}{S_{a1}} 
    = \frac{\dim(b)}{\dim(a)} \ . 
\end{equation}

From the above computation we can read off the pairing \eqref{eq:disc-pairing} on boundary fields. Namely, for the primary fields $\psi_i^{(ab)} \in \mathcal{H}_{ab}$ and $\psi_j^{(ba)} \in \mathcal{H}_{ba}$ we find
\begin{equation}\label{eq:disc-pairing-minmod-primary}
\big(\psi_i^{(ab)},\psi_j^{(ba)}\big)_{ab} \,=\, \delta_{i,j} 
     \, \frac{S_{a1}}{\sqrt{S_{11}}} \,
    F_{b1}\!\begin{bmatrix}a & a \\i & i\end{bmatrix} \ .
\end{equation}
The normalised primary states are
\begin{equation}\label{eq:normState}
    \kappa_i^{(ab)} 
    = 
    \left(\frac{S_{a1}}{\sqrt{S_{11}}} 
    F_{b1}\!\begin{bmatrix}a & a \\i & i\end{bmatrix} \right)^{\!-\frac{1}{2}}
    \psi_i^{(ab)}
    \ .
\end{equation}
These indeed satisfy $\big(\kappa_i^{(ab)},\kappa_j^{(ba)}\big)_{ab} = \delta_{i,j}$ as follows from combining \eqref{eq:F-matrix-S-identity} and \eqref{eq:disc-pairing-minmod-primary}.

\subsection{The clipped triangle amplitude}\label{sec:clipped-amplitude}

The computation of $T_{d,R}^{abc}$ in \eqref{eq:triangle-amplitude-with-anomaly} will be split into two parts: a contribution from the primary states which involves structure constants and the anomaly factor, and a contribution from descendant states relative to the primary ones which is determined solely by a Virasoro mode calculation for a conformal 3-point block.

In more detail, consider the representation $\mathfrak{R}_i$, $i \in \mathcal{I}$ and let $|i\rangle$ be the primary state normalised to $\langle i | i \rangle_{\mathfrak{R}_i} = 1$. Write $|i\alpha\rangle$ for an ON-basis of $\mathfrak{R}_i$ with respect to the (non-standard) bilinear form satisfying $\langle L_m u | v \rangle = (-1)^m \langle u | L_{-m} v \rangle_{\mathfrak{R}_i}$ as in \eqref{eq:disc-pairing-properties}:
\begin{equation}\label{eq:pairing-on-desc}
	\langle i\alpha |i\beta\rangle_{\mathfrak{R}_i} \,=\, \delta_{\alpha,\beta}
\end{equation}
The vectors $|i\alpha\rangle$ can be expressed as $M_{i,\alpha} |i\rangle$, where $M_{i,\alpha}$ is some polynomial in Virasoro modes. 
The ON-basis of $\mathcal{H}_{ab}$ can now be written as
\begin{equation}\label{eq:bnd-on-basis-desc}
	\kappa^{(ab)}_{i\alpha} := M_{i,\alpha} \kappa^{(ab)}_{i} \,\in\,\mathcal{H}_{ab} \ . 
\end{equation}
By construction, it satisfies
\begin{equation}
	\big(\kappa^{(ab)}_{i\alpha},\kappa^{(ba)}_{j\beta}\big)_{ab} = \delta_{i,j}\,\delta_{\alpha,\beta} \ .
\end{equation}
We will decompose the triangle amplitude as
\begin{align}\label{eq:triangle-amplitude-with-anomaly-decomp}
    T_{d,R}^{abc}(\kappa_{i\alpha}^{ba},\kappa_{j\beta}^{cb},\kappa_{k\gamma}^{ac})
    &= e^A \, \mathcal{A}(D;\kappa^{(ba)}_i(1,\phi_0),\kappa^{(cb)}_j(e^{\frac{2 \pi i}{3}},\phi_1),\kappa^{(ac)}_k(e^{-\frac{2 \pi i}{3}},\phi_{-1}))  
    \nonumber\\
    & \qquad \times \, 
\Big(\tfrac{4}{3X}\Big)^{-h_i-h_j-h_k} \,
    B\big( (|i\alpha\rangle,\phi_0) , (|j\beta\rangle,\phi_1) ,  (|k\gamma\rangle,\phi_{-1}) \big)
    \ ,
\end{align}
where $B$ denotes a conformal 3-point block on the complex plane with local coordinates at the field insertions as indicated, normalised such that for the three primaries we have
\begin{equation}\label{eq:3pt-block-triangle-normalisation}
B( (|i\rangle,\phi_0) , (|j\rangle,\phi_1) ,  (|k\rangle,\phi_{-1}) )
\,=\,
\Big(\tfrac{4}{3X}\Big)^{h_i+h_j+h_k} \ .    
\end{equation}
We will now discuss the factors $\mathcal{A}(D,\dots)$ and $B(\dots)$ in turn, which will also explain the normalisation chosen for $B$.

\smallskip
To evaluate $\mathcal{A}(D,\dots)$, we first give the 3-point amplitude on $D$ for the local coordinates $\rho_{e^{i\vartheta}} = e^{i\vartheta} \rho(z)$. 
So, let $0<\vartheta_1<\vartheta_2<2\pi$ and let $\psi_i^{(ab)}$ etc. be primary fields as in \eqref{eq:primary-boundary-OPE-minmod}. We consider the disc with insertion points $1$, $e^{i \vartheta_1}$, $e^{i \vartheta_2}$. The boundary conditions are, starting from $1$ in counter-clockwise direction, $b$, $c$, $a$:
\begin{equation}
    \includegraphics[valign=c,scale=1.5]{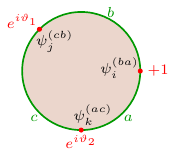}
\end{equation}

\noindent
This three-point amplitude is given by 
\begin{align}
\mathcal{A}(D;\psi_i^{ba}(1,\rho_1)&,\psi_j^{cb}(e^{i \vartheta_1},\rho_{e^{i \vartheta_1}}),\psi_k^{ac}(e^{i \vartheta_2},\rho_{e^{i \vartheta_2}}))  
\nonumber \\[1em]
&= 
\frac{C^{(cba)k}_{ji} C^{(aca)1}_{kk}
\mathcal{A}(D^a)}{
\big( \sin \tfrac{\vartheta_1}2 \big)^{h_j+h_i-h_k}
\big( \sin \tfrac{\vartheta_2}2 \big)^{h_i+h_k-h_j}
\big( \sin \tfrac{\vartheta_2-\vartheta_1}2 \big)^{h_k+h_j-h_i}} \ .
\end{align}
Specialising to $\vartheta_1 = 2\pi/3$, $\vartheta_2 = 4\pi/3$ and substituting the explicit values of the structure constants \eqref{eq:stConst} and the disc amplitude \eqref{eq:discA} gives
\begin{align}\label{eq:disc-3pt-rho-coord-psi-primary}
	&\mathcal{A}\big(\,D; \, \psi_i^{ba}(1,\rho_1) \, , \, \psi_j^{cb}(e^{\frac{2 \pi i}{3}},\rho_{e^{2 \pi i/3}}) \, ,\, \psi_k^{ac}(e^{-\frac{2 \pi i}{3}},\rho_{e^{-2 \pi i/3}})\,  \big)  
	\nonumber \\
	&= \,
	F_{bk}\!\begin{bmatrix}c & a \\j & i\end{bmatrix}
	F_{c1}\!\begin{bmatrix}a & a \\k & k\end{bmatrix}
	\frac{S_{a1}}{\sqrt{S_{11}}} 
	\,
	\left( \frac{\sqrt{3}}{2} \right)^{\!-h_i-h_j-h_k} \ .
\end{align}
This expression is symmetric under cyclic permutation of $a,b,c$ and $i,j,k$ which follows from combining \eqref{eq:F-matrix-S-identity} with the F-matrix identity (see e.g.\ \cite[App.]{Runkel:1998he})
\begin{equation}
	     \Fsym bkcaji \Fsym c1aakk  = \Fsym ciabkj \Fsym b1aaii \ .
\end{equation}

\noindent 
To make the connection to \eqref{eq:disc-3pt-rho-coord-psi-primary} we need to change the local coordinates $\phi_n$, whose expansion for $n=0$ is given in \eqref{eq:phi0-expansion}, to $\rho_{e^{2\pi i n/3}}$, use the transformation rules \eqref{eq:rule-ii-local-coord} in (ii) and use the normalisation given in \eqref{eq:normState}. 

The coordinate change is a holomorphic function $H$ such that
\begin{equation}
    \phi_n(z) = \rho_{e^{2\pi i n/3}} \circ H(z)
    \quad , \qquad z \in D_+  \ ,
\end{equation}
with $H(0) = 0$. As we are considering primary fields, only the first derivative enters in the transformation rule. We get
\begin{equation}
    \partial H(0) = \frac{\partial \phi_n(0)}{\partial \rho_{e^{2\pi i n/3}}(H(0))} =\frac{2}{3 X(t)} \ ,
\end{equation}
which follows from $\partial \rho_1(0)=2i$ and $\partial \phi_0(0) = 4i / (3X(t))$, cf.\ \eqref{eq:rho+-1} and \eqref{eq:phi0-expansion}. For a primary field $\psi$, the transformation rule \eqref{eq:rule-ii-local-coord} gives (for $n=0$)
\begin{equation}
    \psi(1,\phi_0) = \psi(1,\rho_1 \circ H) = \tilde\psi(1,\rho_1) = 
    (\partial H(0))^{h_\psi} \, \psi(1,\rho_1) \ ,
\end{equation}
where in the last step we used $\tilde\psi = \Gamma_H 
\psi = (\partial H(0))^{L_0} \psi$, cf.\ \eqref{eq:Gamma_G-2nd-order}. 
Using the latter and the normalisation \eqref{eq:normState} we obtain
\begin{align}\label{eq:disk-amplitude-primary}
&\mathcal{A}(D;\kappa^{(ba)}_i(1,\phi_0),\kappa^{(cb)}_j(e^{\frac{2 \pi i}{3}},\phi_1),\kappa^{(ac)}_k(e^{-\frac{2 \pi i}{3}},\phi_{-1}))  
\nonumber \\
&=
\Big( \tfrac{2}{3 X(t)} \Big)^{h_i+h_j+h_k}
\mathcal{A}(D;\kappa^{(ba)}_i(1,\rho_1),\kappa^{(cb)}_j(e^{\frac{2 \pi i}{3}},\rho_{e^{2\pi i /3}}),\kappa^{(ac)}_k(e^{-\frac{2 \pi i}{3}},\rho_{e^{-2\pi i /3}}))  
\nonumber \\
&=
\left( \frac{4}{3 \sqrt{3} X(t)} \right)^{h_i+h_j+h_k}
\frac{
    \left(\dim(a) 
    F_{c1}\!\begin{bmatrix}a & a \\k & k\end{bmatrix} \right)^{\!\frac{1}{2}}
~
F_{bk}\!\begin{bmatrix}c & a \\j & i\end{bmatrix}
}{
S_{11}^{\,1/4} \,
    \left( \dim(b) 
    F_{a1}\!\begin{bmatrix}b & b \\i & i\end{bmatrix}
    \dim(c) 
    F_{b1}\!\begin{bmatrix}c & c \\j & j\end{bmatrix} \right)^{\!\frac{1}{2}}
}
\ .
\end{align}

\smallskip
We now turn to the conformal 3-point block $B(\cdots)$ in \eqref{eq:triangle-amplitude-with-anomaly-decomp}. We mostly follow the ideas developed in \cite{Brehm:2021wev} with only slight changes in the setup.
The identity in transformation rule (ii) holds on the level of conformal blocks (this is the setting in \cite[Sec.\,5]{Frenkel:2004}), and we are free to choose the local coordinates in a way which makes the evaluation on descendant states convenient, irrespective of whether they map the real axis to the boundary of the disc. We choose the local coordinates as in \cite[Eqn.\,(6.32)]{Brehm:2021wev},
\begin{equation}
    \sigma_s(z) = e^{2 \pi i s} (1 + i z)\ ,
\end{equation}
for $s = n/3$, $n=0,\pm 1$. As derived in \cite[Sec.\,6.2.2]{Brehm:2021wev}, we get
\begin{align}
&    B\big( (|i\alpha\rangle,\phi_0) , (|j\beta\rangle,\phi_1) ,  (|k\gamma\rangle,\phi_{-1}) \big)
\nonumber \\    
&    =
    B\big( (\Gamma |i\alpha\rangle,\sigma_0) , (\Gamma |j\beta\rangle,\sigma_{1/3}) ,  (\Gamma  |k\gamma\rangle,\sigma_{-1/3}) \big) \ ,
\end{align}
where up to level $-2$ the map $\Gamma$ is given by 
\begin{equation}\label{eq:Gamma-for-clipped-triang}
    \Gamma = \Big(1 + \frac{i}{2} L_1 - \frac{1}{8} L_1^{\,2} - \frac{5}{192} \frac{1+4t^2}{1+t^2} L_2
 + \cdots\Big)
    \left(\frac{4}{3X}\right)^{\!L_0}
\,.
\end{equation}
The conformal block $B$ can be determined recursively. To state the relation concisely, we abbreviate, for $u \in \mathfrak{R}_i$, $v \in \mathfrak{R}_j$, $w \in \mathfrak{R}_k$,
\begin{equation}
    B(u,v,w) := 
B\big( (u,\sigma_0) , (v,\sigma_{1/3}) ,  (w,\sigma_{-1/3}) \big) \ .
\end{equation}
The reason to choose the normalisation of $B$ as in \eqref{eq:3pt-block-triangle-normalisation} is that in terms of the coordinates $\sigma_n$ we have
\begin{equation}\label{eq:tilde-B-norm}
 B\big(|i\rangle,|j\rangle,|k\rangle\big) \,=\, 1 
\end{equation}
for the  primary states $|i\rangle$, $|j\rangle$, $|k\rangle$.
The recursion relation for $B$ is
derived in \cite[Sec.\,6.3.3]{Brehm:2021wev}. For $m \ge 1$ we have
\begin{equation}\label{eq:recusive-relation}
B(L_{-m} u,v,w)
 = 
B(A_0 u,v,w)
+
B(u,A_+ v,w)
+
B(u,v,A_- w)~,
\end{equation}
where
\begin{align}
A_0 &= -i L_{-m+1} + \tfrac13 L_{-m+2}~,
\nonumber \\
A_+ &= -(-1)^m \, 3^{-\frac{m}2} \, \zeta^{m-1} \left[
L_0 + 
\sum_{k=1}^\infty \binom{1-m}{k-1} (-1)^k \, 3^{-\frac{k}2} \, \bar\zeta^k \left( \frac{2-m-k}k - \bar\zeta \right) L_k \right]~,
\nonumber \\
A_- &= - 3^{-\frac{m}2} \, \bar\zeta^{m-1} \left[
L_0 + 
\sum_{k=1}^\infty \binom{1-m}{k-1}  3^{-\frac{k}2} \, \zeta^k \left( \frac{2-m-k}k - \zeta \right) L_k \right]~,
\end{align}
and $\zeta = e^{2\pi i/3}$.
Together with the normalisation \eqref{eq:tilde-B-norm} and cyclic symmetry, this determines $B$ uniquely.

\smallskip
Altogether, the triangle amplitude is
\begin{align}\label{eq:triangle-amplitude-with-anomaly_computed}
    T_{d,R}^{abc}(\kappa_{i\alpha}^{ba},\kappa_{j\beta}^{cb},\kappa_{k\gamma}^{ac})    
    =\, \frac{e^A\ \mathcal{N}_{ijk}^{abc} }{  
    \big( \dim(a)\dim(b)\dim(c) \big)^{\frac12} }
    \,
    B\big( \Gamma |i\alpha\rangle, \Gamma |j\beta\rangle, \Gamma  |k\gamma\rangle \big)\ ,
\end{align}
where $A$ is given in \eqref{eq:triangle-anomaly-factor}, and 
\begin{equation}\label{eq:Nijkabc-expression}
    \mathcal{N}_{ijk}^{abc} 
    \,:=\,
    \frac{\dim(a)}{S_{11}^{1/4} \  3^{(h_i+h_j+h_k)/2}} \, \,
\left(\frac{ F_{c1}\!\begin{bmatrix}a & a \\k & k\end{bmatrix} }{
 F_{a1}\!\begin{bmatrix}b & b \\i & i\end{bmatrix}
    F_{b1}\!\begin{bmatrix}c & c \\j & j\end{bmatrix}
    }
    \right)^{\!\!\frac{1}{2}}
F_{bk}\!\begin{bmatrix}c & a \\j & i\end{bmatrix}
\ .
\end{equation}

\noindent
These constants are cyclically symmetric in the sense that $\mathcal{N}_{ijk}^{abc}=\mathcal{N}_{jki}^{bca}$.

\subsection{Cloaking boundary conditions}\label{sec:cloaking-bcs}

Just as the elementary conformal boundary conditions, the elementary topological defects of the A-series Virasoro minimal model $M(p,q)$ are labelled by $\mathcal{I}$ \cite{Petkova:2000ip,Fuchs:2002cm}. Their fusion rules are the same as for the representations indexed by $\mathcal{I}$. 

As explained in Section~\ref{sec:lattice-from-QFT}, we can construct a cloaking boundary condition for every subset $\mathcal{J} \subset \mathcal{I}$ that closes under fusion. 
The corresponding cloaking boundary condition we consider is given by fusing the cloaking topological defect into the conformal boundary labelled by $1$. The resulting conformal boundary condition is the superposition
\begin{equation}
    \gamma = \bigoplus\limits_{b\in \mathcal{J}} b \ ,
\end{equation}
together with an insertion of the weight zero boundary field
\begin{equation}
    \delta = \delta_0\sum_{b \in \mathcal{J}} \text{dim}(b) 1_b \ , 
\end{equation}
where $1_b$ denotes the identity field on the elementary boundary condition $b$ and $\delta_0 \in \mathbb{C}^\times$ is a normalisation constant which we leave arbitrary for now. The combination of a circular boundary labelled by $\gamma$ together with the $\delta$-insertion will be denoted by $\gamma(\delta)$.

\subsubsection{Open channel: triangle amplitude} 

The field content on the boundary $\gamma$ is given by
\begin{equation}
    \mathcal{H}_{\gamma\gamma} = \bigoplus\limits_{a,b\in \mathcal{J}} \mathcal{H}_{ab}\,,\quad \mathcal{H}_{ab} = \bigoplus_{i\in \mathcal{J}} {N_{ib}}^a \mathfrak{R}_i \ .
\end{equation}
Note that a priori the second sum runs over all of $\mathcal{I}$ as in \eqref{eq:min-mod-bnd-fields}, but as $\mathcal{J}$ closes under fusion, it restricts to $\mathcal{J}$.

\smallskip
Consider a torus with holes labelled $\gamma(\delta)$ as in Figure~\ref{fig:lattice-model-from-punching-holes}\,a). Each circular boundary is cut into six pieces, and we will distribute the insertion of $\delta$ evenly by placing an insertion of
\begin{equation}
    \delta^{1/6} = \delta_0^{1/6} \sum_{b \in \mathcal{J}} \text{dim}(b)^{1/6} \, 1_b  
\end{equation}
on each of the three boundary segements of the clipped triangle for the cloaking boundary $\gamma$.
We denote the resulting triangle amplitude by $T^{\gamma(\delta)}_{R,d}$.
Explicitly, in the ON-basis $\{ \kappa^{(ab)}_{i \alpha} \}$ chosen in each of the $\mathcal{H}_{ab}$, we get
\begin{align}\label{eq:triangle-amplitude-cloaking}
    T_{d,R}^{\gamma(\delta)}(\kappa_{i\alpha}^{(ba)},\kappa_{j\beta}^{(cb)},\kappa_{k\gamma}^{(ac)})
=
    \frac{e^A  \ \delta_0^{\,1/2} \ \mathcal{N}_{ijk}^{abc}}{
     \big(\dim(a)\dim(b)\dim(c)\big)^{\frac13}}
     \,
    B\big( \Gamma |i\alpha\rangle, \Gamma |j\beta\rangle, \Gamma  |k\gamma\rangle \big)
\ .
\end{align}
Here, $a,b,c,i,j,k \in \mathcal{J}$. We also want to introduce a graphical representation of the amplitude:
\begin{equation}\label{eq:graphical_rep_ampl}
    T_{d,R}^{\gamma(\delta)}(\kappa_{i\alpha}^{(ba)},\kappa_{j\beta}^{(cb)},\kappa_{k\gamma}^{(ac)})  = \includegraphics[valign=c]{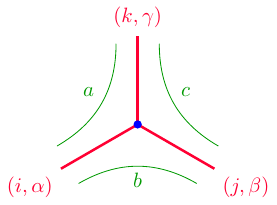}
    .
\end{equation}

\subsubsection{Closed channel: boundary state}\label{sec:cloaking-bnd-states-properties}

We would like to understand for which choices of $\mathcal{J}$ the cloaking boundary condition satisfies the stability condition in Section~\ref{sec:CFT-hexagonal}. To this end, for a given $\mathcal{J}$ we define 
\begin{equation}
    \widetilde{\mathcal{J}}
    \,=\,
    \Big\{ x \in \mathcal{I} \,\Big|\, \frac{S_{xj}}{S_{x1}} = \frac{S_{1j}}{S_{11}} \text{ for all } j \in \mathcal{J} \Big\} \ .
\end{equation}
Basically we ask which rows $x \in \mathcal{I}$ of the $S$-matrix (normalised by their first entry) agree with the first row if we only compare the columns indexed by $\mathcal{J}$.

\bigskip

\noindent
\textbf{Lemma.} For each $x \in \mathcal{I}$ we have
\begin{equation}\label{eq:char-ortho}
    \sum_{j \in \mathcal{J}} \dim(j)  \frac{S_{xj}}{S_{x1}} 
    \,=\, 
    \mathrm{Dim}(\mathcal{J}) \, \delta_{x \in \widetilde{\mathcal{J}}} \ .
\end{equation}

\begin{proof}
The statement follows from orthogonality of characters:
Write $A_{\mathcal{J}}$ for the $\mathbb{C}$-algebra with basis $\{ b_j | j \in \mathcal{J}\}$ and product given by the fusion rules, $b_i b_j = \sum_{k \in \mathcal{J}} N_{ij}^{~k} b_k$. 
Define $\varepsilon : A_{\mathcal{J}} \to \mathbb{C}$ as $\varepsilon(b_i) = \delta_{i,1}$, with $b_1$ denoting the unit of $A_{\mathcal{J}}$. Then $(a,b) \mapsto \varepsilon(ab)$ is a non-degenerate invariant pairing, and the dual basis to $b_i$ is $b_{i^*}$ with $i^*$ being the dual of $i$ (in minimal models $i^*=i$). See \cite[Ch.\,3]{EGNO-book} for more on this type of algebra.
The copairing is $\sum_j b_j \otimes b_{j^*} \in A_{\mathcal{J}} \otimes_{\mathbb{C}} A_{\mathcal{J}}$ and has the property that 
\begin{equation}\label{eq:char-ortho-aux1}
    \sum_j (a b_j) \otimes b_{j^*} = \sum_j b_j \otimes (b_{j^*} a)
    \quad \text{for all} \quad a \in A_{\mathcal{J}} \ .
\end{equation}

Let $V,W$ be two representations of $A_{\mathcal{J}}$ with actions $\rho_V$ and $\rho_W$, and let $f:V \to W$ be any linear map. By \eqref{eq:char-ortho-aux1}, $\sum_j \rho_W(b_j) \circ f \circ \rho_V(b_{j^*})$ is an intertwiner of $A_{\mathcal{J}}$-representations. If $V,W$ are irreducible and non-isomorphic, by Schur's Lemma this sum must be zero. 

Consider a one-dimensional representation with underlying vector space $\mathbb{C}$ and action $\rho : A_{\mathcal{J}} \to \mathbb{C}$ (i.e.\ $\rho$ is a character of $A_{\mathcal{J}}$). Then $(\mathbb{C},\rho)$ is automatically irreducible, and $(\mathbb{C},\rho)$ and $(\mathbb{C},\varphi)$ are isomorphic iff $\rho(a)=\varphi(a)$ for all $a \in A_{\mathcal{J}}$.
Applying the above argument to $\rho \neq \varphi$ and $f = id_{\mathbb{C}}$ gives
\begin{equation}\label{eq:char-ortho-aux2}
    \sum_{j \in \mathcal{J}} \rho(b_j) \varphi(b_{j^*}) = 0 \ .
\end{equation}

For each $x \in \mathcal{I}$, the function
\begin{equation}
    \psi_x : A_{\mathcal{J}} \to \mathbb{C}
    ~~,~~
    b_j \mapsto \frac{S_{xj}}{S_{x1}} \ ,
\end{equation}
is a character of $A_{\mathcal{J}}$, and so by \eqref{eq:char-ortho-aux2} we get
\begin{equation}
    \sum_{j \in \mathcal{J}} \psi_x(b_j) \psi_y(b_{j^*}) 
    = 0
    \quad \text{for} \quad \psi_x \neq \psi_y \ .
\end{equation}
Taking $y=1$ and using that $\psi_x = \psi_1$ iff $x \in \widetilde{\mathcal{J}}$ shows the statement \eqref{eq:char-ortho} for $x \notin \widetilde{\mathcal{J}}$. For $x \in \widetilde{\mathcal{J}}$ we have $S_{xj}/S_{x1} = S_{1j}/S_{11} = \dim(j)$ and so the sum in \eqref{eq:char-ortho} indeed produces $\mathrm{Dim}(\mathcal{J})$.
\end{proof}

\smallskip

The circular boundary $\gamma(\delta)$ of radius $1$ can be represented by a state in (a completion of) the space of bulk fields, which is inserted at the centre of the hole. 
The elementary boundary states $\bndstate{a}$, $a\in \mathcal{I}$, in Virasoro minimal models are given by
\begin{equation}
    \bndstate{a} = \sum_{b\in\mathcal{I}} \frac{S_{ab}}{\sqrt{S_{b1}}} \ket{b}\!\rangle \ ,
\end{equation}
where $\ket{b}\!\rangle$ are the Ishibashi states \cite{Ishibashi:1988kg,Cardy:1989ir}.
Using the above lemma, the boundary state for the cloaking boundary condition $\gamma(\delta)$ can be written as 
\begin{align}\label{eq:cloaking-bnd-state-via-tildeJ}
    \bndstate{\gamma(\delta)} 
    &\,=\, 
    \delta_0
    \sum_{j\in\mathcal{J}}\dim(j) \bndstate{j} 
    \,=\, 
\delta_0 \sum_{x \in\mathcal{I}} \sqrt{S_{x1}} \sum_{j \in\mathcal{J}} 
    \frac{S_{1j}}{S_{11}} \frac{S_{jx}}{S_{x1}} \ket{x}\!\rangle
    \nonumber \\
&\!\!\overset{\eqref{eq:char-ortho}}=\,
\delta_0 \, \mathrm{Dim}(\mathcal{J}) \sum_{x \in\widetilde{\mathcal{J}}} \sqrt{S_{x1}}  \,
\ket{x}\!\rangle
\end{align}
In particular, only bulk fields in the sectors $\mathfrak{R}_x \otimes \mathfrak{R}_x$ with $x \in \widetilde{\mathcal{J}}$ occur in the cloaking boundary state for $\mathcal{J}$.

\smallskip
Using the explicit $S$-matrix of $M(p,q)$ from \eqref{eq:min-mod-Smatrix}, by definition $\widetilde{\mathcal{J}}$ consists of all $(x,y)$ such that
\begin{equation}
(-1)^{(s-1)(x-1)+(r-1)(y-1)} \,
  \frac{\sin( \pi rx/ t ) \sin( \pi s y  t) }{\sin( \pi x/ t ) \sin( \pi y  t) } 
\,=\,
\frac{\sin( \pi r/ t ) \sin( \pi s  t) }{
\sin( \pi / t ) \sin( \pi   t)}
\end{equation}
holds for all $(r,s) \in \mathcal{J}$.
Let us consider three examples in more detail:
\begin{enumerate}
    \item  $\mathcal{J} = \mathcal{I}$, the full set of topological defects.

    \item  $\mathcal{J} = \mathcal{R}$, where $\mathcal{R}$ denotes the first row of the Kac-table,
\begin{equation}
    \mathcal{R} = \{ (1,s) \,|\, s = 1,\dots,q-1 \} \ .
\end{equation}

    \item $\mathcal{J} = \mathcal{Z}$, where $\mathcal{Z}$ consists of the two representations $(1,1)$ and $(1,q-1)=(p-1,1)$. We require $p,q>2$, so that indeed $(1,1) \neq (1,q-1)$. The corresponding topological defects have $\mathbb{Z}_2$-fusion rules and describe the $\mathbb{Z}_2$ symmetry present in $M(p,q)$ with $p,q>2$.
\end{enumerate}
In case 1, by non-degeneracy of the $S$-matrix we obtain
\begin{equation}\label{eq:tilde-I-just-11}
    \widetilde{\mathcal{I}} \,=\, \{ (1,1) \} \ .
\end{equation}
For case 2, we have to find all $(x,y)$ such that
\begin{equation}
(-1)^{(s-1)(x-1)} \,
  \frac{   \sin( \pi s y  t) }{  \sin( \pi y  t) } 
\,=\,
\frac{ \sin( \pi s  t) }{
 \sin( \pi   t)}
\quad \text{for all} ~~
 s=1,\dots,q-1 \ .
\end{equation}
The solutions are $y=1$ and $x$ odd, as well as $y=q-1$ and $x$ such that $p-x$ is odd. These two sets are related by the identification on Kac-labels, so that we can write
\begin{equation}
    \widetilde{\mathcal{R}} \,=\, \big\{ \, (x,1) \, \big| \, 1 \le x \le p-1 \text{ and $x$ odd} \, \big\} \ .
\end{equation}
In case 3, the condition reduces to $(-1)^{q(x-1)+p(y-1)} =1$, and so 
\begin{equation}
    \widetilde{\mathcal{Z}} \,=\, \big\{ \, (x,y) \, \big| \, q(x-1)+p(y-1) \text{ even} \, \big\} \ .
\end{equation}

We now specialise further to the unitary minimal model $M(p,p+1)$ and check the stability condition from Section~\ref{sec:CFT-hexagonal} in each case, that is, we check if $\widetilde{\mathcal{J}}$ contains only irrelevant operators or not:
\begin{enumerate}
    \item $\widetilde{\mathcal{I}}$: Only contains the vacuum representation and the first contribution after the identity field is $T \bar T$, which has total weight $4$. The stability condition is satisfied.

    \item $\widetilde{\mathcal{R}}$: The lowest $L_0$-eigenvalue in the representation $(r,1)$ is $h_{(r,1)} = \frac{r-1}2 \big(\frac{r+1}{2t}-1\big)$, cf.\ \eqref{eq:MM-c-h}. This is $>1$ for $r \ge 3$, and so all bulk fields indexed by $\widetilde{\mathcal{R}}$ are irrelevant -- the stability condition is satisfied.

    \item $\widetilde{\mathcal{Z}}$: We always have $(1,3) \in \widetilde{\mathcal{Z}}$. Since $h_{(1,3)} = 2t-1 < 1$, the stability condition is never satisfied.    
\end{enumerate}
We will look more closely at the lattice model for $\mathcal{Z}$ and $\mathcal{R}$ in Sections~\ref{sec:Ising-lattice} and~\ref{sec:loop-models}, respectively.


\section{One-hole partition function in the Ising CFT}\label{sec:one-hole-comparison}

Following the methodology established in \cite{Brehm:2021wev}, we aim to compute approximations of the one-hole partition function in the Ising CFT for both open and closed channels. In this study, we introduce the cloaking boundary condition and calculate the associated anomaly factors in both channels. These results will serve as a verification of the structure constants and of the anomaly factor. 

This situation was treated numerically already in \cite{Brehm:2021wev} on the geometry shown in Figure~\ref{fig:Torus_with_a_hole} for a ratio of Ising CFT partition functions with elementary boundary conditions (and without including the Weyl anomaly). In \cite{Cheng:2023kxh} the lattice model obtained from the Ising CFT with cloaking boundary condition was investigated for a square lattice (in the limit $R \to 0$), and the corresponding transfer matrix was numerically diagonalised for different choices of cutoff, reproducing the low-lying spectrum of the Ising CFT to good accuracy. This was also done without including the Weyl anomaly (which just produces an overall shift in the spectrum).

\subsection{Ising data and cloaking boundary condition}

The Ising CFT, denoted as $M(3,4)$ in the A-series of minimal models and previously discussed in \cite[Sec. 8.1]{Brehm:2021wev}, represents the simplest unitary non-trivial model in this series. It comprises three primary fields, indexed by Kac-labels:
\begin{equation}
    1 := (1,1)\,,\quad \sigma := (1,2)\,,\quad \epsilon:= (1,3)\,.
\end{equation}
These fields have the following conformal weights and quantum dimensions:
\begin{align}\nonumber
 h_1 &= 0\,,\quad& h_\sigma &= \frac{1}{16}\,,\quad& h_\epsilon&=\frac{1}{2}\,,\\ 
 \dim(1) &= 1\,,\quad& \dim(\sigma) &= \sqrt{2}\,,\quad& \dim(\epsilon) &= 1\,.
\end{align}
The model's three elementary defects and boundary conditions share the same label set. The cloaking boundary condition, central to our current analysis, is defined as the superposition 
\begin{equation}
    \gamma = 1 + \sigma + \epsilon
\end{equation}
with the weight zero boundary field 
\begin{equation}
    \delta = \delta_0 \left(\one_1 + \sqrt{2}\, \one_\sigma + \one_\epsilon\right)\,.
\end{equation}
The non-trivial fusion rules of the chiral Ising CFT are
\begin{equation}
	\epsilon \times \epsilon = \one
	~,~~
	\epsilon \times \sigma = \sigma
~,~~
	\sigma \times \sigma = \one + \epsilon ~.
\end{equation}
and the state space $\mathcal{H}_{ab}$ of open channel states between boundaries $a$ and $b$ is given by $\mathcal{H}_{ab} = \bigoplus_{c}  N_{ab}^{~c}\, \mathfrak{R}_c$, see \cite{Cardy:1989ir}. We will only need
\begin{equation}
    \mathcal{H}_{\one\one} = \mathfrak{R}_{\one}
    \quad , \quad 
    \mathcal{H}_{\epsilon\epsilon} = \mathfrak{R}_{\one}
    \quad , \quad
    \mathcal{H}_{\sigma\sigma} = \mathfrak{R}_{\one} \oplus \mathfrak{R}_{\epsilon} ~.
\end{equation}

\subsection{The torus with a single hole}

\begin{figure}
    \centering
    \includegraphics[width=0.8\linewidth]{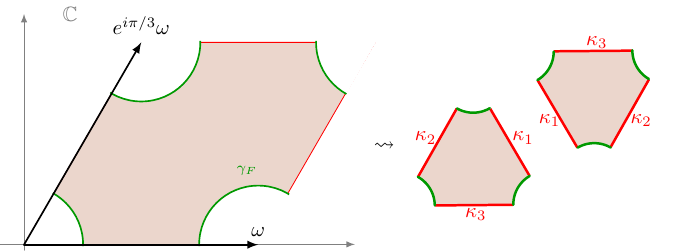}
    \caption{The torus with a single hole gets cut into two pieces }
    \label{fig:Torus_with_a_hole}
\end{figure}

Let us regard the torus as the quotient $\mathbb{C}/\Lambda$, where $\Lambda = \omega \mathbb{Z} + e^{i\pi/3}\omega \mathbb{Z}$ and $|\omega| = d > 2R$. 
Let the hole's center be at position $0$. Such a torus for $\omega\in\mathbb{R}$ is shown on the left of Figure~\ref{fig:Torus_with_a_hole}. In our construction the torus gets cut along the three shortest straight paths  on the torus from the boundary to itself into two triangle amplitudes as shown on the right of Figure~\ref{fig:Torus_with_a_hole}. 

\subsubsection{How to: Closed channel computation}

In the closed channel we use the expression \eqref{eq:cloaking-bnd-state-via-tildeJ} for the cloaking boundary state with $\mathcal{J} = \mathcal{I}$. Together with \eqref{eq:tilde-I-just-11} we get that only the vacuum Ishibashi state remains,
\begin{align}\label{eq:Ising-cloaking-bnd-state-is-vac-Ish}
    \bndstate{\gamma(\delta)} 
    &\,=\, 
    \delta_0 \, \mathrm{Dim}(\mathcal{I}) \sqrt{S_{11}} \, \ket{\one}\!\rangle 
    \,\overset{(*)}= \,
    \ket{\one}\!\rangle 
    \,= \,
    \sum_{\chi}    \ket{\chi}\otimes\overline{\ket{\chi}} 
    \,= \,
    \ket{\one} + \tfrac{2}{c} L_{-2}\bar{L}_{-2} \ket{\one} + \dots
\end{align}
where in $(*)$ we set $\delta_0 = S_{11}^{3/2}$ and used $\mathrm{Dim}(\mathcal{I}) = 1/S_{11}$. The sum is over an ON-basis of $\mathfrak{R}_1$. 

The above boundary state is for hole radius $R=1$. How to obtain the expansion for arbitrary radius via the transformation rules (i)--(iv) is explained in Appendix~\ref{eq:radius-bnd-state}. For a torus $T^2$ with circular hole of radius $R$ this gives
\begin{equation}
    \mathcal{A}\left(T^2,\gamma(\delta)_R\right) 
=
    \sum_{\chi} 
    R^{2h_\chi-\frac{c}{6}} 
    \mathcal{A}\left(T^2,(\chi\otimes\overline{\chi})(0;z)\right) \ ,
\end{equation}
where the sum is over the same ON-basis of $\mathfrak{R}_1$ as in \eqref{eq:Ising-cloaking-bnd-state-is-vac-Ish}, and the field $\chi\otimes\overline{\chi}$ is inserted at 0 with the standard local coordinate on $\mathbb{C}$. 

The unnormalised torus-one point functions $\mathcal{A}(T^2,(\chi\otimes\overline{\chi})(0;z))$
can be computed from the recursive algorithm presented in \cite[Sec. 4.2.3]{Brehm:2021wev}. Note that the result does not explicitly depend on any  structure constants of primary bulk or boundary fields.
However, the structure of the vacuum representation $\mathfrak{R}_1$ depends on the central charge and the result is indeed different for different Virasoro minimal models. Also note that the computation here is simpler than the one in \cite[Sec. 5.2]{Brehm:2021wev} because the cloaking boundary condition projects on the vacuum Ishibashi state in the closed channel and we therefore do only need to compute states in the vacuum representation. In an explicit computation this allows us to go higher in the closed channel cutoff, i.e.\ in the maximum descendant level we perform the sum to. 

If we compensate for the anomaly factor $R^{-\frac c6}$ as in \eqref{eq:zero-radius-limit-is-id}, in the $R \to 0$ limit we obtain the torus partition function of the Ising CFT,
\begin{equation}\label{eq:Ising-numerics-closed-limit}
    \lim_{R\to 0} R^{\frac1{12}} \mathcal{A}\left(T^2,\gamma(\delta)_R\right) 
    = Z_\text{Ising CFT}(\tau = e^{\pi i /3})
    = 1.88.. \ , 
\end{equation}
where the numerical value was e.g.\ given in \cite[Eqn.\,8.11]{Brehm:2021wev}.

\subsubsection{How to: Open channel computation}

\begin{figure}[t]
    \centering
    \includegraphics[width=.8\linewidth]{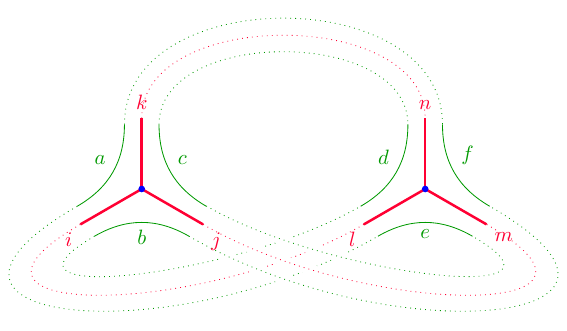}
    \caption{The graphical representation of the contraction of the two amplitudes that produce the partition function of our construction from a torus with a single hole. The graphical representation of the amplitudes was given in \eqref{eq:graphical_rep_ampl}, where for simplicity we omit the indices for the states in the representation. A dotted line stands for a contraction such that indices connected by a line have to be equal. Inspecting the graph shows that all the boundary indices (in green) have to be the same. }
    \label{fig:contraction}
\end{figure}

For a single hole the computation in the open channel simplifies because only boundary preserving fields $\kappa^{(aa)}$ contribute. In other words, no boundary changing fields enter the computation. 

This follows from expanding out the cloaking boundary condition of the single hole in terms of simple boundary conditions, or alternatively from the way in which the two triangle amplitudes are contracted (Figure~\ref{fig:contraction}): 
\begin{align}
    &\sum_{\begin{smallmatrix}ijklmn\\abcdef\\\alpha\beta\gamma\delta\epsilon\varphi\end{smallmatrix}} 
    T_{d,R}^{\gamma(\delta)}(\kappa_{i\alpha}^{ba},\kappa_{j\beta}^{cb},\kappa_{k\gamma}^{ac}) \,
    T_{d,R}^{\gamma(\delta)}(\kappa_{l\delta}^{ed},\kappa_{m\epsilon}^{fe},\kappa_{n\varphi}^{df})
    \nonumber\\[-2em]
    & \hspace{10em} \times ~
    \delta_{il} \, \delta_{jm} \, \delta_{kn}~
    \delta_{\alpha\delta} \, \delta_{\beta\epsilon} \, \delta_{\gamma\varphi}
    ~
    \delta_{af} \, \delta_{cd} \, \delta_{bd} \,  \delta_{ae} \, \delta_{bf} \, \delta_{ce} \nonumber\\[1em]
    &~=\, \sum_a\sum_{\begin{smallmatrix}ijk\\\alpha\beta\gamma\end{smallmatrix}} \left(T_{d,R}^{\gamma(\delta)}(\kappa_{i\alpha}^{aa},\kappa_{j\beta}^{aa},\kappa_{k\gamma}^{aa})\right)^2
    =\,
    \sum_a\sum_{\begin{smallmatrix}ijk\\\alpha\beta\gamma\end{smallmatrix}} \delta_0 \dim(a)\left(T_{d,R}^{aaa}(\kappa_{i\alpha}^{aa},\kappa_{j\beta}^{aa},\kappa_{k\gamma}^{aa})\right)^2
\end{align}

\noindent 
We can, hence, compute the one-hole partition functions for each individual simple boundary condition and add up the results. 
\subsubsection*{Anomaly factor}

The constant $A$ in the anomaly factor $e^A$ that enters the triangle amplitude is given in \eqref{eq:triangle-anomaly-factor}: 
\begin{equation}
    A \,=\, 
    \frac{c}{8}\Big(
    \, L_{\widetilde T_\triangleleft}(\log |\partial E|^2) - L_{D_+}(\log |\partial \phi_0|^2) \, \Big) \ .
\end{equation}
Except for an overall factor of $c$ it is independent of any specifics of the model. The integrals appearing in $L_{\widetilde T_\triangleleft}(\log |\partial E|^2)$ and $L_{D_+}(\log |\partial \phi_0|^2)$ depend on $t$ and can be computed numerically (see Appendix \ref{app:numerics}).

\subsubsection*{The formula}

Write $B_i$ for an ON-basis in the irreducible Virasoro representation $\mathfrak{R}_i$ with respect to the pairing \eqref{eq:pairing-on-desc}.
Using
the boundary state spaces in the Ising CFT, we can write
\begin{align}\nonumber
    \mathcal{A}\left(T^2,\gamma(\delta)_R\right) &= \sum_a\sum_{\begin{smallmatrix}ijk\\\alpha\beta\gamma\end{smallmatrix}} \delta_0 \dim(a)\left(T_{d,R}^{aaa}(\kappa_{i\alpha}^{aa},\kappa_{j\beta}^{aa},\kappa_{k\gamma}^{aa})\right)^2 \\
    &= \delta_0\sum\limits_{\alpha\beta\gamma\in B_\one}\Big( T^{\one\one\one}_{d,R}(\kappa^{\one\one}_{\one\alpha},\kappa^{\one\one}_{\one\beta},\kappa^{\one\one}_{\one\gamma})^2 
    + T^{\epsilon\epsilon\epsilon}_{d,R}(\kappa^{\epsilon\epsilon}_{\one\alpha},\kappa^{\epsilon\epsilon}_{\one\beta},\kappa^{\epsilon\epsilon}_{\one\gamma})^2 \Big) \\
    \nonumber
    &\quad+ \sqrt{2}\,\delta_0\Bigg(\sum\limits_{\alpha\beta\gamma\in B_\one} T^{\sigma\sigma\sigma}_{d,R}(\kappa^{\sigma\sigma}_{\one\alpha},\kappa^{\sigma\sigma}_{\one\beta},\kappa^{\sigma\sigma}_{\one\gamma})^2 + 3 \,\sum\limits_{\begin{smallmatrix}\alpha\beta\in B_\epsilon \\ \gamma\in B_\one\end{smallmatrix}} T^{\sigma\sigma\sigma}_{d,R}(\kappa^{\sigma\sigma}_{\epsilon\alpha},\kappa^{\sigma\sigma}_{\epsilon\beta},\kappa^{\sigma\sigma}_{\one\gamma})^2\Bigg) \ .
\end{align}
We now need to use \eqref{eq:triangle-amplitude-with-anomaly_computed} and  \eqref{eq:Nijkabc-expression}. The F-symbols appearing here are
\begin{align}
    \Fsym{\one}{\one}{\one}{\one}{\one}{\one} &= \Fsym{\epsilon}{\one}{\epsilon}{\epsilon}{\one}{\one}  = \Fsym{\sigma}{\one}{\sigma}{\sigma}{\one}{\one} = 1\, \quad \text{and}
    \nonumber\\
    \Fsym{\sigma}{\one}{\sigma}{\sigma}{\epsilon}{\epsilon} &= \frac12 \ .
\end{align}
For $\delta_0$ we use $\delta_0 = S_{11}^{3/2}$ as we did in \eqref{eq:Ising-cloaking-bnd-state-is-vac-Ish} for the closed channel, with $S_{11} = \frac12$. Altogether, this gives
\begin{equation}\label{eq:Ising-numerics-open-limit}
    \begin{split}
        \mathcal{A}\left(T^2,\gamma(\delta)_R\right) =  \frac{e^{2A}}{2} \Bigg(\,&\sum\limits_{\alpha\beta\gamma\in B_\one} 3  B\big( \Gamma |\one\alpha\rangle, \Gamma |\one\beta\rangle, \Gamma  |\one\gamma\rangle \big) \\
        + &\,\sum\limits_{\begin{smallmatrix}\alpha\beta\in B_\epsilon\\\gamma\in B_\one\end{smallmatrix}} ~ B\big( \Gamma |\epsilon\alpha\rangle, \Gamma |\epsilon\beta\rangle, \Gamma  |\one\gamma\rangle \big)\Bigg)\,.
    \end{split}
\end{equation}
In the limit of $\frac{R}{d} \to 1/2$ every contribution other than the vacuum contribution gets suppressed and without the anomaly factor we obtain
\begin{equation}
    \lim_{R \to d/2} e^{-2A}\mathcal{A}\left(T^2,\gamma(\delta)_R\right) =  \frac{3}{2}\,.
\end{equation}

\subsection{Comparison of the two channels in the Ising CFT}

As in the previous paper we want to compare the results of the one-hole partition function in the open and closed channel. This serves as a check on the structure constants and on the anomaly factor and emphasises how the latter actually affects the result. 

We evaluate the Ising CFT on a torus with periods $\omega_1 =1$ and $\omega_2=e^{i\pi/6}$. We computed the closed channel up to descendants of conformal dimension $h=14$ and in the open channel up until descendants of level $7$.\footnote{
    \textsf{Mathematica} files containing the routines used to arrive at the results presented here can be found as ancillary files to the arXiv submission. This includes the recursive routines to compute the interaction vertex, recursive routines for the torus one-point functions, and the numerical integration of the anomaly action.}

\begin{figure}[bt]
    \raisebox{14em}{a)}\hspace{-1.5em}
    \includegraphics[width=.49\textwidth]{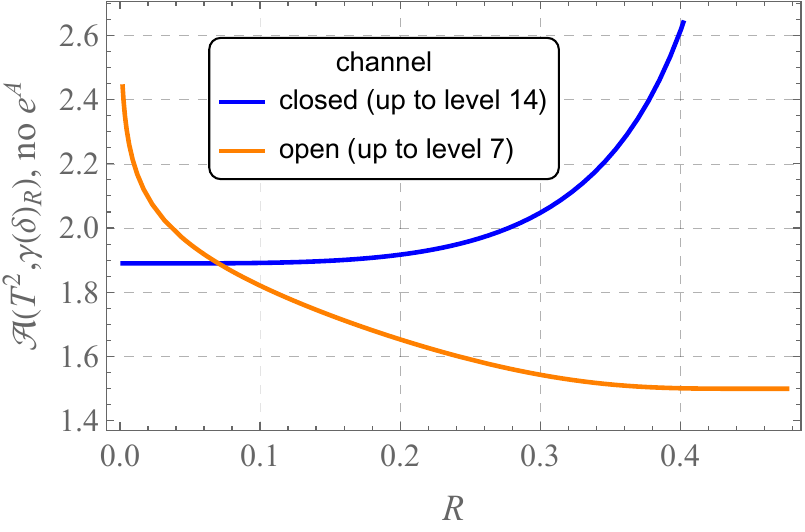}
    \hspace{1.5em}
    \raisebox{14em}{b)}\hspace{-1.5em}
    \includegraphics[width=.49\textwidth]{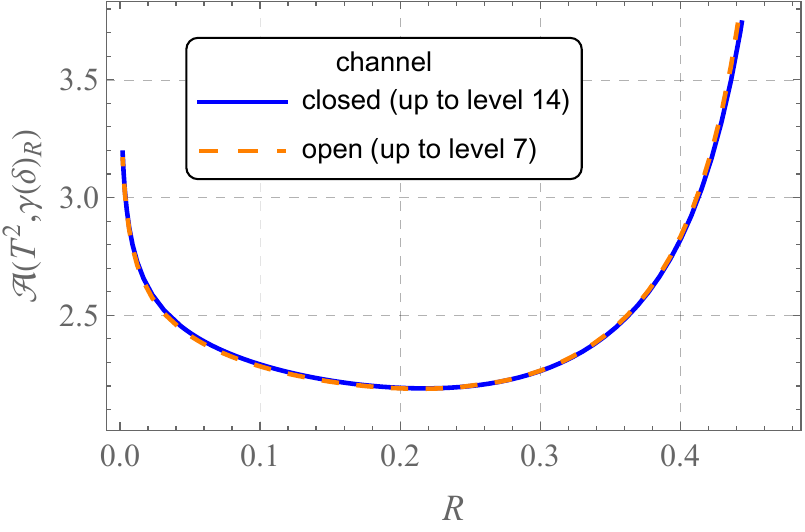}
    \caption{The one-hole partition function on a torus approximated in the closed channel (blue) and the open channel (orange). a) Result without any anomaly factor. b) Result with the anomaly factor $R^{-\frac{1}{12}}$ for the closed channel and with the anomaly factor $e^{2A}$ in \eqref{eq:triangle-amplitude-with-anomaly_computed} with $A$ in 
    \eqref{eq:triangle-anomaly-factor} 
    for the open channel.}
    \label{fig:compare1}
\end{figure}

Figure~\ref{fig:compare1}\,a) shows the result in the two channels without the anomaly factors, and we see that the two curves do not agree and do not even show the same behaviour.
Note that the limits of the closed and open curves are as in \eqref{eq:Ising-numerics-closed-limit} and \eqref{eq:Ising-numerics-open-limit}.
    
\begin{figure}[bt]
    \centering
    \includegraphics[width=.49\textwidth]{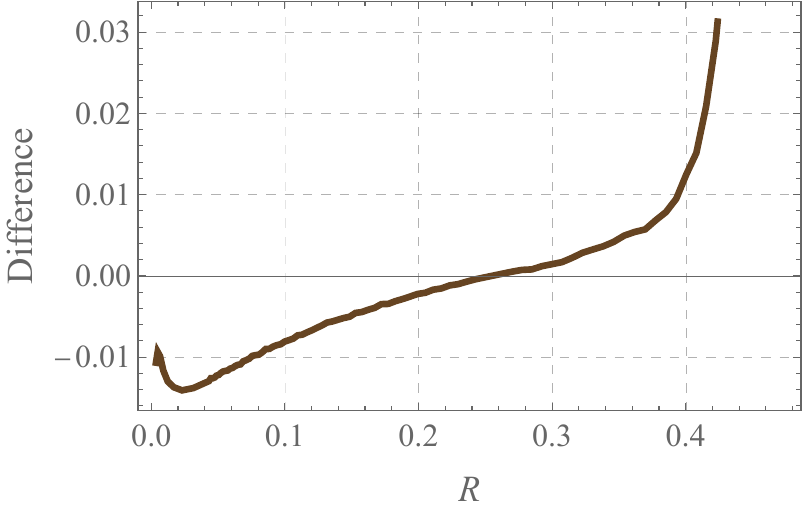}
    \caption{Difference between the closed and open channel approximation in Figure~\ref{fig:compare1}\,b).}
    \label{fig:compare2}
\end{figure}

In clear contrast, Figure~\ref{fig:compare1}\,b) shows the two channels when we include the anomaly factors. Only then the two channels match to good accuracy, which is also visible when plotting the difference as in Figure~\ref{fig:compare2}. The growing but small differences as $R$ goes to $1/2$ and as $R$ goes to $0$ are expected from the approximation and will shrink with increasing levels. 

The good agreement of the numerical results in these two channels provides a strong test of many of ingredients developed in this paper and in \cite{Brehm:2021wev}, as well as of their numerical implementation:
\begin{itemize}
    \item the cutting of amplitudes (Section~\ref{sec:cutting-amplitude}) with a sum over normalised intermediate states \eqref{eq:normState},
    \item the transformation rules of descendants \eqref{eq:Gamma-for-clipped-triang} and the anomaly factors \eqref{eq:triangle-anomaly-factor} for the uniformisation map \eqref{eq:uniform} to the unit disc, 
    \item recursive relations to compute descendant correlators on the disc \eqref{eq:recusive-relation} and on the torus \cite[Sec.\,4.2.3]{Brehm:2021wev},
    \item the clipped triangle amplitude for primary fields \eqref{eq:triangle-amplitude-with-anomaly_computed} as determined by the various structure constants.
\end{itemize}

\section{Relation to the Ising lattice model}\label{sec:Ising-lattice}

In this short section we construct the lattice model obtained from the Ising CFT with preserved topological symmetry $\mathcal{F}$ chosen to be the $\mathbb{Z}_2$-symmetry, i.e.\ the fusion category with simple objects $\mathcal{Z} = \{ 1, \epsilon \}$. In Section~\ref{sec:cloaking-bnd-states-properties} we saw that this choice does not satisfy the stability condition. Indeed, the boundary state \eqref{eq:cloaking-bnd-state-via-tildeJ} of the cloaking boundary condition $\gamma_{\mathcal{Z}}(\delta)$ is 
\begin{equation}
    \bndstate{\gamma_{\mathcal{Z}}(\delta)} 
\,=\,
\sqrt2 \, \delta_0  \big(  \,
\ket{\one}\!\rangle + \ket{\epsilon}\!\rangle \big) \ .
\end{equation}
This contains the bulk field $\epsilon(0)$ of weights $(\frac12,\frac12)$ as a summand, which is relevant and generates the temperature perturbation of the Ising CFT.

The space of boundary fields on $\gamma_{\mathcal{Z}}$ is
\begin{equation}
	\mathcal{H}_{\gamma\gamma} 
 \,=\, 
 \mathcal{H}_{\one\one} \oplus \mathcal{H}_{\one\epsilon} \oplus \mathcal{H}_{\epsilon\one} \oplus \mathcal{H}_{\epsilon\epsilon}
 \,=\, 
 \mathfrak{R}_\one \oplus \mathfrak{R}_\epsilon \oplus \mathfrak{R}_\epsilon \oplus \mathfrak{R}_\one 
 \ .
\end{equation}
We will work at lowest non-trivial cutoff, which in this case is $h_\text{max} = \frac12$. 
The resulting model has four states on each edge, namely the primary boundary (changing) fields in each of the four sectors above,
\begin{equation}\label{eq:Ising-lattice-4-states}
    \{ \kappa^{(\one\one)}_\one ,  \kappa^{(\one\epsilon)}_\epsilon ,  \kappa^{(\epsilon\one)}_\epsilon ,  \kappa^{(\epsilon\epsilon)}_\one   \} \ .
\end{equation}

For the computation of the interaction vertex we will temporarily work in a more general setting as we can re-use the computation in the next section. Namely, we work in $M(p,p+1)$ with two boundary conditions $a,b$ and a boundary changing field $f$ between $a$ and $b$. 
We require the three-point conformal block of $f$ to be zero, $N_{ff}^{~f}=0$.
The Ising CFT situation is recovered for the choices $p=3$, $a,b \in \{\one,\epsilon\}$, $f=\epsilon$. 
The only boundary three-point amplitudes involving $\kappa_\one^{(aa)}$ and $\kappa_f^{(ab)}$ which can be non-zero are:
\begin{equation}\label{eq:only-two-disc-amplitudes}
\includegraphics[valign=c]{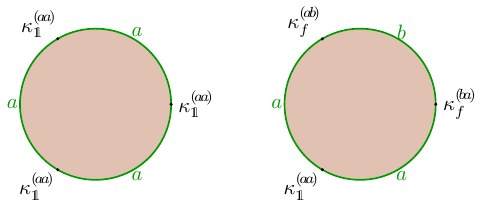}
\end{equation}
Combining 
\eqref{eq:triangle-amplitude-cloaking}, \eqref{eq:Nijkabc-expression}
and
\eqref{eq:F-matrix-S-identity},
the clipped triangle amplitude simplifies in these two cases to 
\begin{align}\label{eq:min-cutoff-vertices}
    T_{aaa} := T_{d,R}^{\gamma(\delta)}\!\left(\kappa_\one^{(aa)},\kappa_\one^{(aa)},\kappa_\one^{(aa)}\right) 
    &= F_{d,R} \ ,
    \nonumber \\
    T_{aab} := T_R^{\gamma(\delta)}\!\left(\kappa_f^{(ba)},\kappa_f^{(ab)},\kappa_\one^{(aa)}\right) 
    &= 
    F_{d,R} \left(\frac{\dim(b)}{\dim(a)}\right)^{1/6} x(R) \ ,
\end{align}
where 
\begin{equation}
F_{d,R} = e^A \delta_0^{\,1/2} S_{11}^{-1/4}
~~,\quad
x(R) = \left({4/(3\sqrt{3}X)}\right)^{2h_f} \ .
\end{equation}
By cyclic symmetry, this determines the $T_{abc}$ to be
\begin{align}\label{eq:Tabc-via-cyclic}
    T_{abc} = \begin{cases} 
    T_{xxx} &; \text{ all indices equal to some $x$ ,}
    \\
    T_{xxy} &; \text{ two indices equal to $x$ and one $y$ for $x \neq y$ ,}
    \\
    0 &; \text{ else .}
\end{cases}    
\end{align}
Recall from \cite[Sec.\,6.1]{Brehm:2021wev} that for $R \to d/2$ (touching hole limit) we have $X \to \infty$, and for $R \to 0$ (small hole limit) we get $X \to 1$. Accordingly we obtain
\begin{equation}\label{eq:x-limit-from-X-limit}
x(\tfrac d2) := \lim_{R \to d/2} x(R) = 0
\quad , \quad
x_\text{max}
:= \lim_{R \to 0} x(R) = \big( 16/27 \big)^{h_f}
\ .
\end{equation}

We now restrict our attention again to the Ising CFT $M(3,4)$. Then \eqref{eq:min-cutoff-vertices} and \eqref{eq:x-limit-from-X-limit} specialise to
\begin{equation}
    T_{aaa} = F_{d,R}
    ~,~~
    T_{aab} = 
    F_{d,R} \, x(R)
    ~,~~
    a,b \in \{\one,\epsilon\}
    ~,~~
    x_\text{max} = \tfrac{4}{3\sqrt{3}} \ .
\end{equation}
Which of the four states in \eqref{eq:Ising-lattice-4-states} is inserted on an edge of the hexagonal lattice is determined uniquely by the boundary conditions $a,b \in \{\one,\epsilon\}$ of the hole on either side of the edge (cf.\ Figure~\ref{fig:intro-lattice-construction}). This is equivalent to assigning a degree of freedom to each vertex of the original triangular lattices which takes values in $\{\one,\epsilon\}$. 

Let $\tilde V$, $\tilde E$, $\tilde F$ be the set of vertices, edges and faces of the triangular lattice  (these correspond to $F,E,V$ in the dual hexagonal lattice). The partition function of the lattice model is given by 
\begin{align}
    Z(R) 
    &= 
    \sum_\varphi \prod_{f \in \tilde F} T_{\varphi(f(1))\varphi(f(2))\varphi(f(3))} \ ,
\end{align}
where the sum is over all functions $\varphi : \tilde V \to \{\one,\epsilon\}$, and $f(1),f(2),f(3) \in \tilde V$ denotes the three vertices of the triangle $f \in \tilde F$. We introduce an additional constant $\lambda \neq 0$ to be adjusted later and rewrite $Z(R)$ as
\begin{align}\label{eq:ZR-Ising-example}
    Z(R) 
    &=
    (F_{d,R} / \lambda)^{|\tilde F|}
    \sum_\varphi \prod_{f \in \tilde F}
    \begin{cases}
        \lambda &:~ \varphi(f(1))=\varphi(f(2))=\varphi(f(3)) \ ,
        \\
        \lambda \, x(R) &:~ \text{else .}
    \end{cases}
\end{align}

We would like to relate $Z(R)$ to the standard Ising model partition function on a triangular lattice,
\begin{equation}
    Z_\text{Ising}(\beta) = \sum_{s} \prod_{\langle ij \rangle} e^{\beta s_i s_j} \ .
\end{equation}
Here, $s$ is an assignment of a spin $s_i \in \{ \pm 1 \}$ to each vertex $i \in \tilde V$ of the triangular lattice, the product $\langle ij \rangle$ is over neighbouring sites, and $\beta$ is the inverse temperature. Each edge is shared by two faces, so that in terms of faces, the partition function can be written as
\begin{align}
    Z_\text{Ising}(\beta) 
    &= \sum_{s} 
    \prod_{f \in \tilde F}
    \exp\big( \tfrac12 \beta( s_{f(1)}s_{f(2)} + s_{f(2)}s_{f(3)} + s_{f(3)}s_{f(1)} ) \big) 
    \nonumber \\
    &=
    \sum_s \prod_{f \in \tilde F}
    \begin{cases}
        e^{3\beta/2} &:~ s_{f(3)} = s_{f(2)} = s_{f(3)}  \ ,
        \\
        e^{-\beta/2} &:~ \text{else .}
    \end{cases}
\end{align}
If we set $\lambda = e^{3 \beta / 2}$ and $x(R) = e^{-2\beta}$, this agrees with $Z(R)$ up to a constant:
\begin{equation}\label{eq:relation-Ising-lattice}
    Z_\text{Ising}(\beta)
    \,=\, \big(x(R)^{\frac34} F_{d,R}\big)^{-|\tilde F|}
    \, Z(R)
    \quad , \quad
    \beta = -\tfrac12 \log x(R) \ .
\end{equation}
Given the range of $x(R)$ in \eqref{eq:x-limit-from-X-limit}, as $R$ varies between $\frac d2$ and $0$, the inverse temperature $\beta$ will cover the range
\begin{equation}
    \beta \in (\beta_\text{min},\infty)
    \qquad \text{where} \quad
    \beta_\text{min} = \tfrac14 \log \tfrac{27}{16} = 0.13..
    \ .
\end{equation}
The phase diagram of the 2d Ising lattice model has an unstable critical point $\beta = \beta_*$ and is massive away from the critical point. On the triangular lattice, the critical point is (see e.g.\ \cite[Sec.\,4.5]{Mussardo-book})
\begin{equation}
    \beta_* = \frac{\log 3}{4} = 0.27..
    \ .
\end{equation}
Thus, 
if we apply our lattice model construction to the Ising CFT for the choice of $\mathcal{F}$ and $h_\text{max}$ as described above, we recover the Ising lattice model on a triangular lattice. The range of inverse temperatures $\beta$ covered by varying $R$ includes the critical point $\beta_*$ of the Ising model on a triangular lattice.

On the other hand, with the present choice of $\mathcal{F}$ the stability condition for the cloaking boundary state is not satisfied. Punching small holes into the worldsheet is thus a relevant perturbation of the Ising CFT and drives it towards a massive theory. This is also the behaviour of the lattice model which is massive near $R=0$. As we would expect, we do not get a region near $R=0$ where the lattice model recovers the Ising CFT as in the phase diagram \eqref{eq:general-phase-diagram}. Of course, we only treated the smallest non-trivial value for the cutoff $h_\text{max}$, but we do not expect the behaviour near $R=0$ to change for larger $h_\text{max}$.

\smallskip
We conclude the discussion of the Ising lattice model example with a number of remarks.
\begin{enumerate}
    \item 
    One
could have build the cloaking boundary condition on top of the free boundary condition of the Ising CFT, i.e.\ we could have taken $\gamma_{\mathcal{Z}}(\delta)\bndstate{\sigma}$ instead of $\gamma_{\mathcal{Z}}(\delta)\bndstate{1}$. Up to normalisation this results just in $\bndstate{\sigma}$ as this boundary state is invariant under the $\mathbb{Z}_2$-symmetry. The resulting model (again at cutoff $h_\text{max}=\frac12$) has a $\mathbb{Z}_2$-degree of freedom on each edge of the hexagonal lattice (rather than four states per edge as in \eqref{eq:Ising-lattice-4-states}) and interaction vertex forces their sum to be even around each vertex.

\item The $\mathbb{Z}_2$-symmetry considered here for the Ising CFT is present in every A-type unitary minimal model $M(p,p+1)$. The corresponding defect line has Kac-label $(1,p)$, and the boundary changing field between the boundary conditions $(1,1)$ and $(1,p)$ has weight $h_{(1,p)} = \frac14 (p-2)(p-1)$. The smallest non-trivial cutoff is $h_\text{max} = h_{(1,p)}$ which is larger than $2$ for $p \ge 5$. Hence for $p \ge 5$ one would need to include vacuum descendants in the truncated state space of the lattice model.

\item
On the other hand, for the tricritical Ising CFT $M_{4,5}$ one has $h_{(1,4)} = \frac32$ and the state space of the lattice model for $h_\text{max} = \frac32$ is again of the form \eqref{eq:Ising-lattice-4-states}.
One obtains the same model $Z(R)$ as in \eqref{eq:ZR-Ising-example}, but with different values for $F_{d,R}$, $\lambda$ and $x(R)$. The relation \eqref{eq:relation-Ising-lattice} to the usual Ising lattice model with $\beta = - \frac12 \log x(R)$
still holds, but now for the new values of $F_{d,R}$ and $x(R)$.
The new value for $x_\text{max}$ in \eqref{eq:x-limit-from-X-limit} is given by $(16/27)^{\frac32}$ and accordingly $\beta_\text{min} = \frac34 \log \frac{27}{16} = 0.39..$, which no longer covers the critical point $\beta_*$.
The resulting lattice model thus lies in the trivial universality class for the entire range of $R$.

In the next section we will see that when preserving a larger topological symmetry than just $\mathbb{Z}_2$, the universality class of the lattice model obtained from $M(4,5)$ has the correct central charge to produce the tricritical Ising CFT.
\end{enumerate}

\section{Relation to loop models}\label{sec:loop-models}

In this section we restrict ourselves to unitary minimal model CFTs $M(p,p+1)$ and consider the lattice model for the fusion-subcategory formed by representations with Kac-labels $(1,s)$ at lowest non-trivial cutoff. It turns out that the resulting lattice model can be exactly rewritten as a loop model. The phase diagram of the loop model supports the idea that for a high enough cutoff there is a critical radius $R_C$ such that for hole radius $R<R_C$, the universality class of the lattice model is given by the CFT one started from. In fact, in the present example already the lowest non-trivial cutoff is high enough.

\subsection{Lattice model at lowest non-trivial cutoff}

Consider the unitary A-type minimal model $M(p,p+1)$ for $p \ge 3$ of central charge 
\begin{equation}\label{eq:unitary-minmod-c}
    c = 1 - \frac{6}{p(p+1)} 
\end{equation}
as in Section~\ref{sec:vir-min-mod}.
Let $\mathcal{R}$ be the fusion category formed by representations in the first row of the Kac-table, i.e.\ by representations with Kac-label $(1,s)$, $s=1,\dots,p$. We have seen in Section~\ref{sec:cloaking-bcs} that the cloaking boundary state for $\mathcal{R}$ satisfies the stability condition from Section~\ref{sec:CFT-hexagonal}.

The cloaking boundary condition for $\mathcal{R}$ is $\gamma = \bigoplus_{s=1}^{p} (1,s)$, and accordingly the boundary fields on $\gamma$ have representation labels of the form $(1,s)$ as well. The conformal weights of the corresponding primary fields are
\begin{equation}
    h_{(1,s)} = \frac{(s-1)\big(p(s-1)-2\big)}{4(p+1)} \ .
\end{equation}

To define the lattice model we need to choose a cutoff $h_\mathrm{max}$ for the open channel state space $\mathcal{H}_{\gamma\gamma}$. 
For $h_\mathrm{max} = 0$ only the $p$ weight zero boundary fields on $\gamma$ survive. These are the projectors onto the $p$ elementary boundary conditions contained in $\gamma$. One obtains the topological field theory described in \cite[Sec.\,3]{Brehm:2021wev}. For the comparison to loop models we choose the smallest non-trivial cutoff
\begin{equation}
    h_\mathrm{max} = h_{(1,2)} = \frac14\,\frac{p-2}{p+1} \ .
\end{equation}

We abbreviate $\one = (1,1)$, $f = (1,2)$ and we write $a$ for the boundary condition $(1,a)$. The basis of normalised states in the cutoff state space $\mathcal{H}_{\gamma\gamma}^{\le h_\mathrm{max}}$ is
\begin{equation}\label{eq:minimal-nontrivial-basis}
\big\{  \kappa_{\one}^{(aa)} \,\big|\, a=1,\dots,p \big\}
    ~\cup~
\big\{  \kappa_{f}^{(a,a+1)} \,\big|\, a=1,\dots,p-1 \big\}
    ~\cup~
\big\{  \kappa_{f}^{(a,a-1)} \,\big|\, a=2,\dots,p \big\} \ .
\end{equation}
The overall dimension of the truncated state space is therefore 
\begin{equation}
    \dim \mathcal{H}_{\gamma\gamma}^{\le h_\mathrm{max}} \,=\, 3p-2 \ .
\end{equation}
For the lattice model this means that each edge can be in one of $3p-2$ states. For example, in the Ising CFT ($p=3$), each edge can be in one of $7$ states (as opposed to 4 states as in the $\mathbb{Z}_2$-symmetry preserving model in Section~\ref{sec:Ising-lattice}).

Denote the basis vectors in $\mathcal{H}_{\gamma\gamma}^{\le h_\mathrm{max}}$ by $\{\kappa_\alpha \,|\, \alpha = 1,\dots,3p-2 \}$. Let $\Lambda$ be a hexagonal lattice, and write $E$ for its set of edges and $V$ for its vertices. For a vertex $V$ denote by $v_1,v_2,v_3 \in E$ the three adjacent edges in counter-clockwise order. 
A configuration of lattice spins is a function $\alpha : E \to \{1,2,\dots,3p-2\}$. 
The partition function of the lattice model is
\begin{equation}\label{eq:original-partition-function}
    Z = \sum_{\alpha} \prod_{v \in V} T_{d,R}^{\gamma(\delta)}(\kappa_{\alpha(v_1)},\kappa_{\alpha(v_2)},\kappa_{\alpha(v_3)}) ~,
\end{equation}
where the sum is over all spin configurations $\alpha : E \to \{1,\dots,3p-2\}$ and $T_{d,R}^{\gamma(\delta)}$ was given in \eqref{eq:triangle-amplitude-cloaking}.
We will now proceed to simplify and rewrite this partition function to eventually arrive at a representation as a loop model.

\smallskip
Since the states $\kappa^{(ab)}_i$ with representation $i=\one$ must have the same boundary condition on either side (i.e.\ $a=b$), and since for $i=f$ the boundary label must change by $\pm1$, there are just two types of boundary three-point amplitudes we have to consider, namely those given in \eqref{eq:only-two-disc-amplitudes} with $b = a \pm 1$.
We will use the notation $T_{abc}$ as defined in \eqref{eq:min-cutoff-vertices} and \eqref{eq:Tabc-via-cyclic}. The maximal value $x_\text{max}$ for $x(R)$ in \eqref{eq:x-limit-from-X-limit} now reads
\begin{equation}\label{eq:x-max-loopmodel}
x_\text{max}
=
\big( 2 / 3^{3/4} \big)^{(p-2)/(p+1)}
\ .
\end{equation}

We set $d_{b,a} := \dim(b)/\dim(a)$ and represent the vertices in \eqref{eq:min-cutoff-vertices} as
\begin{align}\label{eq:three-vertices-for-loops}
    T_{aaa} &=  
    \includegraphics[valign=c]{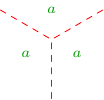}
    = F_{d,R}
    \nonumber\\
    T_{aab} &= 
    \includegraphics[valign=c]{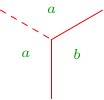}
    = 
    F_{d,R} \, x(R)\, d_{b,a}^{1/6}
    \nonumber\\
    T_{bba}  &= 
    \includegraphics[valign=c]{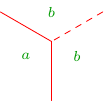}
    = 
    F_{d,R} \,x(R) \, d_{b,a}^{-1/6} 
\end{align}
The solid line indicates that the representation $i$ of the corresponding state $\kappa^{(ab)}_i$ is $i=f$, and the dashed line stands for $i=\one$. 
Of course, the third vertex is just the second one with permuted labels and appropriately rotated, but this way of presenting it emphasises that one incurs a factor of $d_{b,a}^{1/6}$ when the solid line turns to the right and $d_{b,a}^{-1/6}$ when it turns to the left.

\begin{figure}[tb]
    \centering
    \includegraphics[width=.7\linewidth]{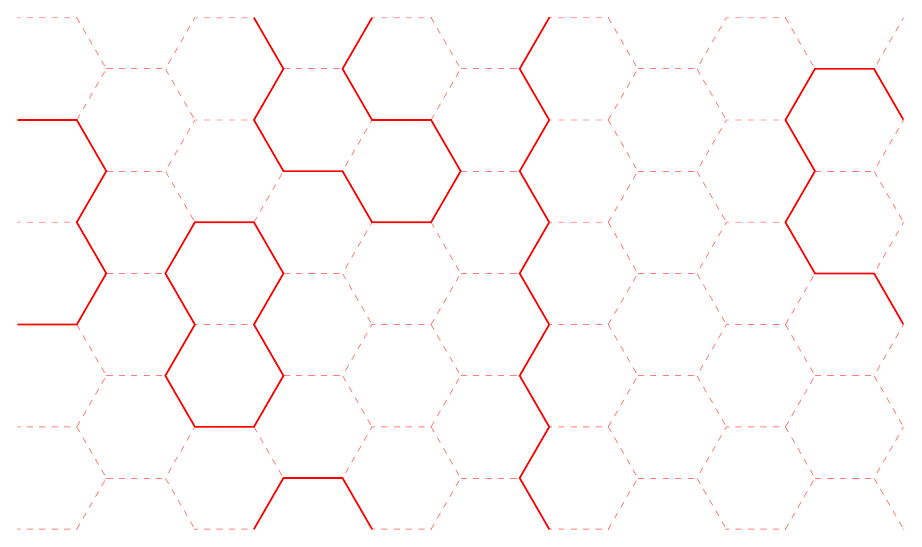}

\caption{Doubly periodic hexagonal lattice with loops indicating the boundaries between different values of the height variable.}
\label{fig:loops-on-hex-lattice}
\end{figure}

A sample configuration on a lattice with periodic boundary conditions is shown in Figure~\ref{fig:loops-on-hex-lattice}.

\subsection{Mapping to a loop model}

The $3p-2$ states $\kappa^{(ab)}_i$ in \eqref{eq:minimal-nontrivial-basis} are uniquely characterised by the values of $a,b$, subject to the constraint that $b \in \{a-1,a,a+1\}$. 
We can use this to define weights $T_{abc}$ as in \eqref{eq:Tabc-via-cyclic}.
Let $F$ be the set of faces of a doubly-periodic hexagonal lattice $\Lambda$, and let $V$ the set of vertices.
The partition function \eqref{eq:original-partition-function} of the lattice model can now be rewritten by assigning a value $a \in \{1,2,\dots,p\}$ to each face of the hexagonal lattice, so that configurations are functions $\varphi : F \to \{1,2,\dots,p\}$. For a vertex $v \in V$ denote by  $v(1),v(2),v(3) \in F$ the three adjacent faces in anticlockwise order. Then
\begin{equation}\label{eq:RSOS-rewrite}
    Z = \sum_{\varphi} \prod_{v \in V} T_{\varphi(v(1))\varphi(v(2))\varphi(v(3))} ~.
\end{equation}
where the sum is over all functions $\varphi : F \to \{1,\dots,p\}$.

As indicated in Figure~\ref{fig:loops-on-hex-lattice}, each configuration $\varphi$ gives rise to a configuration of non-intersecting closed loops on the edges of the hexagonal lattice. Since across each loop, the face index has to change by $\pm 1$, we see that \eqref{eq:RSOS-rewrite} is an RSOS (restricted solid-on-solid) height model for the Dynkin-diagram $A_p$ on the dual triangular lattice with weights $T_{abc}$ assigned to the triangular faces.
The height model can be rewritten as a loop model in the usual way, see e.g.\ \cite{Nienhuis:1982fx,Pasquier:1987,Cardy:2006yt}, as well as the detailed derivation leading to a densely packed loop model (in the Fortuin-Kasteleyn representation) in \cite[App.\,A]{He:2020mhb}.
We will go over the steps quickly to extract the relevant weight factors.

\medskip

\noindent
\textit{Step 1:} For a hexagonal lattice as in Figure~\ref{fig:loops-on-hex-lattice} but without periodic boundary conditions, every loop configuration can occur for some choice of $\varphi$. On the torus, the only forbidden configurations are those where an odd number of loops winds around some cycle of the torus, as in an odd number of $\pm1$ steps one cannot return to the initial value.

Write $\mathcal{L}$ for the set of allowed loop configurations on the hexagonal lattice, and for $L \in \mathcal{L}$ write $\varphi\sim L$ for a function $\varphi : F \to \{1,\dots,p\}$ whose corresponding loop configuration is $L$. Then \eqref{eq:RSOS-rewrite} can be rewritten as
\begin{equation}\label{eq:RSOS-rewrite-loop}
    Z = \sum_{L \in \mathcal{L}} \sum_{\varphi \sim L} \prod_{v \in V} T_{\varphi(v(1))\varphi(v(2))\varphi(v(3))} ~.
\end{equation}

\noindent
\textit{Step 2:} Let $L$ be a loop configuration, and let $\ell \subset L$ be a loop bounding a disc (i.e.\ a contractible loop on the torus) such that this disc does not contain any other loops. Keeping the outer value $a$ fixed, there are at most two choices for the inner value $b$, namely those for which $N_{af}^{~b} \neq 0$. As we explain in a moment, the sum over $b$ can be carried out explicitly. The result is
\begin{equation}\label{eq:omit-disc-loop}
        \sum_b 
        \includegraphics[width=.3\linewidth,valign=c]{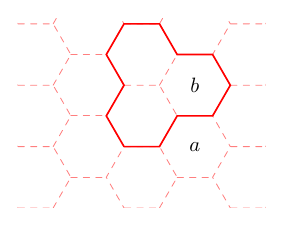}
 ~= 
 \dim(f) \, x(R)^{|\ell|} 
     ~~
     \includegraphics[width=.3\linewidth,valign=c]{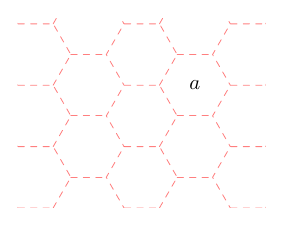}
\end{equation}
where $|\ell|$ is the number of edges traversed by the loop $\ell$, and the quantum dimension is given by (recall \eqref{eq:min-mod-Smatrix})
\begin{equation}\label{eq:min-mod-dimf}
    \dim(f) = \frac{S_{f1}}{S_{11}} = 2 \cos \tfrac{\pi}{p+1} ~. 
\end{equation}
To see this, first note that there are the same number of vertices on both sides, so that both sides contribute the same power of $F_{d,R}$. Second, the loop $\ell$ altogether makes one full turn, so that the factors of $d_{b,a}^{\pm 1/6}$ in \eqref{eq:three-vertices-for-loops} multiply together to give precisely $d_{b,a}$. Third, each vertex on the loop contributes a factor of $x(R)$. But the number of vertices is equal to the number of edges $|\ell|$. We can now carry out the sum over $b$,
\begin{align}
    \sum_{b} d_{b,a} \, x(R)^{|\ell|} \, N_{af}^{~b}
    &=
    \frac{x(R)^{|\ell|}}{\dim(a)} \sum_{b} N_{af}^{~b} \dim(b)
    = 
    \frac{x(R)^{|\ell|}}{\dim(a)} \dim(a \otimes f)
\nonumber\\    
    &= 
    x(R)^{|\ell|} \dim(f)~,
\end{align}
and we arrive at \eqref{eq:omit-disc-loop}.

\medskip

\noindent
\textit{Step 3:} After removing all contractible loops, one may be left with $n$ loops that wind around the torus. As loops cannot intersect, all loops then wind the same cycle. The weight of such a configuration is

\begin{equation}
\includegraphics[valign=c]{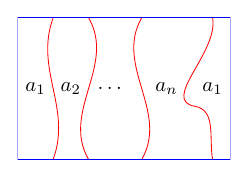}
~=~ 
F_{d,R}^{|V|} \, (\dim f)^n \, x(R)^{|\ell_1| + \cdots + |\ell_n|}\, \sum_{a=1}^p \xi_a^{\,n}
~.
\end{equation}
where
\begin{equation}\label{eq:xa-def}
    \xi_a = 
    \frac{\cos( \pi \tfrac{a}{p+1} )}{\cos( \pi \tfrac{1}{p+1} )} = -\xi_{p+1-a}
    ~ \in [-1,1] ~.
\end{equation}
In more detail, since a loop winding around the torus involves the same number of left and right turns, the factors $d_{a,b}^{\pm 1/6}$ cancel. 
The $n$ loops thus contribute a combined factor of $x(R)^{|\ell_1| + \cdots + |\ell_n|}$, and it remains to carry out the sum over $a_1,\dots,a_n$, subject to the constraint that $a_{i+1} = a_i \pm 1$, i.e.\ that $N_{a_i f}^{~a_{i+1}} \neq 0$. This gives
\begin{equation}\label{eq:winding-loop-factor}
    \sum_{a_1,\dots,a_n} N_{a_1 f}^{~a_2} N_{a_2 f}^{~a_3} \cdots N_{a_n f}^{~a_1} 
    = \mathrm{tr}(N_f)^n
    = \sum_{a = 1}^p \left( \frac{s^{\mathrm{su}(2)}_{fa}}{s^{\mathrm{su}(2)}_{1a}}\right)^{\!n}
    = (\dim f)^n \sum_{a = 1}^p \xi_a^n ~.
\end{equation}
Here, $N_f$ is the fusion rule matrix with entries $(N_f)_a^b = N_{fa}^{~b}$, and we used that the fusion rule matrix has eigenvalues $s^{\mathrm{su}(2)}_{fa} / s^{\mathrm{su}(2)}_{1a}$ with $a = 1,\dots,p$,
where
\begin{equation}
    s^{\mathrm{su}(2)}_{xy} = \sin \tfrac{\pi xy}{p+1}
    \quad , ~~1 \le x,y \le p \ ,
\end{equation}
is the $\widehat{su}(2)_{p-1}$ $S$-matrix (up to an overall factor and an index shift).
The reason that $s^{\mathrm{su}(2)}$ appears rather than the minimal model $S$-matrix \eqref{eq:min-mod-Smatrix} is that the sum in \eqref{eq:winding-loop-factor} only involves the subcategory with simple objects $\mathcal{R}$, rather than all of $\mathcal{I}$.
Note also that due to the anti-symmetry of $\xi_a$, \eqref{eq:winding-loop-factor} is automatically zero for $n$ odd. 
For example, in the case of the Ising CFT $p=3$ we have $\xi_1 = 1$, $\xi_2 = 0$  and $\xi_3 = -1$.

%
%
%

\medskip

Altogether, the lattice model partition function $Z$ in \eqref{eq:original-partition-function} becomes
\begin{equation}\label{eq:loop-rewrite}    
Z = F_{d,R}^{|V|} \, \sum_{a=1}^p \sum_{L} x(R)^{|L|} \, (\dim f)^{\#(L)} \, \xi_a^{\,w(L)}
 ~.
\end{equation}
Here, $L$ is a configuration of non-intersecting loops on the hexagonal lattice, $|L|$ is the total number of edges contained in the configuration $L$ (the total length of all loops in $L$), $\#(L)$ is the number of connected components of $L$ (the number of individual loops, irrespective of whether they wind around the torus or not), and $w(L)$ is the number of loops that wind around the torus. 
The sum $\sum_L$ is over all non-intersecting loop configurations. The condition that only an even number of loops may wind around the torus is imposed automatically since we saw above that \eqref{eq:winding-loop-factor} is zero for odd $n$.

If $d(L)$ denotes the number of individual loops in $L$ that bound a disc, we have $\#(L) = d(L) + w(L)$. We can hence think of \eqref{eq:loop-rewrite} as a superposition of $p$ loop models indexed by $a \in \{1,\dots,p\}$, where a loop that bounds a disc gets weight $\dim(f)$ and a loop that wraps the torus gets weight $\dim(f) \xi_a$:
\begin{equation}\label{eq:loop-rewrite-2}    
Z = \sum_{a=1}^p F_{d,R}^{|V|} \sum_{L} x(R)^{|L|} \, (\dim f)^{d(L)} \, (\dim(f) \xi_a)^{\,w(L)}
 ~.
\end{equation}
We stress that \eqref{eq:loop-rewrite-2} is exactly equal to the truncated CFT lattice model \eqref{eq:original-partition-function}, and not an approximation in some parameter regime.

\subsection{Phase diagram of the loop model}

We start with a quick overview of loop models following \cite[Sec.\,2.1]{Dubail:2009mb} and \cite[Sec.\,2]{Estienne:2015sua}.
Let $\Lambda$ be a hexagonal toroidal lattice. The $O(n)$-loop model on $\Lambda$ has the partition function
\begin{equation}
Z_{O(n)} = \sum_{L} x^{|L|} \, n^{d(L)} \, \widetilde n{}^{w(L)} ~,
\end{equation}
where $x \ge 0$ and $n,\widetilde n \in (-2,2)$. As in \eqref{eq:loop-rewrite-2}, the sum $\sum_L$ is over all configurations of non-intersecting loops on $\Lambda$, $|L|$ is the total number of edges covered by $L$, $d(L)$ is the number of contractible loops, and $w(L)$ is the number of non-contractible loops. 

The phase diagram was first conjectured in \cite{Nienhuis:1982fx} (for $n = \widetilde n$). Loops that wind a non-trivial cycle are discussed in \cite{diFrancesco:1987,Estienne:2015sua}. 
For an overview of exact results see \cite{Peled:2017}.
The phase diagram is independent of $\widetilde n$ and for a fixed value of $n$ takes the form
\begin{equation*}
\includegraphics[valign=c]{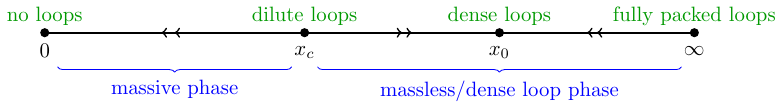}
\end{equation*}
The two critical points are
\begin{equation}\label{eq:loop-critical}
    x_c = \frac{1}{\sqrt{2+\sqrt{2-n}}}
    ~~, \qquad
    x_0 = \frac{1}{\sqrt{2-\sqrt{2-n}}} ~,
\end{equation}
and the central charges at these two points are, for $n = 2 \cos(\pi r)$,
\begin{equation}
    c_c = 1- 6 \frac{r^2}{1+r}
    ~~, \qquad
    c_0 = 1- 6 \frac{r^2}{1-r} ~.
\end{equation}
In our example, $n = 2 \cos(\pi/(p+1))$ from \eqref{eq:min-mod-dimf}, so that
\begin{equation}
    c_c = 1- 6 \frac{1}{(p+1)(p+2)}
    ~~, \qquad
    c_0 = 1- 6 \frac{1}{p(p+1)} ~.
\end{equation}
The value $c_0$ is equal to the central charge of the minimal model $M(p,p+1)$ from which we constructed the lattice model, and $c_c$ is the central charge of the next minimal model up, i.e.\ of $M(p+1,p+2)$.
The three values $x_c$, $x_0$ from \eqref{eq:loop-critical} and $x_\text{max}$
from \eqref{eq:x-max-loopmodel} depend on $p$ as follows:
\begin{equation*}
\includegraphics[width=.6\linewidth,valign=c]{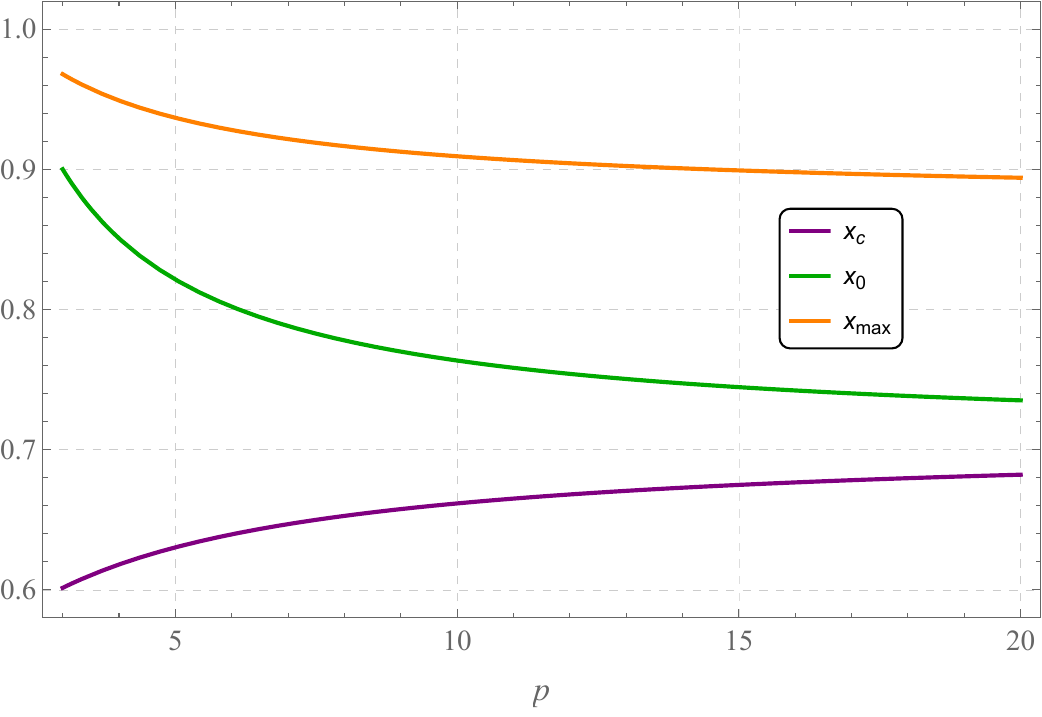}   
\end{equation*}
In particular, for all values of $p$ we have $x_c < x_0 <  x_\text{max}$.

\medskip

Thus, in terms of the hole radius $R$ of the lattice model, there is a critical radius $R_C$ defined by $x(R_C) = x_c$.
For $R > R_C$ the model is gapped and hence flows to a 2d\,TFT. For $R<R_C$ the model flows to a critical model of central charge $c_0$, which conjecturally (and before the rewriting as a loop model) is the A-type minimal model $M(p,p+1)$.  
For $R=R_0$ with $x(R_0)=x_0$, the lattice model already is at this critical point.
At the phase transition $R=R_C$ there is an unstable critical point which is conjecturally described by $M(p+1,p+2)$.

These observations support our expectation that if the cloaking boundary condition for the preserved symmetry $\mathcal{F}$ satisfies the stability condition, the phase diagram of the lattice model is of the form \eqref{eq:general-phase-diagram}. In the present example even the smallest non-trivial cutoff is already large enough to achieve this result.

\section{Summary and outlook}

The main aim of this article was the construction of a family of lattice models from the input of a 2d CFT $C$ and a choice of fusion category symmetry $\mathcal{F}$, building on the results in the first paper \cite{Brehm:2021wev}.

As a basis for this construction, we systematically reviewed how to compute with amplitudes in 2d\,CFT, including a careful treatment of the Weyl anomaly via the anomaly action. We are not aware of another such treatment in the literature which includes boundaries and provides detailed example computations. In the numerical verification of our formalism we have illustrated that the anomaly contributions are crucial (Figure~\ref{fig:compare1}).

We proposed a stability condition for the lattice models constructed from $C$ and $\mathcal{F}$ under which we expect the universality class of the lattice model to be $C$ for a finite interval of $R$-values near zero and large enough cutoff $h_\text{max}$. The stability condition was that the $\mathcal{F}$-cloaking boundary state only contains irrelevant bulk fields as summands.

This condition fails for the Ising CFT with $\mathcal{F}$ the $\mathbb{Z}_2$ symmetry, but is satisfied if we take $\mathcal{F}$ to include all defects. At lowest non-trivial cutoff, the resulting lattice model was massive near $R=0$ in the first case, and conjecturally was in the Ising universality class in the second case (it had the correct central charge -- this was the case $p=3$ in the loop model examples).

\smallskip
There are still many directions for further investigation, and we hope to return to some of these in the future:
\begin{itemize}

\item A more detailed comparison of our lattice models to those constructed in \cite{Vanhove:2018wlb,Aasen:2020jwb,Lootens:2020mso} which only take $\mathcal{F}$ as an input needs to be carried out. Furthermore, rather than discretising just the 2d\,CFT as we do now, one could also discretise the coupled system of the 2d\,CFT and the 3d\,TFT which encodes its topological symmetries. 

\item In this paper and in \cite{Brehm:2021wev} we focused on partition functions. There is a canonical way to include lattice observables by placing bulk fields -- local fields or defect fields -- in the centres of some of the triangles. Investigating lattice correlators of these observables should help to identify the universality class of the lattice model. It would also allow for a more detailed investigation of dualities between lattice models which originate from topological defects.

\item The numerical work in this paper and in \cite{Brehm:2021wev} was concerned with approximating the exact CFT answer by considering large cutoffs for a fixed choice of lattice (a torus with one hole). An important aspect to study would be the behaviour as the lattice size increases, i.e.\ moving towards the continuum limit. In particular, the behaviour of the phase diagram of the lattice model as one increases the cutoff $h_\text{max}$ would be very interesting.

 \end{itemize}

 \newpage

\appendix

\section{The anomaly action}\label{app:AnomalyAction}

\subsection{Motivation for the anomaly action}\label{app:LiouvilleMotivation}

There are several ways to arrive at the anomaly action \eqref{eq:anomaly-action} and its role in the behaviour of amplitudes under Weyl transformations in \eqref{eq:AmplLiou}. 

Mathematically, it arises in the behaviour of regularised determinants of Laplacians under Weyl transformations (see e.g.\ \cite[Sec.\,2.5.3]{Benoist:2014} for a short summary and references), as well as in the behaviour of SLE loop measures \cite[Thm.\,1.1]{Benoist:2014}. Furthermore, one can show that amplitudes of Liouville theory on the sphere produce this anomaly factor under Weyl transformations \cite[Thm.\,1.1]{Kupiainen:2019fhf}.

Here we instead review the physical motivation, which uses the path integral approach to quantum field theory amplitudes. This approach is described e.g.\ in \cite[Sec.\,13.3]{Zamolodchikov:2007}, which is our main reference in this subsection.

\smallskip
The anomaly action follows from the \emph{anomaly equation}, namely the normalised correlator of the trace $\theta = \sum_{\mu} T^\mu_\mu$ of the energy momentum tensor on a closed Riemann surface $\Sigma_g$ with metric $g$:
\begin{equation}\label{eq:WeylAnomaly}
    \langle \theta(p) \cdots \rangle = \frac{\mathcal{A}(\Sigma_g;\theta(p), \dots )}{\mathcal{A}(\Sigma_g;\dots )} = -\frac{c}{12} R(p) \langle \cdots \rangle + \text{only contact terms}\,,
\end{equation}
where $R(p)$ is the Ricci scalar for $g$ at the insertion point $p$ of $\theta$. The field $\theta$ is regarded as the source for variations of the conformal factor of the metric. Let us consider some action $W[g,\phi]$ that governs the dynamics of the theory with field content $\phi$, such that we can write amplitudes as functional integrals
\begin{equation}
    \mathcal{A}(\Sigma_g;\dots) = \int \dots e^{-W[g,\phi]} D[\phi] 
\end{equation}
for some not further specified measure $D[\phi]$. Partition functions -- i.e.\ amplitudes with no field insertion -- are then simply given by 
\begin{equation}
    \mathcal{A}(\Sigma_g;\one) = \int e^{-W[g,\phi]}D[\phi]\,.
\end{equation}
For changes 
\begin{equation}
    g_{ab}(x) \to \left(1+\delta \sigma(x)\right) g_{ab}(x)
\end{equation}
we have 
\begin{equation}\label{eq:ActionChange}
    \delta W = \frac{1}{4 \pi} \int \sqrt{g} \,\delta\sigma(x) \theta(x) d^2x\,,
\end{equation}
which quantifies the response of the theory to a change of the conformal factor. Now take two metrics $g$ and $\hat{g}$ on $\Sigma$ that are related by a scaling field $\Omega$, 
\begin{equation}\label{eq:gTrafo}
    g = e^{\Omega}\hat{g}\,,
\end{equation}
with some smooth real function $\Omega$ on $\Sigma$. For the scalar curvature one has
\begin{equation}
    \sqrt{g}R = \sqrt{\hat{g}} \left(\hat{R} - \Delta_{\hat{g}} \Omega \right)\,,
\end{equation}
where $\hat{R}$ and $\Delta_{\hat{g}}$ are the curvature and the Laplace operator associated with the metric $\hat{g}$. Let us regard $\hat{g}$ as some fixed background metric. Then we can write
\begin{equation*}
\begin{split}
    \delta\mathcal{A}\left(\Sigma_g,\one\right) &= -\int  \delta W\, e^{-W[g,\phi]} D[\phi] = - \frac{1}{4\pi}\int \sqrt{g}\, \delta\Omega \left(\int \theta(x) \,e^{-W} D[\phi] \right) d^2x\\
     &= -\frac{1}{4\pi}\int \sqrt{g}\, \delta\Omega \,\mathcal{A}(\Sigma_g;\theta) \, d^2x = \mathcal{A}(\Sigma_g;\one) \, \frac{c}{48\pi}\int \sqrt{g}\, R\,\delta\Omega \,d^2x  \\
\end{split}
\end{equation*}
or
\begin{align}
    \delta \log \mathcal{A}(\Sigma_g,\one) &= \frac{\delta \mathcal{A}(\Sigma_{g};\one)}{\mathcal{A}(\Sigma_g;\one)} 
    \nonumber
    \\
    &= \frac{c}{48\pi}\int_\Sigma \sqrt{g}R\, \delta \Omega\, d^2x= \frac{c}{48\pi} \int_{\Sigma} \left(\sqrt{\hat{g}}\hat{R} - \Delta_{\hat{g}}\Omega\right) \delta \Omega d^2x
    \nonumber
    \\
    &= \frac{c}{48\pi} \delta \left(\int _{\Sigma} \sqrt{\hat{g}}\left(\hat{R}\,\Omega + \frac{1}{2} \hat{g}^{\mu\nu}\partial_\mu\Omega\partial_\nu\Omega\right)  d^2x \right)\,,
\end{align}
where in the last step we use that $\hat{g}$ is defined as fixed background metric, integrate by parts and use that there is no boundary to $\Sigma$. Then we can integrate the latter to
\begin{equation}
    \mathcal{A}(\Sigma_{g},\one) = e^{\frac{c}{24}L_{\Sigma_{\hat{g}}}(\Omega)} \mathcal{A}(\Sigma_{\hat{g}},\one)
\end{equation}
with the anomaly action
\begin{align}\label{eq:anomaly-derivation-closed}
    L_{\Sigma_{\hat{g}}}(\Omega) = \frac{1}{2\pi} \int_\Sigma \sqrt{\hat{g}} \left( \hat{R}\, \Omega + \frac{1}{2}\hat{g}^{\mu\nu}\partial_\mu\Omega \partial_\nu \Omega \right)\,d^2x\,.
\end{align}
When we consider surfaces with boundaries, the integration by parts picks up a boundary term that can be written as
\begin{equation}\label{eq:anomaly-derivation-bnd}
    \frac{1}{\pi} \int_{\partial\Sigma} k_{\hat{g}}(l)\, \Omega\!\left(x^\mu(l)\right) dl
\end{equation}
where $l$ parameterises the boundary with geodesic curvature $k_{\hat{g}}(l)$. 
Combining \eqref{eq:anomaly-derivation-closed} and \eqref{eq:anomaly-derivation-bnd} gives \eqref{eq:anomaly-action}.

\smallskip
In the presence of field insertions, $\theta$ has contact terms which lead to an additional factor for each field insertion.
Let us illustrate this in the example of spin-less primary fields $\phi_i$ with left/right conformal weight $(h_i,h_i)$. In this case, \eqref{eq:WeylAnomaly} becomes
\begin{equation}\label{eq:contact-aux1}
\Big\langle \theta(x) \prod_i\phi_i(x_i) \Big\rangle = \Big( - \frac{c}{12} R(x) +
\sum_i h_i\frac{4\pi}{\sqrt{g}}\delta^{(2d)}(x-x_i) \Big)
\, \Big\langle \prod_i\phi_i(x_i) \Big\rangle \ .
\end{equation}
An analogous computation as above gives, for $g = e^{\Omega} \hat g$,
\begin{equation}\label{eq:contact-aux2}
\mathcal{A}(\Sigma_{g},\prod_i \phi_i) = e^{\frac{c}{24}L_{\Sigma_{\hat{g}}}(\Omega) - \sum_i h_i \Omega(x_i)} \mathcal{A}(\Sigma_{\hat{g}},\prod_i \phi_i)
\ ,
\end{equation}
cf.\ also \cite[Eqn.\,(13.50)]{Zamolodchikov:2007} and \cite[Eqn.\,(1.3)]{Kupiainen:2019fhf}.

The setup in Section~\ref{sec:amplitute} produces the same answer, even though the extra factors $e^{-h_i\Omega(x_i)}$ are missing from \eqref{eq:anomaly-action}.
The key point is that in \eqref{eq:contact-aux1} and \eqref{eq:contact-aux2}, it is implicitly assumed that the local coordinates around field insertions are adapted to the metric. In the case of spin-less primaries it is enough to demand that the local coordinates $\varphi$ are chosen such that the pullback metric at the origin is just the standard metric, $(\varphi^* g)(0)_{\mu\nu} = \delta_{\mu\nu}$. 
A Weyl transformation then also entails a change of local coordinates, leading to an extra factor from transformation rule (ii). Namely, if $\varphi_i$ satisfies $\varphi_i^* g(0) = \delta$ and $\hat\varphi_i$ satisfies $\hat\varphi_i^* \hat g(0) = \delta$, we may choose $\varphi_i(z) = \hat\varphi_i \circ G_i$ with $G_i(z) = e^{-\Omega(x_i)/2} z$ and so $\Gamma_{G_i} = e^{-\frac12\Omega(x_i)(L_0+\overline L_0)}$. Rule (ii) then states
\begin{equation}
    \phi_i(x_i,\varphi_i) = (\Gamma_{G_i} \phi_i)(x_i,\hat\varphi_i)
    = e^{-h_i \Omega(x_i)} \phi_i(x_i,\hat\varphi_i) \ .
\end{equation}
Altogether, the amplitude transforms according to the rules in Section~\ref{sec:amplitute} as
\begin{align}
\mathcal{A}\!\left(\Sigma_{g};\phi_1(x_1;\varphi_1),\dots \right) 
&\overset{\text{(iii)}}=
e^{\frac{c}{24}L_{\Sigma_{\hat{g}}}(\Omega)}\,
\mathcal{A}\!\left(\Sigma_{\hat g};\phi_1(x_1;\varphi_1),\dots \right) 
\nonumber \\
&\overset{\text{(ii)}}=
e^{\frac{c}{24}L_{\Sigma_{\hat{g}}}(\Omega)- \sum_i h_i \Omega(x_i)}\,
\mathcal{A}\!\left(\Sigma_{\hat g};\phi_1(x_1;\hat\varphi_1),\dots \right) \ .
\end{align}

\subsection{Some properties of the anomaly action}\label{app:LProperties}

Let $\Sigma^1, \Sigma^2, \Sigma^3 \subset \mathbb{C}$ be surfaces with piecewise smooth boundary and induced metric $\delta$. Let $F : \Sigma^1 \to \Sigma^2$ and $G : \Sigma^2 \to \Sigma^3$ be orientation preserving conformal bijections (i.e.\ biholomorphic maps), and set $H := G \circ F : \Sigma^1 \to \Sigma^3$. Let $\Omega:\Sigma^2 \to \mathbb{R}$ be smooth. We abbreviate 
\begin{equation}
    \Omega_F := \log|\partial_z F(z)|^2 \ ,
\end{equation}
etc. Then the anomaly action has the properties: 
\begin{subequations}
\begin{align}
    L_{\Sigma^1_\delta}(\Omega_{H}) &= L_{\Sigma^2_\delta}(\Omega_G) + L_{\Sigma^1_\delta}(\Omega_F)\ ,
    \label{eq:Lid1}
    \\    
    L_{\Sigma^1_\delta}(\Omega_F) &= - L_{\Sigma^2_\delta}(\Omega_{F^{-1}}) \ ,
    \label{eq:Lid2}
    \\    
    L_{\Sigma^1_\delta}(\Omega\circ F+\Omega_F) &= L_{\Sigma^2_\delta}(\Omega) + L_{\Sigma^1_\delta}(\Omega_F)\,.
    \label{eq:Lid3}
\end{align}
\end{subequations}
Note that \eqref{eq:Lid1} is just a special case of \eqref{eq:Lid3} since $\Omega_H = \Omega_G \circ F + \Omega_F$. 
Furthermore, \eqref{eq:Lid2} is a special case of \eqref{eq:Lid1} where $G = F^{-1}$ so that $H=\mathrm{id}$ and $\Omega_H=0$.
It remains to show \eqref{eq:Lid3}. 

To do so, let us first state a special case of the cocycle identity (see e.g.\ \cite[Lem.\,2.1]{Kontsevich:2006} for surfaces without boundary). We give a proof here for the convenience of the reader.

\smallskip
\noindent
\textbf{Lemma.} 
Let $\Sigma \subset \mathbb{C}$ a surface with piecewise smooth boundary, and with metric $\delta$ induced by $\mathbb{C}$. Let $\Omega_1,\Omega_2:\Sigma \to \mathbb{R}$ be smooth functions and set $g_1 = e^{\Omega_1} \delta$. Then
\begin{equation}\label{eq:LO1+O2}
    L_{\Sigma_\delta}(\Omega_1 + \Omega_2) \,=\, L_{\Sigma_{g_1}}(\Omega_2) + L_{\Sigma_\delta}(\Omega_1) \ .
\end{equation}

\begin{proof}
The l.h.s. is 
\begin{align}
    L_{\Sigma_\delta}(\Omega_1 + \Omega_2) &= \frac{1}{2\pi} \int_\Sigma \frac{1}{2}\,\delta^{\mu\nu} \partial_\mu (\Omega_1+\Omega_2)\partial_\nu (\Omega_1+\Omega_2) d^2x 
    \nonumber
    \\
    &\quad+ \frac{1}{\pi}\int_{\partial\Sigma} k_\delta(l) (\Omega_1+\Omega_2) dl
\end{align}
On the r.h.s.\ we have
\begin{align}
    L_{\Sigma_{g_1}}(\Omega_2) &= \frac{1}{2\pi} \int_\Sigma \sqrt{g_1} \left(R_1\Omega_2 +\frac{1}{2} g_1^{\mu\nu} \partial_\mu \Omega_2\partial_\nu\Omega_2\right) d^2x
    \nonumber
    \\
    &\quad+\frac{1}{\pi} \int_{\partial\Sigma} k_{g_1}(l) \Omega_2 dl
    \nonumber
    \\
    L_{\Sigma_{\delta}}(\Omega_1) &=\frac{1}{2\pi} \int_\Sigma \frac{1}{2} \delta^{\mu\nu} \partial_\mu \Omega_1\partial_\nu\Omega_1 \,d^2x
    \nonumber
    \\
    &\quad+\frac{1}{\pi} \int_{\partial\Sigma} k_{\delta}(l) \Omega_1 dl
\end{align}
Here we use $\sqrt{g_1} = e^{\Omega_1}$\,, $R_1 = -e^{-\Omega_1}\Delta\Omega_1$ and $g_1^{\mu\nu} = e^{-\Omega_1}\delta^{\mu\nu}$. The geodesic curvature is given by $k_{g}(l) = t^a n_b \nabla_a t^b$, where $t^a$ are unit tangent vector at the boundary, i.e. $t^a t^b g_{ab} = 1$, $n^a$ is the unit normal vector to the curve pointing inwards, and $\nabla_a$ is the covariant derivative w.r.t. the metric $g$. For $g = g_1 = e^{\Omega_1}\delta$ we can write 
\begin{align}
    \nabla_a t^b = \partial_a t^b + {\Gamma^b}_{ac}t^c = \partial_a t^b +\frac{1}{2}\left(\frac{\partial\Omega_1}{\partial x^a} \delta^b_c + \frac{\partial\Omega_1}{\partial x^c} \delta^b_a - \frac{\partial\Omega_1}{\partial x^l} \delta^{lb}\delta_{ac} \right)t^c
\end{align}
such that
\begin{equation}
    k_{g_1}(l) = t^a n_b \partial_a t^b -\frac{1}{2} \frac{\partial \Omega_1}{\partial x^l} n_b \delta^{lb}\, \delta_{ac} t^at^c = t^a n_b \partial_a t^b -\frac{1}{2} \frac{\partial \Omega_1}{\partial x^l} n^l
    \ .
\end{equation}
The vector $\tilde{t}^a = e^{\Omega_1/2} t^a$ is tangent to the boundary and $\tilde{n}^a = e^{\Omega_1/2}n^a$ is perpendicular to the boundary and both are of unit norm w.r.t.\ $\delta$. So
\begin{equation}
    k_{g_1}(l) = e^{-\Omega_1/2}\tilde{t}^a \tilde{n}_b \partial_a \tilde{t}^b  -\frac{1}{2} e^{-\Omega_1/2} \frac{\partial \Omega_1}{\partial x^l} \tilde{n}^l = e^{-\Omega_1/2} \left(k_\delta(l) - \frac{1}{2} \frac{\partial \Omega_1}{\partial x^l} \tilde{n}^l \right)\,.
\end{equation}
With the rescaled line element $d\tilde{l} = e^{-\Omega_1/2} dl$ we get 
\begin{align}
    L_{\Sigma_{g_1}}(\Omega_2) &= \frac{1}{2\pi} \int_{\Sigma} \left(\delta^{\mu\nu}\partial_\mu\Omega_1\partial_\nu\Omega_2 +\frac{1}{2}\delta^{\mu\nu}\partial_\mu\Omega_2\partial_\nu\Omega_2\right)d^2x 
\nonumber
    \\
    &\quad - \frac{1}{2\pi} \int_{\Sigma} \delta^{\mu\nu} \partial_\mu\left(\partial_\nu\Omega_1\,\Omega_2\right)\,d^2x
\nonumber
    \\
    &\quad + \frac{1}{\pi} \int_{\partial\Sigma} \left(- \frac{1}{2}\tilde{n}^k \partial_k\Omega_1\, \Omega_2 + k_\delta \Omega_2\right)d\tilde{l}\,.
\end{align}
For the second line we can use the divergence theorem: 
\begin{equation}
    - \frac{1}{2\pi} \int_{\Sigma} \delta^{\mu\nu} \partial_\mu\left(\partial_\nu\Omega_1\,\Omega_2\right)\,d^2x = \frac{1}{2\pi} \int_{\Sigma} \tilde{n}^k \partial_k\Omega_1 \Omega_2 d\tilde{l}
    \ ,
\end{equation}
which cancels the first term in the boundary integral. Adding up all remaining terms shows \eqref{eq:LO1+O2}\,.
\end{proof}

We now return to \eqref{eq:Lid3}. Applying the lemma to the left hand side gives, for $g = e^{\Omega_F}\delta$,
\begin{equation}
    L_{\Sigma^1_\delta}(\Omega\circ F+\Omega_F)
    =
    L_{\Sigma^1_g}(\Omega\circ F) 
    + L_{\Sigma^1_\delta}(\Omega_F)
\end{equation}
But $F : \Sigma^1_g \to \Sigma^2_\delta$ is actually an isometry, i.e.\ $g = F^*\delta$, and so $L_{\Sigma^1_g}(\Omega) = L_{\Sigma^2_\delta}(\Omega)$. This shows \eqref{eq:Lid3}.

\subsection{Examples of anomaly factors}\label{app:anomalyExamples}

\subsubsection{Disc with radius R}\label{app:discfactor}

Let $D_R$ be unit disc of radius $R$ and let $F: D_R \to D_1$ be the map $F(z)=z/R$.
As before, we denote the standard metric on $\mathbb{C}$ by $\delta$. The push-forward of $\delta$ along $F$, i.e.\ the pull-back along $F^{-1}$, is $(F^{-1})^* \delta = |\partial F^{-1}|^{2} \delta = e^{\Omega} \delta$ with
\begin{equation}
    \Omega(w) = \Omega_{F^{-1}}(w) =  2 \log R \ .
\end{equation}
In the anomaly action on $D_1$, only the boundary term gives a contribution with $k(l) dl = d\varphi$%
\begin{equation}
   L(\Omega_{F^{-1}}) = \frac{1}{\pi}\int_{0}^{2\pi} 2\log(R) d\varphi
    =4 \log(R) \ .
\end{equation}
It follows that the amplitude of a disc $D_R$ with standard metric $\delta$ and boundary condition $b$ is
related to that on $D_1$ by 
\begin{equation}
    \mathcal{A}\big((D_R)_\delta;b\big) 
    \overset{\text{(i)}}= 
    \mathcal{A}\big((D_1)_{e^{\Omega}\delta};b\big) 
    \overset{\text{(iii)}}=
    e^{\frac{c}{24}L(\Omega)}\mathcal{A}\big((D_1)_\delta;b\big) = R^{\frac{c}{6}} \mathcal{A}\big((D_1)_\delta;b\big)\ ,
\end{equation}
where above the equalities we indicate the transformation rules from Section~\ref{sec:amplitute}.

\subsubsection{Cylinder and annulus}\label{app:cyl-ann}

Let $\mathrm{Cyl}(R,L)$ be a cylinder of radius $R$ and length $L$. We write $\mathrm{Cyl}(R,L)$ as a rectangle in the complex plane consisting of the points $z = x+ i y$ with $x\in [-\pi R,\pi R]$ and $y\in[-\tfrac{L}{2},\tfrac{L}{2}]$, and where $\{-\pi R\} \times [-\tfrac{L}{2},\tfrac{L}{2}]$ is identified with $\{\pi R\} \times [-\tfrac{L}{2},\tfrac{L}{2}]$. The metric on $\mathrm{Cyl}(R,L)$ is the standard metric $\delta$ on $\mathbb{C}$.

Write $\mathrm{Ann}(r_+,r_-) = \{ z \in \mathbb{C} \,|\, r_- \le |z| \le r_+ \}$ for the annulus in $\mathbb{C}$ with inner radius $r_-$ and outer radius $r_+$.  The metric on $\mathrm{Ann}(r_+,r_-)$ is again the standard metric $\delta$ on $\mathbb{C}$.

We have a conformal bijection
\begin{equation}
    F : \mathrm{Cyl}(R,L) \longrightarrow \mathrm{Ann}(r_+,r_-) ~,~ z \mapsto e^{i z /R} 
    ~ ,
    \quad \text{where} ~~ r_\pm= e^{\pm \frac{ L}{2R}} \ .
\end{equation}
This becomes an isometry if on $\mathrm{Ann}(r_+,r_-)$ we choose the metric $e^{\Omega}\delta$ with
\begin{equation}
    \Omega(w) = \Omega_{F^{-1}}(w) = \log \left|\partial_w (F^{-1}(w))\right|^2 = \log \left|\frac{R}{w}\right|^2\,.
\end{equation}
The bulk integral of the anomaly action then is (note that $dw d\bar w = 2 dxdy = 2 r dr d\varphi$)
\begin{equation}\label{eq:ann-Lbulk}
    L^\text{bulk}  = \frac{1}{2\pi}\int dw d\bar{w} \frac{1}{w\bar{w}}  
    = \frac{1}{\pi} \int_{r_-}^{r_+} r dr \int_{0}^{2\pi} d\varphi\, \frac{1}{r^2} 
     = 2 \log \frac{r_+}{r_-} = \frac{2L}{R}\,.
\end{equation}
The boundary term is
\begin{align}\label{eq:ann-Lbnd}
    L^\text{bdy} &=\frac{1}{\pi}\left( \int_{0}^{2\pi} d\phi\,  r_+ k_+ \log \left(\frac{R}{r_+}\right)^2 + \int_{0}^{2\pi} d\phi\,  r_- k_- \log \left(\frac{R}{r_-}\right)^2 \right)
    \nonumber
    \\
    &= 4 \log \frac{r_-}{r_+} = - \frac{4L}{R} \ .
\end{align}
where we used $k_+ =  \tfrac{1}{r_+}$ and $k_- = -\tfrac{1}{r_-}$. Putting both results together we obtain the following relation between the amplitude on the cylinder and the annulus with boundary conditions $b_1$, $b_2$:
\begin{align}\label{eq:cylinder-to-annulus}
    \mathcal{A}\big(\mathrm{Cyl}(R,L)_\delta;b_1,b_2\big) 
    &\overset{\text{(i)}}= 
    \mathcal{A}\big(\mathrm{Ann}(r_+,r_-)_{e^{\Omega}\delta};b_1,b_2\big) 
        \nonumber
    \\
    &\overset{\text{(iii)}}=
    e^{\frac{c}{24}L(\Omega)}\mathcal{A}\big(\mathrm{Ann}(r_+,r_-)_\delta;b_1,b_2\big) 
    \nonumber
    \\
    &= e^{-\frac{c L}{12 R}} \mathcal{A}\big(\mathrm{Ann}(r_+,r_-)_\delta;b_1,b_2\big)\ .
\end{align}

\subsubsection{Hole in a torus}\label{sec:MoebiusUDLD}

Consider the torus $T(\omega_1,\omega_2,R)$ with periods $\omega_1,\omega_2 \in \mathbb{C}$ and with a hole of size $R<\min\{|\omega_1|,|\omega_2|\}$ and boundary condition $b$. The metric on $T(\omega_1,\omega_2,R)$ is simply $\delta$ as induced from $\mathbb{C}$. Using the scaling transformation $z\mapsto \tfrac{z}{R}$ we obtain 
\begin{equation}\label{eq:ZBR}
\mathcal{A}\big(T(\omega_1,\omega_2,R)_\delta;b\big) 
=
R^{-c/6}\,
\mathcal{A}\big(T(\omega_1',\omega_2',1)_\delta;b\big) 
\ .
\end{equation}%
with $\omega'_i = \frac{\omega_i}{R}$ and the factor $R^{-c/6}$ again follows from the boundary part of the anomaly action. Note that for the modular parameter we have $\tau = \omega_2/\omega_1 =\omega'_2/\omega'_1 = \tau'$. 

\subsubsection{Upper half disc and left half disc}

Consider the closed upper unit half disc $D_+$ and the closed right unit half disc $D_r$ as given in Figure~\ref{fig:clipped-triangle-and-uniform}\,c) and d), as well as the conformal bijection 
\begin{equation}
F := \rho_{+1} : D_+ \to D_r
~~,\quad 
z \mapsto (i-z)/(i+z)
\end{equation}
as in \eqref{eq:rho+-1}. Note that in both cases, the boundary consists of two smooth pieces, one a straight line and the other a half-circle.
The map $F$ becomes an isometry $(D_+)_{e^\Omega\delta} \to (D_r)_{\delta}$ if we take $\Omega(z) = \Omega_F(z) = \log |\partial F(z)|^2$. We have
$\partial F(z) = -2i \, (z+i)^{-2}$
and, with $z=r e^{i\varphi}$,
\begin{equation}
    \Omega(r,\varphi) = \Omega_F(z) = \log \frac{4}{(1+r^2 +2r\sin(\varphi))^2} \ .
\end{equation}

In the boundary term of the anomaly action only the boundary segment of $D_+$ with $|z| = 1$ contributes, so that 
\begin{equation}
    L^\text{bdy} = \frac{1}{\pi}\int\limits_0^\pi \Omega(1,\varphi) d\varphi
    = 
    -\frac8\pi G + 2\log(2)\,,
\end{equation}
where $G=0.915..$ is Catalan's constant. 
For the bulk term we first compute
\begin{equation}
\partial_z \Omega\, \partial_{\bar z} \Omega
= \left| \frac{F''(z)}{F'(z)} \right|^2
= \left| \frac{2}{z+i} \right|^2=\frac{4}{1+r^2+2r\sin(\varphi)} \ ,
\end{equation}
and with that
\begin{align}
L^\text{bulk} &= \frac{1}{2\pi}
    \int dzd\bar{z} \partial_z \Omega \partial_{\bar z} \Omega = \frac{1}{2\pi}\int\limits_0^1 dr \int\limits_0^\pi d\varphi \,\frac{2r \cdot 4}{1+r^2+2r\sin(\varphi)} 
    \nonumber
    \\
    &= \frac{8}{\pi} G -2 \log(2)\ . 
\end{align}
The bulk and boundary term cancel each other, so that altogether
\begin{equation}
 L_{(D_+)_\delta}(\Omega)=0 \ .  
\end{equation}
Note that in this example it does not make sense to write an equality of amplitudes, as these are only defined for smooth physical boundaries. The situation treated here involves a physical boundary and a gluing boundary and  appears as an auxiliary term in the computation of the clipped triangle amplitude. 

\subsection{Corners on the physical boundary} \label{app:nonSmoothBdy}

The contribution of corners on the physical boundary to the behaviour of the free energy under a global rescaling has been derived in \cite[Eqn.\,(4.4)]{Cardy:1988tk}. We propose that the same behaviour applies for a smooth scaling field $\Omega$. This motivates the following definition for the anomaly action on a surface with piecewise smooth physical boundary, with corners at points $x_i$ on the boundary with internal opening angles $0<\alpha_i<2\pi$,
\begin{align}
    L_{\Sigma_g}(\Omega) = &~~~\,\frac{1}{2\pi} \int_\Sigma \sqrt{g} \left( R\, \Omega + \frac{1}{2}g^{\mu\nu}\partial_\mu\Omega \partial_\nu \Omega \right)\,d^2x
    \nonumber
    \\
    &+ \frac{1}{\pi} \int_{\partial\Sigma} k_g(l)\, \Omega\!\left(x^\mu(l)\right) dl 
    \nonumber
    \\
    &+ \frac1{2\pi}  \sum_i \frac{\pi^2-\alpha_i^2}{\alpha_i} \, \Omega(x_i) \ .
    \label{eq:anomaly-with-corner}
\end{align}
This agrees with the behaviour of $\zeta$-regularised determinants of Laplacians on curvilinear polygonal domains (physically a free boson with $c=1$), see \cite[Thm.\,1.6]{Aldana:2010}.

We note that this is \emph{not} the behaviour one would find when approximating a corner by a segment of a circle of radius $\epsilon$ and then taking the limit $\epsilon \to 0$. This would result in a contribution $(1-\frac{\alpha_i}\pi) \Omega(x_i)$ for each corner. But both expressions have the same leading behaviour for the angle $\alpha = \pi - \delta$, i.e.\ as one approaches a smooth boundary.

In line with the above observation, we can evaluate \eqref{eq:anomaly-with-corner} on a regular $n$-gon embedded in $\mathbb{C}$ with corner points located on the unit circle. Let us take $\Omega$ constant for simplicity. The corner angles are $\alpha = \pi - (2\pi)/n$. The only contribution to \eqref{eq:anomaly-with-corner} comes from the corner terms and is $2 (1-\frac1n)/(1-\frac2n) \, \Omega$. In the limit $n \to \infty$ this reproduces precisely the contribution from the unit disc with smooth circular boundary, where only the second term in  \eqref{eq:anomaly-with-corner} contributes.

\smallskip
The anomaly action \eqref{eq:anomaly-with-corner} with corner term still satisfies the cocycle identity \eqref{eq:LO1+O2}. Indeed, the only way the metric enters into the corner term is to determine the opening angle $\alpha$, and that angle does not change under conformal rescalings of the metric. Hence, the contribution of the corner terms to both sides of \eqref{eq:LO1+O2} is simply
\begin{align}
    &\frac1{2\pi}  \sum_i \frac{\pi^2-\alpha_i^2}{\alpha_i} \, \big( \Omega_1(x_i) + \Omega_2(x_i) \big)
\nonumber\\    
    &=~
    \frac1{2\pi}  \sum_i \frac{\pi^2-\alpha_i^2}{\alpha} \, \Omega_1(x_i)
\,+\,
    \frac1{2\pi}  \sum_i \frac{\pi^2-\alpha_i^2}{\alpha_i} \, \Omega_2(x_i) \ .
\end{align}
This implies that the anomaly action with corner term \eqref{eq:anomaly-with-corner} also satisfies the identities \eqref{eq:Lid1}--\eqref{eq:Lid3}.

\section{Examples of sums over intermediate states}\label{app:state-sum-examples}

\subsection{Sum over intermediate bulk states}\label{app:bulk-state-sum}

We did not discuss cutting along circles and summing over bulk states in detail in Section~\ref{sec:cutting-amplitude}, so we provide a sketch here. 

To define the orthonormal basis in the space of bulk states $\mathcal{H}$,
consider the unit sphere $S^2$ with its metric $g_{us}$ induced by the embedding in $\mathbb{R}^3$. We can cover it with two charts $\chi_0 : \mathbb{C} \to S^2$, $\chi_\infty : \mathbb{C} \to S^2$ obtained by stereographic projection, and which misses either the north or the south pole of $S^2$. The two charts are related by $\chi_\infty^{-1} \circ \chi_0(z) = 1/z$ on $\mathbb{C}^\times$.
The pullback metric is the same on each chart, $\chi_0^* g_{us} = \chi_\infty^* g_{us} = e^{\Omega_{us}} \delta$. Explicitly, $\Omega_{us}(z) = 2 \log \frac{2}{1+|z|^2}$, which is a multiple of the Fubini-Study metric on $\mathbb{CP}^1$ (the explicit expression will not actually be needed below).

We place field insertions on $S^2$ at the south and north pole, with local coordinates $\chi_0$ and $\chi_\infty$. 
The CFT amplitude defines the pairing 
\begin{equation}\label{eq:sphere-pairing}
(-,-) : \mathcal{H} \times \mathcal{H} \to \mathbb{C}
\quad,\quad
(\zeta,\xi) = \mathcal{A}\big(S^2;\zeta(N;\chi_{\infty}),\xi(S;\chi_{0})\big)
\ ,
\end{equation}
where $S$ and $N$ denote the poles of $S^2$.
This pairing satisfies\footnote{
    The relative sign as compared to \eqref{eq:disc-pairing-properties} arises as there the change of coordinate is $\rho_{-1}^{\,-1} \circ \rho_1 (z) = -1/z$. Note that $z \mapsto 1/z$ does not map the upper half plane to itself.
}
\begin{equation}\label{eq:sphere-pairing-properties}
(\zeta,\xi) = (\xi,\zeta)
~~ , \quad
(L_m \zeta,\xi) = (\xi,L_{-m} \zeta)
~~ , \quad
(\overline L_m \zeta,\xi) = (\xi,\overline L_{-m} \zeta)
\ .
\end{equation}
We fix an ON-basis $\{ \kappa_i \}$ in $\mathcal{H}$ with respect to this pairing.

Below we will work with the one-point compactification $\mathbb{C}_* := \mathbb{C} \cup \{\infty\}$ instead of $S^2 \subset \mathbb{R}^3$. The corresponding metric on $\mathbb{C}_*$ is $g_*(z) = e^{\Omega_{us}(z)} \delta$, and the local coordinates are 
$\varphi_0 , \varphi_\infty: \mathbb{C} \to \mathbb{C}_*$, $\varphi_0(z)=z$, $\varphi_\infty(z)=1/z$. The pairing is then
\begin{equation}
    (\zeta,\xi) = \mathcal{A}\big((\mathbb{C}_*)_{g_*};\zeta(\infty;\varphi_{\infty}),\xi(0;\varphi_{0})\big) \ .
\end{equation}
 
We can now state the behaviour of amplitudes under cutting along an circle $S$ obtained by conformally embedding an annular neighbourhood $U \subset \mathbb{C}$ of the unit circle, $\gamma : U \to \Sigma$ (cf.\ Section~\ref{sec:open-sum-states}):
\begin{equation}
     \mathcal{A}(\Sigma;\dots) 
     =
     e^{-\frac{c}{24}L_{S^2}(\Omega)} 
     \sum_{i=1}^\infty \mathcal{A}\big(\Sigma(\gamma,\Omega); \kappa_i(p_\infty,\sigma_\infty), \kappa_i(p_0,\sigma_0) , \dots \big) \ .
\end{equation}
Here, 
\begin{itemize}
    \item $\Omega : S^2 \to \mathbb{R}$ gives a smooth extension $e^{\Omega}\delta$ of the pullback metric $\gamma^*g$ from $U$ to $S^2$, 
\item $\Sigma(\gamma,\Omega)$ is the surface obtained by cutting along $S = \gamma(S^1)$ and gluing two half spheres to the cut via the map $\gamma$. 
\item $(p_\infty,\sigma_\infty)$ and $(p_0,\sigma_0)$ are the two field insertion points and local coordinates $(N,\varphi_\infty)$ and $(S,\varphi_0)$ on the two half spheres, now seen as part of the glued surface $\Sigma(\gamma,\Omega)$.
\end{itemize}

\smallskip
For the sake of this appendix, let us make the following extension of the formalism presented in Section~\ref{sec:anomaly-factor}: we only require the metric on $\Sigma$ to be piecewise smooth, in the sense that $\Sigma$ is decomposed into patches $P$ with piecewise smooth boundary, and the metric is smooth on each patch $P$ and represents the conformal class of $\Sigma$ (restricted to $P$). This is consistent due to the sum property of the anomaly action in \eqref{eq:anomaly-sum-patches}. 
The advantage is that we no longer need to choose a smooth extension of the metric: Let $\Sigma(\gamma)$ be the surface obtained by gluing the two half-spheres as above, but now equipped with three patches: The image of the original surface $\Sigma$, with its original metric, and the images of the two half-spheres, with their original metric $g_{us}$. The amplitudes on $\Sigma(\gamma,\Omega)$ and $\Sigma(\gamma)$ are related by \eqref{eq:AmplLiou}:
\begin{equation}
    \mathcal{A}\big(\Sigma(\gamma,\Omega); \dots \big) 
=
    e^{\frac{c}{24}L_{S^2}(\Omega)} 
    \mathcal{A}\big(\Sigma(\gamma); \dots \big)
\end{equation}
so that 
the sum over intermediate states above takes the simpler form
\begin{equation}
     \mathcal{A}(\Sigma;\dots) 
     =
     \sum_{i=1}^\infty \mathcal{A}\big(\Sigma(\gamma); \kappa_i(p_\infty,\sigma_\infty), \kappa_i(p_0,\sigma_0) , \dots \big) \ .
\end{equation}

The analogous reasoning for the sum over intermediate states for a cut along an arc allows to simplify \eqref{eq:cut-amplitude-via-sum} to 
\begin{equation}
     \mathcal{A}(\Sigma;\dots) 
=
     \sum_{i=1}^\infty \mathcal{A}\big(\Sigma(\gamma); \kappa^{(ab)}_i(p_-,\sigma_-), \kappa^{(ba)}_i(p_+,\sigma_+) , \dots \big) \ .
    \label{eq:cut-amplitude-via-sum-simpl}
\end{equation}
Here, $\Sigma(\gamma)$ is obtained by gluing two half discs along the cut, and they both contain their original metric $\delta$.

\subsection{Torus from sphere with two bulk insertions}

Consider the torus $T = \mathbb{C} / (2 \pi R \mathbb{Z} + i L \mathbb{Z})$ with metric $\delta$ as induced from the complex plane. We can alternatively write
\begin{equation}
    T = [0,2 \pi R] \times [0,L] / \sim
\end{equation}
where $(x,0) \sim (x,L)$ and $(0,y) \sim (2 \pi R,y)$. The conformal modulus of this torus is $\tau = i L / (2 \pi R)$.

We want to cut along $S = \{ (x,0) \,|\, x\in [0,2\pi R) \} \subset T$, and we choose, for a suitable neighbourhood $U \subset \mathbb{C}$ of $S^1$, $\gamma : U \to T$, $\gamma(z) = \frac{R}{i}\log z$. The cut surface with half spheres glued in can be described as
\begin{equation}
    T(\gamma) = D_{1+\varepsilon}^\infty \sqcup A_{R,L} \sqcup D_{1+\varepsilon}^0  \,/ \sim_\text{glue} \ ,
\end{equation}
where
\begin{itemize}
    \item $A_{R,L}$ is a cylinder of height $L$ and radius $R$, i.e.\ $A_{R,L} = (\mathbb{R} / (2 \pi R \mathbb{Z}) \times [0,L] = [0,2 \pi R] \times [0,L] / \sim$ with $(0,y) \sim (2 \pi R,y)$, but no identification between $(x,0)$ and $(x,L)$.
    The metric on the patch $A_{R,L}$ of $T(\gamma)$ is $\delta$, as induced by $\mathbb{C}$.

    \item $D_{1+\varepsilon}^0$ and $D_{1+\varepsilon}^\infty$ are each the same disc $\{ z \in \mathbb{C} \,|\, |z| < 1+\varepsilon \}$. $D_{1+\varepsilon}^0$ is glued to $A_{R,L}$ via $z \sim \gamma(z) = \frac{R}{i}\log z+iL$ for $1 \le |z| < 1+\varepsilon$, and  $D_{1+\varepsilon}^\infty$ is glued to $A_{R,L}$ via $z \sim \gamma(1/z) = -\frac{R}{i}\log z$. 

    \item The metric on each unit disc patch $D_1 \subset D_{1+\varepsilon}^{0} \subset T(\gamma)$ and $D_1 \subset D_{1+\varepsilon}^{\infty} \subset T(\gamma)$ is $g_* = e^{\Omega_{us}}\delta$.
\end{itemize}

We will now uniformise $T(\gamma)$ by mapping to the standard unit sphere $\mathbb{C}_*$ as described by the two charts $\varphi_{0,\infty} : \mathbb{C} \to \mathbb{C}_*$ in Section~\ref{app:bulk-state-sum}. 
Let $F : T(\gamma) \to \mathbb{C}_*$ be given by
\begin{align}
    w \in A_{R,L} &: F(w) = e^{i w/R} \ ,
    \nonumber
    \\
    w \in  D_{1+\varepsilon}^0 &: F(w) = e^{-L/R} w \ ,
    \nonumber
    \\
    w \in  D_{1+\varepsilon}^\infty &: F(w) = 1/w  \ .
\end{align}
These agree on the overlaps by construction. The pullback metric $g$ in the three patches is
\begin{align}
    &P_0 : |z| \le e^{-L/R} &&g_0 = e^{\Omega_0} \delta &&; ~\Omega_0 = \Omega_{us}(e^{\frac{L}{R}}z) + \tfrac{2L}R \ ,
    \nonumber
    \\
    &P_A : e^{-L/R} \le z \le 1  &&g_A = e^{\Omega_A}\delta &&; ~\Omega_A(z) = \log\left|\frac{R}{z}\right|^2 \ ,
    \nonumber
    \\
    &P_\infty : 1 \le |z| &&g_\infty = e^{\Omega_\infty}\delta &&; ~\Omega_\infty = \Omega_{us} \ .
\end{align}
We will compute the anomaly factors relative to the standard metric $g_*$ on $\mathbb{C}_*$: $g = e^{\hat\Omega} g_*$. For example, $\hat\Omega|_{P_0}(z) = \Omega_0(z) - \Omega_{us}(z)$, etc. Due to the cocycle identity \eqref{eq:LO1+O2}, on each patch we can write
\begin{equation}
    L_{P_{g_*}}(\hat\Omega|_P) 
=    
    L_{P_\delta}(\Omega_P) - L_{P_\delta}(\Omega_{us})\ .
\end{equation}
We already know that $\hat\Omega|_{P_\infty}=0$. For the other patches, we compute
\begin{align}
    L_{(P_0)_{\delta}}(\Omega_0)
    &\overset{(1)}= L_{(D_1)_\delta}(\Omega_{us}) - L_{(D_1)_\delta}(-\tfrac{2L}R)  
    = L_{(D_1)_\delta}(\Omega_{us}) + \tfrac{4L}R \ ,
\nonumber
\\
    L_{(P_A)_{\delta}}(\Omega_A)
    &\overset{(2)}= -\tfrac{2L}R  \ ,
\end{align}
where in (1) we used \eqref{eq:Lid3}, applied to the map $f : D_1 \to D_{e^{-L/R}}$, $f(z) = e^{-\frac{L}{R}} z$, and (2) is obtained by a computation analogous to the one in Section~\ref{app:cyl-ann}.
Altogether,
\begin{align}\label{eq:torus-anomaly-comp}
    L_{(\mathbb{C}_*)_{g_*}}(\hat\Omega)
    &=
    L_{(P_0)_{g_*}}(\hat\Omega|_{P_0})
    +
    L_{(P_A)_{g_*}}(\hat\Omega|_{P_A})
    +
    L_{(P_\infty)_{g_*}}(\hat\Omega|_{P_\infty})
    \nonumber
    \\
    &= \Big(L_{(D_1)_\delta}(\Omega_{us}) + \tfrac{4L}R  - L_{(P_0)_\delta}(\Omega_{us}) \Big)
    + \Big(-\tfrac{2L}R  - L_{(P_A)_\delta}(\Omega_{us})  \Big)
    + 0
        \nonumber
    \\
    &= \frac{2L}R \ .
\end{align}
In the last step we used that $D_1 = P_0 \cup P_A$.

Combining all this, we can compute the amplitude on the torus as
\begin{align}     
\mathcal{A}(T) 
     &=
     \sum_{i=1}^\infty \mathcal{A}\big(T(\gamma); \kappa_i(p_\infty,\sigma_\infty), \kappa_i(p_0,\sigma_0) \big)
\nonumber
\\
     &=
     \sum_{i=1}^\infty \mathcal{A}\big((\mathbb{C}_*)_g; \kappa_i(\infty,\varphi_\infty) , \kappa_i(0,F \circ \varphi_0) \big) 
\nonumber
\\
     &= e^{\frac{c}{24}L_{(\mathbb{C}_*)_{g_*}}\!(\hat\Omega)} 
     \sum_{i=1}^\infty 
     e^{-\frac{L}{R}(h_i+\bar h_i)}
     \mathcal{A}\big((\mathbb{C}_*)_{g_*}; \kappa_i(\infty,\varphi_\infty) , \kappa_i(0,\varphi_0) \big)
\nonumber
\\
     &=  
     \sum_{i=1}^\infty e^{-\frac{L}{R}(h_i + \bar h_i - \frac{c}{12})} \ .
\end{align}
In the third line we used \eqref{eq:rule-ii-local-coord} in the form $\kappa_i(0,F \circ \varphi_0) = (e^{-\frac{L}{R}(L_0+\bar L_0)}\kappa_i)(0,\varphi_0)$.

Since $q = e^{2 \pi i \tau} = e^{-L/R}$, this is of course the expected result for the torus partition function. We stress that the contribution $\frac{c}{12}$ was obtained by an anomaly factor computation, and not by transporting the Hamiltonian from the torus to the plane as usually done in textbooks. That the two computations agree provides a non-trivial check.

\subsection{Cylinder from two discs with a bulk insertion}\label{sec:cylinder-closed-factorisation}

Let $\mathrm{Cyl}(R,L)$ and $\mathrm{Ann}(r_+,r_-)$ be as in Section~\ref{app:cyl-ann}. By \eqref{eq:cylinder-to-annulus} we  have
\begin{equation}\label{eq:cylinder-2disc-step1}
    \mathcal{A}\big(\mathrm{Cyl}(R,L)_\delta;a,b\big) 
= e^{-\frac{c L}{12 R}} \mathcal{A}\big(\mathrm{Ann}(r_+,r_-)_\delta;a,b\big) \ ,
\quad r_\pm= e^{\pm \frac{ L}{2R}}\ .
\end{equation}
On the right hand side, we will cut along the unit circle $S^1$, i.e.\ for $\gamma$ we will use the identity map. This cuts the annulus into two pieces, and after gluing in the half-spheres, both are conformally equivalent to discs:
\begin{itemize}
    \item 
$\mathrm{Ann}(r_+,1) \subset \mathrm{Ann}(r_+,r_-)$: After gluing the half-sphere, we get $(D_{r_+})_{g_+}$, were the metric $g_+(z)$ is $\delta$ on the patch $|z|\ge 1$ and $g_*(z)$ on the patch $|z| \le 1$.

\item $\mathrm{Ann}(1,r_-) \subset \mathrm{Ann}(r_+,r_-)$: The half sphere $\mathbb{C}_* \setminus D_{<1}$ (with $D_{<1}$ the open unit disc) is glued via the identity map along $S^1$. We apply the conformal transformation $f(z) = 1/z$ to get an isometry
\begin{equation}\label{eq:ann-closed-fact-aux1}
    \mathrm{Ann}(1,r_-)_\delta
    \sqcup (\mathbb{C}_* \setminus D_{<1})_{g_*}
    ~\cong~ (D_{r_+})_{g_-}
\end{equation}
where $g_-(z)$ is $(f^* \delta)(z) = |z|^{-4} \delta$ on the patch $|z| \ge 1$, and $g_*(z)$ on the patch $|z| \le 1$.
\end{itemize}
Continuing from \eqref{eq:cylinder-2disc-step1}, we get
\begin{align}
    \eqref{eq:cylinder-2disc-step1}
    &=
 e^{-\frac{c L}{12 R}} \sum_i 
 \mathcal{A}\big(\mathrm{Ann}(r_+,r_-)(\gamma);a,b;\kappa_i(p_\infty,\sigma_\infty), \kappa_i(p_0,\sigma_0) \big)
\nonumber\\
&=
 e^{-\frac{c L}{12 R}} \sum_i 
 \mathcal{A}\big((D_{r_+})_{g_-};a;\kappa_i(0,\varphi_0) \big)
 \mathcal{A}\big((D_{r_+})_{g_+};b;\kappa_i(0,\varphi_0) \big) \ .
\end{align}
To proceed, we apply a Weyl transformation to change the metric $g_-$ to $g_+$. Namely, $g_- = e^{\Omega_-} g_+$, with $\Omega_-(z) = 2 \log|z|^{-2}$ in the patch $|z|\ge 1$ and $\Omega_-(z)=0$ in the patch $|z| \le 1$.

A computation analogous to \eqref{eq:ann-Lbulk} and \eqref{eq:ann-Lbnd}, but for $\mathrm{Ann}(r_+,1)$, gives
\begin{equation}
    L_{\mathrm{Ann}(r_+,1)_\delta}(\Omega_-)
    = L^\text{bulk} + L^\text{bdy} = \frac{4L}{R} - \frac{4L}{R}  = 0\ ,
\end{equation}
and so simply
\begin{equation}
    \mathcal{A}\big((D_{r_+})_{g_-};a;\kappa_i(0,\varphi_0)
    \big)
    =
    \mathcal{A}\big((D_{r_+})_{g_+};a;\kappa_i(0,\varphi_0)
    \big)
    \ .
\end{equation}
Next we relate the amplitude on $(D_{r_+})_{g_+}$ to that on $(D_1)_{g_*}$ by a rescaling and a Weyl transformation. Let $f(z) = r_+ z$ and write $f^*g_+ = e^{\Omega} \delta$ with $\Omega(z) = \Omega_A(z) := 2 \log r_+$ on $P_A := \mathrm{Ann}(1,r_-) \subset D_1$ and $\Omega(z) = \Omega_0(z) := \Omega_{us}(r_+ z) + 2 \log r_+$ on $P_0 := D_{r_-} \subset D_1$. Set $\hat\Omega = \Omega - \Omega_{us}$ and compute as in \eqref{eq:torus-anomaly-comp} (but replacing $R \leadsto 2R$ relative to that computation and noting the different expression for $\Omega_A$),
\begin{align}
    L_{{(D_1)}_{g_*}}(\hat\Omega)
    &=
    L_{(P_0)_{g_*}}(\hat\Omega|_{P_0})
    +
    L_{(P_A)_{g_*}}(\hat\Omega|_{P_A})
    \nonumber
    \\
    &=
    \big(
        L_{(P_0)_{\delta}}(\Omega_0)
        -
        L_{(P_0)_{\delta}}(\Omega_{us})
    \big)
    +
    \big(
        L_{(P_A)_{\delta}}(\Omega_A)
        -
        L_{(P_A)_{\delta}}(\Omega_{us})
    \big)
    \nonumber
    \\
    &= \big(L_{(D_1)_\delta}(\Omega_{us}) + \tfrac{2L}R  - L_{(P_0)_\delta}(\Omega_{us}) \big)
    + \Big(0  - L_{(P_A)_\delta}(\Omega_{us})  \big)
        \nonumber
    \\
    &= \frac{2L}R \ .
\end{align}
The field insertion $\kappa_i(0,\varphi_0)$ on $D_{r_+}$ becomes $\kappa_i(0,\frac{1}{r_+}\varphi_0)$ on $D_{1}$, which by \eqref{eq:rule-ii-local-coord} is equal to $(e^{-\frac{L}{2R}(L_0+\bar L_0)}\kappa_i)(0,\varphi_0)$. 
This results in
\begin{equation}\label{eq:ann-closed-fact-aux2}
    \mathcal{A}\big((D_{r_+})_{g_+};b;\kappa_i(0,\varphi_0)\big)
    =
    e^{-\frac{L}{2R}(h_i+\bar h_i-\frac{c}{6})}
    \mathcal{A}\big((D_{1})_{g_*};b;\kappa_i(0,\varphi_0)
    \big)
    \ .
\end{equation}
Finally,
\begin{align}\label{eq:cylinder-bulk-channel}
    &\mathcal{A}\big(\mathrm{Cyl}(R,L)_\delta;a,b\big) 
\nonumber \\
& 
= 
\sum_i e^{-\frac{L}{R}(h_i + \bar h_i - \frac{c}{12})} 
\mathcal{A}\big((D_{1})_{g_*};a;\kappa_i(0,\varphi_0)
    \big)
\mathcal{A}\big((D_{1})_{g_*};b;\kappa_i(0,\varphi_0)
    \big)    
 \ .
\end{align}
The disc amplitude is non-zero only for $h_i = \bar h_i$.
For $\tau = i \frac{L}{\pi R}$,
the prefactor then takes the form $e^{2 \pi i \tau (h_i - \frac{c}{24})}$.

\subsection{Cylinder from a disc with two boundary insertions}

For this section we represent $\mathrm{Cyl}(R,L)$ as $[0,2 \pi R] \times [0,L] / {\sim}$ with $(0,y) \sim (2 \pi R,y)$. The two boundary components are $[0,2 \pi R] \times \{0\}$ and $[0,2 \pi R] \times \{L\}$.

It will be convenient to work with the one-point compactification of the closed upper half plane $\mathbb{H}_* := \mathbb{H} \cup \infty$. 
The map $\rho_{+1} : (D_1)_\delta \to (\mathbb{H}_*)_{g_{ud}}$ from \eqref{eq:rho+-1} is an isometry for $g_{ud}(z) = e^{\Omega_{ud}(z)} = |\partial_z (\rho_{+1}^{\,-1})(z)|^2$. By invariance under isometries, we can write the boundary pairing \eqref{eq:disc-pairing} as
\begin{align}
    &\mathcal{A}\big((D_1)_\delta;a,b;\zeta(-1;\rho_{-1}),\xi(1;\rho_{1})\big)
    \nonumber \\
    & \quad =
    \mathcal{A}\big((\mathbb{H}_*)_{g_{ud}};a,b;\zeta(\infty;-\varphi_{\infty}),\xi(1;\varphi_{0})\big) \ ,
\end{align}
where $\varphi_0(z)=z$ as before and $-\varphi_\infty(z) = -1/z$.

We cut $\mathrm{Cyl}(R,L)$ along $\{0\} \times [0,L]$, that is, we choose the map $\gamma$ from a neighbourhood of the upper half of the unit circle to $\mathrm{Cyl}(R,L)$ to be 
$\gamma(z) = \frac{L}{\pi} \log(z)$. Thus the cut $\{0\} \times [0,L]$ is parametrised by $\gamma(e^{i \alpha})$ with $\alpha \in [0,\pi]$. We can map the cut surface $\mathrm{Cyl}(R,L)(\gamma)$ with glued-in half discs isometrically to $(\mathbb{H}_*)_{g}$, where $g$ is defined on three patches analogously to the torus computation. Namely, for $z \in \mathbb{H}_*$,
\begin{align}
    &Q_0 : |z| \le 1 &&g_0 = e^{\Omega_0} \delta &&; ~\Omega_0 = \Omega_{ud} \ ,
    \nonumber
    \\
    &Q_A : 1 \le z \le e^{2 \pi^2 R/L}  &&g_A = e^{\Omega_A}\delta &&; ~\Omega_A(z) = \log\left|\tfrac{L}{\pi z}\right|^2 \ ,
    \nonumber
    \\
    &Q_\infty :  e^{2 \pi^2 R/L} \le |z| &&g_\infty = e^{\Omega_\infty}\delta &&; ~\Omega_\infty =  \Omega_{ud}(e^{-2 \pi^2 R/L}z) - 4 \pi^2 \tfrac{R}L  \ .
\end{align}
Define in addition $Q_1 := \{ z \in \mathbb{H}_* |\, |z| \ge 1 \}$.
We will need
\begin{align}
    L_{(Q_A)_\delta}(\Omega_A)
    &= L^\text{bulk} + L^\text{bdy}
    = \tfrac{2 \pi^2 R}L - \tfrac{4 \pi^2 R}L = -\tfrac{2 \pi^2 R}L
    \ ,
    \nonumber
    \\
    L_{(Q_\infty)_\delta}(\Omega_\infty)
    &= 
    L_{(Q_1)_\delta}(\Omega_{ud})
    -
    L_{(Q_1)_\delta}(\tfrac{4 \pi^2 R}L)
    =
    L_{(Q_1)_\delta}(\Omega_{ud})
    +
    \tfrac{4 \pi^2 R}L
    \ .
\end{align}
We set $g = e^{\hat\Omega} g_{ud}$ and compute the anomaly action
\begin{align}
    &L_{(\mathbb{H}_*)_{g_{ud}}}(\hat\Omega)
    =
    L_{(Q_0)_{g_{ud}}}(\hat\Omega|_{Q_0})
    +
    L_{(Q_A)_{g_{ud}}}(\hat\Omega|_{Q_A})
    +
    L_{(Q_\infty)_{g_{ud}}}(\hat\Omega|_{Q_\infty})
    \nonumber
    \\
    &= 
    0 + 
    \Big(-\tfrac{2 \pi^2 R}L  - L_{(Q_A)_\delta}(\Omega_{ud})  \Big)
    +
    \Big(L_{(Q_1)_\delta}(\Omega_{ud})
    +
    \tfrac{4 \pi^2 R}L 
    - L_{(Q_\infty)_\delta}(\Omega_{ud}) \Big)
        \nonumber
    \\
    &= \tfrac{2 \pi^2 R}L \ .
\end{align}
Using the variant \eqref{eq:cut-amplitude-via-sum-simpl} of the cutting relation, we get
\begin{align}
    &\mathcal{A}\big(\mathrm{Cyl}(R,L)_\delta;a,b\big) 
\nonumber \\
& 
= 
\sum_j \mathcal{A}\big(\mathrm{Cyl}(R,L)(\gamma);a,b; \kappa^{(ab)}_j(p_-,\sigma_-), \kappa^{(ba)}_j(p_+,\sigma_+) \big) 
\nonumber \\
& 
= 
\sum_j \mathcal{A}\big((\mathbb{H}_*)_{g}; \kappa^{(ab)}_j(\infty,\tilde\varphi), \kappa^{(ba)}_j(0, \varphi_0) \big) 
\nonumber \\
& 
= 
\sum_j 
e^{-2 \pi^2 \frac{R}{L} (h_j- \frac{c}{24})}
 \ .
\end{align}
where the coordinate $\tilde\varphi$ at $\infty$ is $\tilde\varphi(z) = - e^{2 \pi^2 R/L} z^{-1} = - (e^{-2 \pi^2 R/L}z)^{-1}$ so that
\begin{equation}
\kappa^{(ab)}_j(\infty,\tilde\varphi_\infty)
\,=\,
e^{-2 \pi^2 \frac{R}{L} h_i}
\kappa^{(ab)}_j(\infty,-\varphi_\infty)
\ .
\end{equation}
For $\tau = i \frac{L}{\pi R}$,
the summand takes the form $e^{2 \pi i (-1/\tau) (h_j - \frac{c}{24})}$, and hence is related to \eqref{eq:cylinder-bulk-channel} by a modular transformation, as it should be.

\subsection{Radius dependence of boundary states}
\label{eq:radius-bnd-state}

The computation in Section~\ref{sec:cylinder-closed-factorisation} implicitly already contained the boundary state and its radius dependence, we just have to extract the relevant pieces. To make comparison easier we use $r_-$ for the radius of the boundary circle.

Consider the surface $H_{r_-} := \mathbb{C}_* \setminus D_{<r_-}$, i.e.\ the Riemann sphere with an open disc of radius $r_-$ removed. We will take $r_-<1$ but the computation below could be adapted for any value of $r_-$.
On $H_{r_-}$ we consider the metric
$g_*$ for $|z| \ge 1$, and the flat metric $\delta$ for $r_- \le |z| \le 1$. We place a field insertion $\phi(\infty,\varphi_\infty)$ at $\infty$.

Our aim is to replace the unit disc with a hole of radius $r_-$ by a unit disc with a sum over field insertions at the origin -- the boundary state -- in such a way that the amplitude with respect to any field placed at $\infty$ does not change.

To this end, we cut $H_{r_-}$ along the unit circle $\gamma = S^1$. This cuts $H_{r_-}$ into two pieces, and after gluing in the half-spheres, one piece  becomes a 2-sphere and the other is again $H_{r_-}$ with its original metric,
\begin{equation}
    H_{r_-}(\gamma) = (\mathbb{C}_*)_{g_*} \sqcup H_{r_-}
\end{equation}
In terms of amplitudes, with boundary condition $b$ placed on the circle of radius $r_-$,
\begin{equation}\label{eq:bnd-state-aux1}
    \mathcal{A}\big(H_{r_-};b;\phi(\infty,\varphi_\infty)\big) 
= 
\sum_i 
\mathcal{A}\big((\mathbb{C}_*)_{g_*};\phi(\infty,\varphi_\infty),\kappa_i(0,\varphi_0)
    \big)
\mathcal{A}\big(H_{r_-};b;\kappa_i(\infty,\varphi_\infty)
    \big)    
 \ .
\end{equation}
This identity is of course trivial, the first factor in the sum just produces the coefficients to write $\phi$ as a linear combination of the $\kappa_i$. But it is still useful in order to exhibit the boundary state.

As in \eqref{eq:ann-closed-fact-aux1} we use the isometry $H_{r_-} \cong (D_{r_+})_{g_-}$. The computation leading to \eqref{eq:ann-closed-fact-aux2} now gives
\begin{equation}
    \mathcal{A}\big(H_{r_-};b;\kappa_i(\infty,\varphi_\infty)
    \big) 
    = (r_-)^{h_i+\bar h_i-\frac{c}{6}}
    \mathcal{A}\big((D_{1})_{g_*};b;\kappa_i(0,\varphi_0)
    \big)
    \ .
\end{equation}
On $\mathbb{C}_*$ we change the metric to $g$ which is $g_*$ for $|z| \ge 1$ and $\delta$ for $|z| \le 1$, i.e.\ we want the flat metric on the unit disc. Then 
\begin{equation}
\mathcal{A}\big((\mathbb{C}_*)_{g_*};\phi(\infty,\varphi_\infty),\kappa_i(0,\varphi_0)
    \big)
= e^A\,   
\mathcal{A}\big((\mathbb{C}_*)_{g};\phi(\infty,\varphi_\infty),\kappa_i(0,\varphi_0)
    \big)
\end{equation}
for some constant $A$ which we could compute, but of which we only need to know that it does not depend on $r_-$.

So far we have rewritten \eqref{eq:bnd-state-aux1} as
\begin{align}
\mathcal{A}\big(H_{r_-};b;\phi(\infty,\varphi_\infty)\big) 
&= 
\sum_i 
\mathcal{A}\big((\mathbb{C}_*)_{g};\phi(\infty,\varphi_\infty),\kappa_i(0,\varphi_0)
    \big)
    \nonumber\\
&    \hspace{5em}
\times ~
(r_-)^{h_i+\bar h_i-\frac{c}{6}}
\underbrace{e^A \mathcal{A}\big((D_{1})_{g_*};b;\kappa_i(0,\varphi_0)
    \big)}_{=: \, c_i} 
\ .
\end{align}

Altogether, 
this shows that if a surface $\Sigma$ contains a region isometric to $(D_1 \setminus D_{<r})_\delta$ for some $r<1$ and boundary condition $b$ on the circle of radius $r$, we can replace this region with $(D_1)_\delta$ and insert the sum (i.e.\ the boundary state)
\begin{equation}
    \sum_i r^{h_i+\bar h_i-\frac{c}{6}} \, c_i \, \kappa_i(0,\varphi_0) \ .
\end{equation}
In this expression, neither $c_i$ nor $\kappa_i$ depend on $r$. It is understood that the sum is taken outside the amplitude.

\section{Numerical integration for the anomaly factors}
\label{app:numerics}

In this appendix we briefly describe the approach to obtain the anomaly action \eqref{eq:triangle-anomaly-factor} numerically. 


\subsubsection*{Computing $L_{\widetilde T_\triangleleft}(\Omega_E)$}

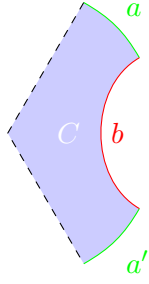
\begin{figure}
    \centering
    \begin{tikzpicture}[scale=2]
        \fill[blue!20] (.5,.866) -- (0,0) -- (.5,-.866) to[out=30,in=-120] (.866,-.5) to[out=-210,in=210] (.866,.5) to[out=120,in=-30] (.5,.866);
        \node[white] at (.4,0) {$C$};
        \draw[dashed]  (.5,.866) -- (0,0) -- (.5,-.866);
        \draw[green] (.5,.866) to[out=-30,in=120] node[above right] {$a$} (.866,.5);
        \draw[green] (.5,-.866) to[out=30,in=-120] node[below right] {$a'$} (.866,-.5);
        \draw[red] (.866,.5) to[out = 210, in = -210] node[right] {$b$} (.866,-.5);
    \end{tikzpicture}
    \caption{The region $C$ and its boundary components $a$, $a'$ and $b$ we have to integrate over.
    The shape of $b$ and hence of $C$ is obtained numerically. The area integral on $C$ and the line integrals along $a$, $a'$ and $b$ are then also performed numerically.}
    \label{fig:int_region}
\end{figure}

The integration region for the integral is shown in Figure~\ref{fig:int_region} with the bulk called $C$, the physical boundary regions on the unit circle are $a$ and $a'$, and the state boundary is $b$. The latter is the preimage of the cut between the boundaries under the uniformisation map $E$  (see Figure~\ref{fig:clipped-triangle-and-uniform} and \cite[Fig.\,15]{Brehm:2021wev}). Its shape, in particular, depends on the parameter $t$ and we determine it numerically. This means that for every value of the parameter $t$ we approximate $b$ by a collection of consecutive points $\{z_j\, | \, j= 0,1,\dots, N\}$. Then for every point $z_j = x_j + i y_j $ with $0<j<N$ we can approximate the tangent and normal vectors by 
\begin{equation}
    t_j = \frac{1}{|z_{j}-z_{j-1}|} \left(\begin{matrix}
        x_j -x_{j-1}\\ y_j - y_{j-1}
    \end{matrix}\right)
    \quad , \quad
    n_j = \frac{1}{|z_{j}-z_{j-1}|} \left(\begin{matrix}
        y_{j-1} - y_{j} \\  x_j -x_{j-1}
    \end{matrix}\right) \ ,
\end{equation}
s.t. the curvature can be written as
\begin{equation}
    k_j =   \frac{2\,t_{j+1}\cdot n_j}{|z_{j+1}-z_{j-1}|}\,.
\end{equation}
This allows to compute the boundary integral along the preimage of the cut numerically. 

The region $C$ depends on the parameter $t$ as well. Again, we obtain its shape  and perform the integration numerically. We want to mention that for small $t$, i.e when the holes get close to touch and the cut $b$ becomes small, the integrand becomes very large close to the cut. This has to be considered when performing the numerical integration. 

\subsubsection*{Computing $L_{D^+}(\Omega_{\phi_0})$}

To compute the bulk integral we chose to expand $\phi_0(z)$ up to order 30. This can be done relatively quickly as $\phi^{-1}_0$ has the explicit expression \eqref{eq:phi0-expansion} and its expansion around $u=0$ is easy to compute.
The integration on the upper unit disk is then straightforward and can be done analytically. Note that for small $t$, i.e. when $R\to d/2$, the radius of convergence of $\phi_0$ approaches $1$ and the error from the finite expansion becomes larger.

\newpage

\newcommand\arxiv[2]      {\href{http://arXiv.org/abs/#1}{#2}}
\newcommand\jdoi[2]        {\href{http://dx.doi.org/#1}{#2}}

\end{document}